\documentclass[prd,aps,twocolumn,10pt,letterpaper,superscriptaddress]{revtex4-2}
\usepackage[utf8]{inputenc}
\usepackage[english]{babel}
\usepackage{bm}
\usepackage{graphicx}
\usepackage{amsmath}
\usepackage{amssymb}
\usepackage{amsfonts}
\usepackage{xcolor}
\usepackage{braket}
\usepackage{hyperref}
\usepackage{multirow}
\usepackage{float}
\hypersetup{
    pdfstartview={FitH},    
    colorlinks=true,       
    linkcolor=blue,          
    citecolor=blue,        
    filecolor=blue,      
    urlcolor=blue           
}

\begin{document}

\title{Influence of the nuclear magnetic field on electron-positron pair production in low-energy heavy-nuclei collisions}

\author{N. K. Dulaev}
\email{st069071@student.spbu.ru}
\affiliation{Department of Physics,
St.~Petersburg State University, 7-9 Universitetskaya nab., St.~Petersburg 
199034, Russia}

\author{D. A. Telnov}
\email{d.telnov@spbu.ru}
\affiliation{Department of Physics,
St.~Petersburg State University, 7-9 Universitetskaya nab., St.~Petersburg 
199034, Russia}

\author{V. M. Shabaev}
\email{v.shabaev@spbu.ru}
\affiliation{Department of Physics,
St.~Petersburg State University, 7-9 Universitetskaya nab., St.~Petersburg 
199034, Russia}
\affiliation{Petersburg Nuclear Physics Institute named by B.~P.~Konstantinov of National Research Center ''Kurchatov Institute'', Orlova roscha 1, 188300 Gatchina, Leningrad region$,$ Russia}

\author{Y.~S.~Kozhedub}
\affiliation{Department of Physics,
St.~Petersburg State University, 7-9 Universitetskaya nab., St.~Petersburg 
199034, Russia}

\author{X.~Ma}
\affiliation{Institute of Modern Physics,
Chinese Academy of Sciences, 730000 Lanzhou, China}

\author{I. A. Maltsev}
\affiliation{Department of Physics,
St.~Petersburg State University, 7-9 Universitetskaya nab., St.~Petersburg 
199034, Russia}

\author{R. V. Popov}
\affiliation{Petersburg Nuclear Physics Institute named by B.~P.~Konstantinov of National Research Center ''Kurchatov Institute'', Orlova roscha 1, 188300 Gatchina, Leningrad region$,$ Russia}
\affiliation{Department of Physics,
St.~Petersburg State University, 7-9 Universitetskaya nab., St.~Petersburg 
199034, Russia}

\author{I. I. Tupitsyn}
\affiliation{Department of Physics,
St.~Petersburg State University, 7-9 Universitetskaya nab., St.~Petersburg 
199034, Russia}

\begin{abstract}
Total and energy–angle differential probabilities of positrons created in slow collisions of two identical nuclei are calculated within relativistic two-center approach. The time-dependent Dirac equation is solved in the rotating frame using the generalized pseudospectral method in modified prolate spheroidal coordinates. The magnetic interaction induced by the motion of the nuclei is included in the Hamiltonian. The rotational coupling term is also taken into account. Angle-integrated and angle-resolved energy spectra of the emitted positrons are calculated by projecting the propagated wave function onto positive-energy plane-wave states. Our results show that the magnetic interaction leads to a slight increase in the critical internuclear distance and enhances the total positron yield by up to several percent. However, it does not qualitatively alter the energy or angular distributions of emitted positrons. The angular distributions remain nearly isotropic. The characteristic supercritical regime signatures, found in previous works, are preserved.
\end{abstract}

\maketitle

\section{Introduction}
For nearly a hundred years it has been known that the presence of a sufficiently strong electromagnetic field  requires a nonperturbative quantum electrodynamics (QED) framework to accurately describe the resulting physical phenomena. In particular, if the field strength exceeds the Schwinger limit of $10^{16}$ V/cm, the vacuum becomes unstable and undergoes spontaneous electron-positron pair production. This process is referred to as spontaneous vacuum decay \cite{Greiner_1985_Quantum}. The experimental observation of vacuum decay would significantly advance understanding of vacuum structure and strong-field QED. Naturally, high-intensity laser fields are considered as a promising tool for such experiments. However, despite remarkable progress in laser technology in recent years \cite{Fedotov_2023_Advances, Kostyukov_2024_Physics}, the attainable field strengths still fall several orders of magnitude short of the Schwinger limit.

An alternative approach involves exploiting strong electromagnetic field in the vicinity of highly charged atomic nuclei. As first pointed out by  Pomeranchuk and Smorodinsky \cite{Pomeranchuk_1945}, the Coulomb field of a finite-radius nucleus with the charge greater than some critical value $Z_{\mathrm{cr}}$ is strong enough for the lowest bound-state energy level to reach the onset of the negative-energy Dirac continuum. Subsequent theoretical investigations by Soviet and German physicists \cite{Gershtein_1969, Pieper_1969_Interior, Popov_1970_1, *Popov_1970_2, *Popov_1970_3, *Popov_1971_1,  Zeldovich_1971_Electronic, Muller_1972_Solution, *Muller_1972_Electron, Mur_1976, Popov_1976,  Muller_1976_Positron, Reinhardt_1977_Quantum, Soff_1977_Shakeoff, Rafelski_1978_Fermions, Popov_1979, Greiner_1985_Quantum} demonstrated that for $Z > 173$ the lowest unoccupied bound state enters the negative-energy continuum becoming an unstable resonance state. This state may decay via spontaneous creation of electron-positron pairs when the electron fills the vacancy and the positron escapes. Since stable nuclei with $Z > 173$ are not currently available, heavy-ion collisions involving nuclei with a combined charge $Z_1+Z_2>Z_{\mathrm{cr}}$ (e.g., U-U collisions) have been proposed to probe the supercritical regime where spontaneous electron-positron pair creation takes place. However, extensive studies by the Frankfurt group in the 1970s and 1980s \cite{Smith_1974_Induced, Reinhardt_1981_Theory, Muller_1988_Positron, Rafelski_1978_Fermions, Greiner_1985_Quantum, Bosch_1986_Positron, Muller_1994_Electron, Reinhardt_2005_Supercritical, Rafelski_2017_Probing} revealed that spontaneous pair creation in such collisions is strongly masked by dynamical pair production, which arises from the time dependence of the electromagnetic field during the collision. The spontaneous and dynamical contributions add up coherently and are, in practice, indistinguishable. This fundamental difficulty hinders direct experimental detection of spontaneous vacuum decay, and dedicated experiments conducted at GSI (Darmstadt) did not show signatures of spontaneous pair production \cite{Greiner_1985_Quantum, Muller_1994_Electron, Reinhardt_2005_Supercritical, Rafelski_2017_Probing}.

With the advent of new experimental capabilities at upcoming facilities in Germany (GSI/FAIR)~\cite{Gumberidze_2009_X, Lestinsky_2016_Physics, Lestinsky_2022_GSI}, China (HIAF)~\cite{Ma_2017_HIAF, Zhou_2022_HIAF}, and Russia (NICA)~\cite{Akopian_2015_Layout, Trubnikov_2023_NICA}, theoretical investigations of spontaneous vacuum decay have gained renewed relevance. The superctritical resonance was studied in Refs. \cite{Ackad_2007_Numerical, Godunov_2017_Resonances, Ackad_2007_Supercritical, Marsman_2011_Calculation}. The QED-vacuum polarization effects in supercritical Coulomb fields were analyzed in Refs.~\cite{Grashin_2022_Vacuum, Krasnov_2022_Non}. The instability of the electron-positron vacuum within a semiclassical framework was addressed in Ref.~\cite{Voskresensky_2021_Electron}.

\begin{figure}[]
   \includegraphics[width=\columnwidth]{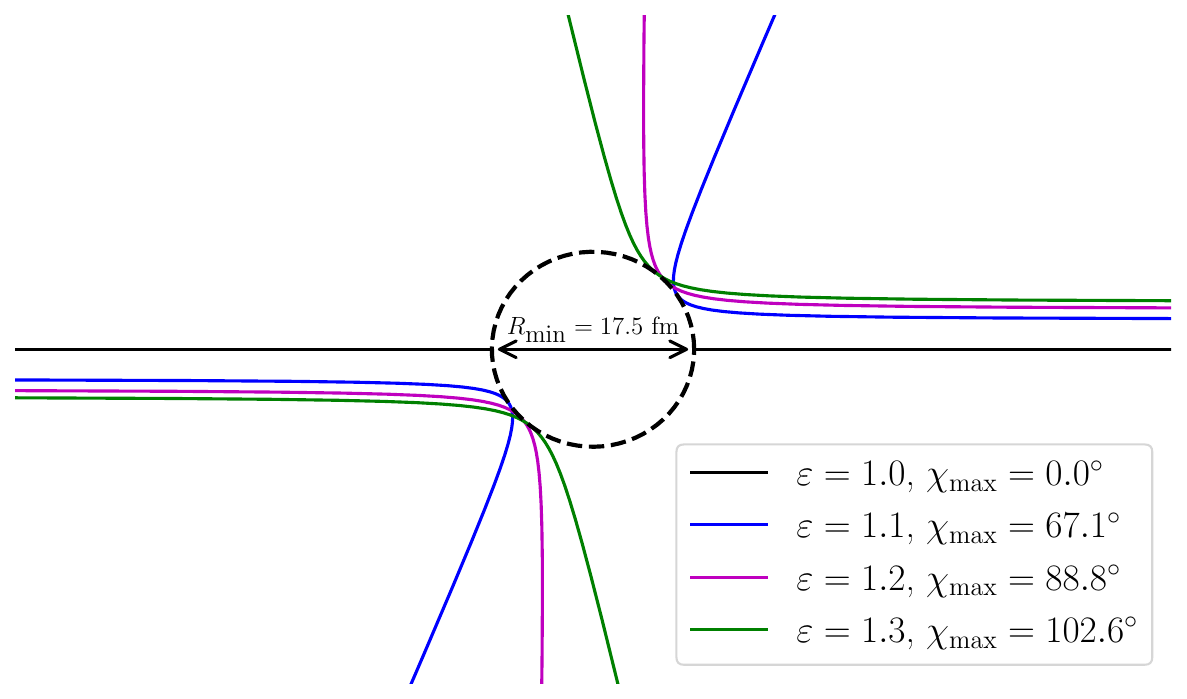}
    \caption{Schematic illustration of the nuclear collision. Solid lines represent the trajectories of the nuclei. The dashed black circle indicates the distance of closest approach, $R_{\mathrm{min}} = 17.5$~fm. The angle $\chi_{\mathrm{max}}$ denotes the final rotation of the internuclear axis after the collision.}
    \label{fig:trj}
\end{figure}

Over the past decade, our research group has extensively studied both spontaneous and dynamical mechanisms of electron-positron pair creation in collisions of heavy nuclei. The properties of supercritical resonances formed in such collisions were investigated in Refs.~\cite{Maltsev_2020_Calculation, Telnov_positron_2024}. Within the monopole approximation, the contributions of spontaneous and dynamical channels to pair production were evaluated in Refs.~\cite{Maltsev_2015_Electron, Maltsev_2019_How, Popov_2020_How}; extensions beyond the monopole approximation were studied in Refs.~\cite{Maltsev_2017_Pair, Maltsev_2018_Electron, Popov_2018_One, Popov_2023_Spontaneous}. This series of works identified novel signatures of spontaneous vacuum decay in supercritical collisions. In particular, a class of nuclear trajectories was employed in which the minimal internuclear distance $R_{\mathrm{min}}$ is fixed while the scaled collision energy parameter $\varepsilon = E / E_0$  varies (with $E$ being the collision energy and $E_0$ the energy of the head-on collision, $\varepsilon \geq 1$). A drawing of such nuclei trajectories is presented in Fig.~\ref{fig:trj}. In Refs.~\cite{Maltsev_2019_How, Popov_2020_How}, it was found that for collisions with $Z_1+Z_2>Z_{\mathrm{cr}}$, the total electron-positron pair creation probability increased as $\varepsilon \rightarrow 1$, while the dynamical channel contribution diminished due to slower nuclear motion. Such an increase in the total positron creation probability with decreasing collision energy signals a growing contribution from the spontaneous vacuum decay. The same trend was observed near the maximum in the energy distribution of emitted positrons. Further calculations beyond the monopole approximation confirmed these findings, revealing similar signatures of spontaneous vacuum decay in the angular distributions of emitted positrons~\cite{Dulaev_2024_Angular, Dulaev_Three_2025}. In Refs.~\cite{Dulaev_Three_2025, Popov_2025_Influence}, it is shown that rotation of the internuclear axis in the course of collision, which was neglected in earlier works, has a minor effect on the electron-positron pair creation and does not alter the identified signatures of the supercritical regime. Additional studies in the monopole approximation involving collisions of partially stripped ions demonstrate that electron-positron pair creation is strongly suppressed and signatures of the supercritical regime disappear if low-lying bound-state energy levels are filled with electrons~\cite{Dulaev_2025_The}.

Besides electric fields, colliding nuclei also produce magnetic fields, which cannot be removed by transition to any inertial frame of reference~\cite{Rafelski_1976_Magnetic}. The role of the magnetic field of the nuclei in heavy-ion collisions and its influence on the process of electron-positron pair creation was examined in Refs.~\cite{Greiner_1978_Delta, Rafelski_1976_Magnetic, Soff_1981_Spin, Soff_1s_1981, Soff_1988_Positron}. In particular, it was demonstrated that, for slow Pb-Pb collisions, the magnetic field had a negligible effect on both the total pair creation probability and the energy spectra of emitted positrons~\cite{Soff_1988_Positron}. However, it is unclear whether this conclusion is still valid for collisions involving nuclei with higher charge numbers, such as U-U collisions, where the magnetic field strength is significantly greater. Furthermore, the question remains whether the magnetic field might modify the angular distribution of the emitted positrons, which was found to be nearly isotropic in the calculations neglecting the influence of the magnetic field~\cite{Dulaev_2024_Angular, Dulaev_Three_2025}.

The present work aims to investigate the influence of the nuclear magnetic field on the positron creation in low-energy heavy-ion collisions. In contrast to our earlier studies~\cite{Dulaev_2024_Angular, Dulaev_Three_2025}, the magnetic interaction term is included in the Hamiltonian. The time-dependent Dirac equation is solved in the rotating frame of reference using the generalized pseudospectral method in modified prolate spheroidal coordinates. The total probabilities, energy spectra as well as angular spectra of outgoing positrons are obtained by projecting the propagated wave function onto relativistic plane waves.

The paper is organized as follows. In Sec.~\ref{methods}, we describe the theoretical framework and numerical methods used in the present work. Section~\ref{results} presents the results of the numerical calculations. In Sec.~\ref{sec:magnetic_splitting}, we analyze the magnetic splitting of quasimolecular energy levels and compare our results with previous findings from the literature. Section~\ref{sec:total_probabilities} discusses the total positron creation probabilities and the influence of the magnetic field on their values. In Sec.~\ref{sec:energy_spectra}, we present the angle-integrated energy spectra of the emitted positrons and examine their modification due to the magnetic field. Section~\ref{sec:angle_resolved} is devoted to the angle-resolved positron spectra with focus on the anisotropy of the distributions.

Atomic units ($\hbar=|e|=m_{e}=1$) are used throughout the paper unless specified otherwise.

\section{Methods}\label{methods}

The theoretical approach we use to solve the problem of electron-positron pair creation in collision of two nuclei has been described in detail elsewhere~\cite{Dulaev_2024_Angular, Dulaev_Three_2025}. In this approach, Dirac's hole picture based on the Dirac equation for positrons is used. In the present paper, we briefly summarize the key expressions, with particular emphasis on the magnetic field of the moving nuclei not taken into account in our previous work~\cite{Maltsev_2015_Electron, Maltsev_2019_How,Popov_2020_How,Maltsev_2017_Pair,Maltsev_2018_Electron,Popov_2018_One,Popov_2023_Spontaneous,Dulaev_2024_Angular,Dulaev_Three_2025}. 
We begin with the time-dependent Dirac equation (TDDE) for a positron in the center-of-mass frame of two bare nuclei:
\begin{equation}
    i \frac{\partial}{\partial t} \Psi(\bm{r}, t) = H\Psi(\bm{r},t), \label{eq:tdde}
\end{equation}
where $\Psi(\bm{r},t)$ is the four-component positron wave function, and the Hamiltonian $H$ takes the form
\begin{equation}
    H = c(\bm{\alpha} \cdot \bm{p}) + c^2 \beta 
    + U_{\mathrm{ne}} + U_{\mathrm{nm}},
    \label{eq:hamiltonian}
\end{equation}
with $c$ denoting the speed of light, $\bm{p}$ the momentum operator, and $\bm{\alpha}$ and $\beta$ the standard Dirac matrices. The terms $U_{\mathrm{ne}}$ and $U_{\mathrm{nm}}$ represent interaction of the positron with the electric and magnetic fields of the moving nuclei, respectively. The electric term $U_{\mathrm{ne}}$ is a sum of the scalar electrostatic potentials of the nuclei:
\begin{equation}
 U_{\mathrm{ne}} = U_{\mathrm{n}}(|\bm{r}+\bm{a}(t)|) + U_{\mathrm{n}}(|\bm{r}-\bm{a}(t)|).
 \label{eq:electric}
\end{equation}
Here the potential $U_{\mathrm{n}}(r)$ describes a spherically symmetric nuclear potential based on the extended nucleus model. We adopt the Fermi distribution for the nuclear charge density~\cite{Elton_1961_Nuclear} and use root-mean-square charge radii from Ref.~\cite{Angeli_2013_Table}. The instantaneous positions of the nuclei are given by $\pm\bm{a}(t)$, where $2a(t) = R(t)$ is the internuclear separation. The nuclear motion is treated classically, which is a valid approximation for low-energy heavy-ion collisions near the Coulomb barrier, where the nuclear velocities are typically on the order of $0.1c$~\cite{Muller_1972_Auto, Heinz_1981_Electron, Greiner_1985_Quantum}. During the collision, both the magnitude and direction of the vector $\bm{a}(t)$ evolve with time. We assume that the internuclear axis of the quasimolecular system is initially aligned with the $z_0$-axis and the collision occurs in the $z_0$–$x_0$ plane. As the nuclei move, the internuclear axis rotates about the $y$-axis, which is perpendicular to the collision plane.

The magnetic term $U_{\mathrm{nm}}$ is expressed through the vector potential $\bm{A}$:
\begin{equation}
\begin{split}
U_{\mathrm{nm}}&=- (\bm{\alpha} \cdot\bm{A}),\\
\bm{A} &= -\frac{\bm{V}}{c} \left[ U_{\mathrm{n}}(|\bm{r}+\bm{a}(t)|)  - U_{\mathrm{n}}(|\bm{r}-\bm{a}(t)|) \right],
\end{split}
\label{eq:magnetic}
\end{equation}
where $\bm{V}$ is the velocity of the nucleus located at the position $\bm{a}(t)$. Obviously, in the center-of-mass frame of two identical nuclei, the velocity of the nucleus at $-\bm{a}(t)$ is equal to $-\bm{V}$. The scalar potential (\ref{eq:electric}) and vector potential (\ref{eq:magnetic}) of the nuclear electromagnetic field satisfy the Lorenz gauge condition~\cite{jackson1999}:
\begin{equation}
 (\nabla\cdot\bm{A}) + \frac{1}{c}\frac{\partial U_{\mathrm{ne}}}{\partial t} = 0.
\end{equation}
We should note that the expressions (\ref{eq:electric}) and (\ref{eq:magnetic}) for the scalar and vector potentials are approximate since they do not include the retardation effects. However, the corrections are of the order of $V^{2}/c^{2}$~\cite{jackson1999,landau1980} and can be safely neglected in the case of collisions of heavy nuclei near the Coulomb barrier.

Applying a unitary transformation to the wave function,
\begin{equation}
    \Psi(\bm{r}, t) = \exp{\left[ -i\chi(t) J_y \right]} \Psi^{(r)}(\bm{r},t), \label{eq:wv_in_rot}
\end{equation}
we perform a transition to the rotating frame of reference. Here $\chi(t)$ is the time-dependent rotation angle of the internuclear axis and $J_y$ is the $y$-component of the total angular momentum operator composed of the orbital $L_y$ and spin $S_y$ angular momentum operators,
\begin{equation}
    J_y = L_y + S_y. \label{eq:ang_mom_proj}
\end{equation}
The electric and magnetic interaction terms are transformed as follows:
 \begin{equation}
   \begin{gathered}
   \exp[i\chi(t) J_y] U_{\mathrm{ne}} \exp[-i\chi(t) J_y] \\
   = U_{\mathrm{n}}(|\bm{r} + a(t) \bm{e}_z|) + U_{\mathrm{n}}(|\bm{r} - a(t) \bm{e}_z|),
   \end{gathered}
 \end{equation}
 \begin{equation}
   \begin{gathered}
   \exp[i\chi(t) J_y] U_{\mathrm{nm}} \exp[-i\chi(t) J_y] \\
   = \frac{(\bm{\alpha} \cdot \bm{V})}{c} [ U_{\mathrm{n}}(|\bm{r} + a(t) \bm{e}_z|) - U_{\mathrm{n}}(|\bm{r} - a(t) \bm{e}_z|) ].
   \end{gathered}
   \label{eq:u_magnetic}
 \end{equation}
Here $\bm{e}_{z}$ denotes the unit vector along the instantaneous internuclear axis, which is rotated by the angle $\chi(t)$ with respect to the fixed laboratory axis $z_0$. The velocity vector $\bm{V}$ is expanded on the basis of orthogonal unit vectors $\bm{e}_{z}$ and $\bm{e}_{x}$ in the collision plane:
\begin{equation}
 \bm{V} = \dot{a}(t) \bm{e}_z + a(t)\dot{\chi}(t) \bm{e}_x.
\end{equation}
Then the scalar product $(\bm{\alpha} \cdot \bm{V})$ is calculated as follows:
\begin{equation}
(\bm{\alpha} \cdot \bm{V}) = 
\begin{pmatrix}
0 & 0 & \dot{a} & a\dot{\chi} \\
0 & 0 & a\dot{\chi} & -\dot{a} \\
\dot{a} & a\dot{\chi} & 0 & 0 \\
a\dot{\chi} & -\dot{a} & 0 & 0
\end{pmatrix}.
\label{eq:alpha*V_matrix}
\end{equation}

After transforming all the terms in Eq.~(\ref{eq:hamiltonian}) to the rotating frame of reference, we obtain the following equation for the wave function $\Psi^{(r)}(\bm{r},t)$:
\begin{equation}
\begin{aligned}
    & i \frac{\partial \Psi^{(r)}(\bm{r}, t)}{\partial t} = \Big[ c(\bm{\alpha} \cdot \bm{p}) + c^2 \beta - \dot{\chi}(t) J_y \\
    + & U_{\mathrm{n}}(|\bm{r} - a(t)\bm{e}_{z}|) + U_{\mathrm{n}}(|\bm{r} + a(t)\bm{e}_{z}|) +\frac{(\bm{\alpha} \cdot \bm{V})}{c}  \\
     \times [& U_{\mathrm{n}}(|\bm{r} + a(t) \bm{e}_z|) - U_{\mathrm{n}}(|\bm{r} - a(t) \bm{e}_z|)]\Big] \Psi^{(r)}(\bm{r}, t).
\end{aligned}
\label{eq:eq_rotated}
\end{equation}
It should be noted that we refer to Eq.~(\ref{eq:eq_rotated}) as the Dirac equation in the rotating frame, although it slightly differs from the standard Dirac equation in the relativistic rotating frame; for further discussion, see Ref.~\cite{Muller_1976_The}.

The TDDE in the rotating frame, Eq.~(\ref{eq:eq_rotated}), contains the rotational coupling term $-\dot{\chi} J_y$. This operator couples positron states whose angular momentum projections onto the internuclear axis differ by $\pm 1$. The rotational coupling term is present irrespective of any physical interaction; its origin is kinematic, and it emerges in Eq.~(\ref{eq:eq_rotated}) because the rotating frame of reference is noninertial. The effect of this term was analyzed in our recent study~\cite{Dulaev_Three_2025}. In the present work, we also account for the magnetic interaction term (\ref{eq:u_magnetic}) which was not included in our previous research. The part of the magnetic term, which is proportional to the angular velocity $\dot{\chi}$ (see Eq.~(\ref{eq:alpha*V_matrix})), couples positron states with the angular momentum projections onto the internuclear axis differing by $\pm 1$, just like the rotational coupling term does. On the contrary, the other part, which is proportional to the radial velocity $\dot{a}$, preserves the angular momentum projection onto the $z$ axis. In the present study, we perform calculations based on Eq.~(\ref{eq:eq_rotated}) and assess the influence of the magnetic term on the positron creation process.

For convenience of the calculations, we apply an additional unitary transformation to the wave function in Eq.~(\ref{eq:eq_rotated}):
\begin{equation}
    \Psi^{(r)}=\begin{pmatrix}
        1&0&0&0\\
        0&\exp(i\varphi)&0&0 \\
        0&0&i&0 \\
        0&0&0&i\exp(i\varphi)
    \end{pmatrix}\psi ,
    \label{eq:phi_transform}
\end{equation}
where the angle $\varphi$ describes rotation about the $z$ axis. Upon this transformation, positron states with the angular momentum projection $J_{z}=m + 1/2$ onto the $z$ axis are represented by wave functions of the form $\psi_{m} \exp(im\varphi)$, where $\psi_{m}$ depends on time and on the cylindrical coordinates $\rho$ and $z$, but not on the azimuthal angle $\varphi$. The total wave function $\psi(\bm{r}, t)$ is expanded in the Fourier series:
\begin{equation}
 \psi(\bm{r}, t) = \sum\limits_{m=-\infty}^{\infty} \psi_{m}(\rho, z, t) \exp(im\varphi),
 \label{eq:m_expansion}
\end{equation}
and the time-dependent Dirac equation for $\psi(\bm{r}, t)$ reduces to a set of coupled equations for the angular momentum components $\psi_{m}$:
\begin{equation}
\begin{aligned}
    i \frac{\partial}{\partial t} \psi_{m} &= \mathsf{H}_{m} \psi_{m} \\
    &+ \sum\limits_{m' = m \pm 1}\left( [\mathsf{U}_{\mathrm{nm}}]_{mm'} - \dot{\chi} [\mathsf{J}_{y}]_{mm'} \right)\psi_{m'},
    \label{eq:tdde_aft_trans}
\end{aligned}
\end{equation}
The Hamiltonian $\mathsf{H}_{m}$ contains the rest and kinetic energy operators as well as the Coulomb interaction with the nuclei and the part of the magnetic interaction, which is diagonal with respect to the angular momentum projection, for the states with the total angular momentum projection $J_{z}=m + 1/2$:
\begin{equation}
\begin{split}
 \mathsf{H}_{m} &= c^{2}\begin{Vmatrix}
 1_{2} & 0_{2}\\
 0_{2} & -1_{2}
 \end{Vmatrix}
+ \frac{c}{\sqrt{\rho}}\frac{\partial}{\partial\rho}\sqrt{\rho}\begin{Vmatrix}
 0_{2} & \sigma_{x}\\
 -\sigma_{x} & 0_{2}
 \end{Vmatrix}\\
 & + c\frac{\partial}{\partial z}\begin{Vmatrix}
 0_{2} & \sigma_{z}\\
 -\sigma_{z} & 0_{2}
 \end{Vmatrix} +c\frac{m+1/2}{\rho}\begin{Vmatrix}
 0_{2} & i\sigma_{y}\\
 -i\sigma_{y} & 0_{2}
 \end{Vmatrix}\\
 & + \left[U_{\mathrm{n}}(\sqrt{\rho^{2}+(z+a)^{2}})\right.\\ 
 &+ \left. U_{\mathrm{n}}(\sqrt{\rho^{2}+(z-a)^{2}})\right]
 \begin{Vmatrix}
 1_{2} & 0_{2}\\
 0_{2} & 1_{2}
 \end{Vmatrix} +\left[\mathsf{U}_{\mathrm{nm}}\right]_{mm},
\end{split}
      \label{eq195}
\end{equation}
where $\sigma_{x}$, $\sigma_{y}$, $\sigma_{z}$ are the Pauli matrices, and $1_{2}$, $0_{2}$ are the unit and zero $2\times 2$ matrices, respectively. The magnetic interaction operator diagonal with respect to the angular momentum projection is defined as
\begin{equation}
\begin{split}
\left[\mathsf{U}_{\mathrm{nm}}\right]_{mm} &= \frac{\dot{a}}{c}\left[U_{\mathrm{n}}(\sqrt{\rho^{2}+(z+a)^{2}})\right.\\
&\left.- U_{\mathrm{n}}(\sqrt{\rho^{2}+(z-a)^{2}})\right]
\begin{Vmatrix}
 0_{2} & i\sigma_{z}\\
 -i\sigma_{z} & 0_{2}
 \end{Vmatrix}.
 \end{split}
 \label{eq:Unm_diagonal}
\end{equation}

Couplings between different $m$-components are due to the nondiagonal part of the magnetic interaction and angular momentum operator $\mathsf{J}_y$. The angular momentum operator $\mathsf{J}_y$ is a sum of the orbital and spin contributions:
\begin{equation}
    \mathsf{J}_y = \mathsf{L}_y + \mathsf{S}_y, \label{eq:jy}
\end{equation}
with the following expressions for the coupling terms:
\begin{equation}
\begin{split}
\left[\mathsf{L}_{y}\right]_{m,m+1} &=-\frac{i}{2} \left[\frac{z}{\sqrt{\rho}}\frac{\partial}{\partial\rho}\sqrt{\rho} -\rho\frac{\partial}{\partial z}\right]
-i\left(m+\frac{1}{2}\right)\frac{z}{2\rho}\\
&-i\frac{z}{4\rho}\begin{Vmatrix}
 1_{2}-\sigma_{z} & 0_{2}\\
 0_{2} & 1_{2} -\sigma_{z}
 \end{Vmatrix},
\end{split}
\label{eq340}
\end{equation}
\begin{equation}
\begin{split}
\left[\mathsf{L}_{y}\right]_{m,m-1} &=-\frac{i}{2} \left[\frac{z}{\sqrt{\rho}}\frac{\partial}{\partial\rho}\sqrt{\rho} -\rho\frac{\partial}{\partial z}\right]
+i\left(m-\frac{1}{2}\right)\frac{z}{2\rho}\\
&+i\frac{z}{4\rho}\begin{Vmatrix}
 1_{2}-\sigma_{z} & 0_{2}\\
 0_{2} & 1_{2} -\sigma_{z}
 \end{Vmatrix},
\end{split}
\label{eq350}
\end{equation}
\begin{equation}
\left[\mathsf{S}_{y}\right]_{m,m+1}
 =\frac{1}{2}\begin{Vmatrix}
 \sigma_{y} +i\sigma_{x} & 0_{2}\\
 0_{2} & \sigma_{y} +i\sigma_{x}
 \end{Vmatrix} ,
 \label{eq360}
\end{equation}
\begin{equation}
\begin{split}
\left[\mathsf{S}_{y}\right]_{m,m-1}
 =\frac{1}{2}\begin{Vmatrix}
 \sigma_{y} -i\sigma_{x} & 0_{2}\\
 0_{2} & \sigma_{y} -i\sigma_{x}
 \end{Vmatrix} .
\end{split}
\label{eq370}
\end{equation}
The nondiagonal terms of the magnetic interaction are as follows:
\begin{equation}
\begin{split}
 \left[\mathsf{U}_{\mathrm{nm}}\right]_{m,m+1} &= \frac{a\dot{\chi}}{2c}\left[U_{\mathrm{n}}(\sqrt{\rho^{2}+(z+a)^{2}})\right.\\ 
 &-\left. U_{\mathrm{n}}(\sqrt{\rho^{2}+(z-a)^{2}})\right]\\
&\times\begin{Vmatrix}
0_{2}&\sigma_{y}+i\sigma_{x}\\
-\sigma_{y}-i\sigma_{x}&0_{2}
 \end{Vmatrix} ,
\end{split}
\label{eq295}
\end{equation}
\begin{equation}
\begin{split}
 \left[\mathsf{U}_{\mathrm{nm}}\right]_{m,m-1} &= \frac{a\dot{\chi}}{2c}\left[U_{\mathrm{n}}(\sqrt{\rho^{2}+(z+a)^{2}})\right.\\ 
 &-\left. U_{\mathrm{n}}(\sqrt{\rho^{2}+(z-a)^{2}})\right]\\
&\times\begin{Vmatrix}
0_{2}&-\sigma_{y}+i\sigma_{x}\\
\sigma_{y}-i\sigma_{x}&0_{2}
 \end{Vmatrix} .
\end{split}
\label{eq310}
\end{equation}
Equations~(\ref{eq195})--(\ref{eq310}) are written in cylindrical coordinates for compactness. However, two-center quantum systems are more naturally described in prolate spheroidal coordinates. In the case of close collisions, these coordinates can be modified to smoothly transition into a one-center system in the united-atom limit. The modified prolate spheroidal coordinates $\lambda$ and $\eta$ are related to the cylindrical coordinates $\rho$ and $z$ as follows~\cite{Dulaev_2024_Angular}:
\begin{equation}
\begin{split}
    \rho  &= \sqrt{ \left[ (\lambda + a)^2 - a^2 \right] \left[ 1 - \eta^2 \right] }, \\
    z  &= (\lambda + a) \eta, \\
    &\quad (0 \le \lambda < \infty, \; -1 \le \eta \le 1).
    \label{eq:mpsc}
\end{split}
\end{equation}
The azimuthal angle $\varphi$ has the same definition in both coordinate systems. In the limit $a \to 0$, the variable $\lambda$ becomes the spherical radial coordinate $r$, and $\eta$ becomes $\cos\vartheta$, where $\vartheta$ is the polar angle in the spherical coordinate system aligned with the internuclear $z$ axis.

Since the transformation~(\ref{eq:mpsc}) is time-dependent through the parameter $a(t)$, the set of equations~(\ref{eq:tdde_aft_trans}) acquires an additional term:
\begin{equation}
\begin{split}
    &i \frac{\partial}{\partial t} \psi_{m}(\lambda, \eta, t) = \left[ \mathsf{H}_{m}  - \frac{\dot{a}}{a} S_a\right]\psi_{m}(\lambda, \eta, t) \\
    &+ \sum\limits_{m' = m \pm 1}\left( [\mathsf{U}_{\mathrm{nm}}]_{mm'} - \dot{\chi} [\mathsf{J}_{y}]_{mm'} \right)\psi_{m'}(\lambda, \eta, t),
\end{split}
\label{eq:tdde_final}
\end{equation}
where the scaling operator $S_a$ is defined as
\begin{equation}
    S_a = i a \frac{\partial \lambda}{\partial a} \frac{\partial}{\partial \lambda}
        + i a \frac{\partial \eta}{\partial a} \frac{\partial}{\partial \eta}.
    \label{eq:scaling}
\end{equation}
The partial derivatives $\frac{\partial \lambda}{\partial a}$ and $\frac{\partial \eta}{\partial a}$ in Eq.~(\ref{eq:scaling}) are calculated at fixed $\rho$ and $z$ using the transformation relations in Eq.~(\ref{eq:mpsc}).

To numerically solve the final set of equations~(\ref{eq:tdde_final}), we use the generalized pseudospectral (GPS) method, which has proven to be highly accurate and efficient for both relativistic and nonrelativistic calculations~\cite{Telnov_2007_Ab, Telnov_2009_Effects, Telnov_2018_Multiphoton, Telnov_2021_Hyd}. After GPS discretization, the operators on the right-hand side of Eq.~(\ref{eq:tdde_final}) form a block-tridiagonal matrix. The diagonal blocks ($m'=m$) contain the matrices for $\mathsf{H}_{m}$ and $S_a$, while the off-diagonal blocks ($m'=m\pm 1$) are filled with the rotational coupling ($[\mathsf{J}_{y}]_{mm'}$) and magnetic interaction ($[\mathsf{U}_{\mathrm{nm}}]_{mm'}$) matrices. The linear size of each block is $4 \times N_\lambda \times N_\eta$, where $N_\lambda$ and $N_\eta$ denote the number of grid points in $\lambda$ and $\eta$, respectively. In the present work, we use $N_\lambda = 384$ and $N_\eta = 20$, and retain angular momentum projections $m = -1, 0, 1$. As a result, the full matrix in the discretized system~(\ref{eq:tdde_final}) has a linear dimension of $92160$. The radial grid extends to a maximum $\lambda$ value of $60/Z$~a.u., where $Z$ is the nuclear charge number. Time propagation is performed using the Crank--Nicolson algorithm~\cite{Crank_1947_A} with a nonuniform time step; the smallest time step corresponds to the point of closest approach between the nuclei. The total number of time steps is set to $2048$.

To extract the energy–angle distribution of the emitted positrons, we project the unbound positron wave packet obtained at the end of the time propagation onto the relativistic positive-energy plane–waves:  \cite{bere1982}:
\begin{equation}
    \Psi_{\bm{k}}^{(1)} = 
    \begin{pmatrix}
    \sqrt{E_{k} + 2c^2} \\
    0 \\
    \sqrt{E_{k}} \cos{\vartheta_k} \\
    \sqrt{E_{k}} \sin{\vartheta_k} \, \exp(i \varphi_k)
    \end{pmatrix}
    \frac{\exp{[i (\bm{k}\cdot\bm{r}) - iE_{k}t]}}{(2\pi)^{3/2}\sqrt{2E_{k} + 2c^2}},
    \label{eq:pw1}
\end{equation}
\begin{equation}
    \Psi_{\bm{k}}^{(2)} = 
    \begin{pmatrix}
    0 \\
    \sqrt{E_{k} + 2c^2} \, \exp(i \varphi_k)\\
    \sqrt{E_{k}}  \sin{\vartheta_k} \\
    -\sqrt{E_{k}} \cos{\vartheta_k}\, \exp(i \varphi_k)
    \end{pmatrix}
    \frac{\exp{[i (\bm{k}\cdot\bm{r}) - iE_{k}t]}}{(2\pi)^{3/2}\sqrt{2E_{k} + 2c^2}}.
    \label{eq:pw2}
\end{equation}
Here $\bm{k}$ is the positron momentum; $\vartheta_k$ and $\varphi_k$ denote the polar and azimuthal angles of $\bm{k}$. The absolute value of the momentum $k$ and the kinetic energy $E_{k}$ are related to each other as
\begin{equation}
    k = \frac{1}{c}\sqrt{E_{k}\left(E_{k}+2c^2\right)}\,. 
\end{equation}
In the  rest frame of the positron, the wave functions (\ref{eq:pw1}) and (\ref{eq:pw2}) correspond to the spin projections $+1/2$ and $-1/2$ on the $z$–axis, respectively. The differential probabilities of positron creation in these states are
\begin{align}
    \frac{dP^{(1)}}{dE_{k}d\Omega}
    &= \frac{1}{c^3}\,\sqrt{E_{k}(E_{k}+2c^2)}\,(E_{k}+c^2)\,
      \bigl|\langle \Psi_{\bm{k}}^{(1)} \mid \Psi^{(c)}\rangle\bigr|^2,
    \label{eq:p1}\\[6pt]
    \frac{dP^{(2)}}{dE_{k}d\Omega}
    &= \frac{1}{c^3}\,\sqrt{E_{k}(E_{k}+2c^2)}\,(E_{k}+c^2)\,
      \bigl|\langle \Psi_{\bm{k}}^{(2)} \mid \Psi^{(c)}\rangle\bigr|^2,
    \label{eq:p2}
\end{align}
where $\Psi^{(c)}$ is the positive continuum component of the propagated positron wave packet. Since the spin projection ceases to be a good quantum number in frames other than the rest frame of the positron, only the total differential probability without spin selection makes sense:
\begin{equation}
    \frac{dP}{dE_{k}d\Omega}
    \;=\;
    \frac{dP^{(1)}}{dE_{k}d\Omega}
    +\;
    \frac{dP^{(2)}}{dE_{k}d\Omega}\,.
    \label{eq:distr}
\end{equation}

\section{Results}\label{results}

We have performed calculations of positron production in slow collisions of two identical bare nuclei with the charge numbers $Z = 83$, $87$, $92$, and $96$. The collisions are analyzed for a range of scaled collision energies $\varepsilon = 1.0$, $1.1$, $1.2$, and $1.3$. The dominant contribution to the positron production comes from the vacuum positron states whose energy levels are shifted close to the onset of the positive-energy positron continuum as the nuclei approach each other. These are the pairs of degenerate discrete vacuum states with the angular momentum projections $J_z = \pm 1/2$. Since the states with $J_z = 1/2$ and $J_z = -1/2$ contribute equally to the positron creation probability, we choose $16$ initial discrete vacuum states with $J_z = 1/2$, lying closest to the onset of the positive-energy continuum, to solve the time-dependent Dirac equation, and then multiply the resulting probability by $2$.
To investigate the transitions induced by the rotational coupling and magnetic interaction between positron states with different angular momentum projections, we include the components with the angular momentum projections $J_z = -1/2$, $1/2$, and $3/2$ into the propagated wave functions. 

Like in our previous calculations~\cite{Dulaev_2024_Angular,Dulaev_Three_2025}, the initial and final internuclear separations are set to $R_{\mathrm{max}} = 5.5/Z$~a.u., ensuring that all important nonadiabatic transitions between the propagated positron states occur inside the simulated domain. The minimum internuclear separation was chosen to be $R_{\mathrm{min}} = 17.5$~fm. At this distance, the highest-lying discrete vacuum positron state for $Z = 92$ and $Z = 96$ already penetrates deeply in the positive-energy continuum as a supercritical resonance, while the collisions involving $Z = 83$ and $Z = 87$ remain subcritical. This setup allows a direct comparison between subcritical and supercritical regimes of positron creation.

\subsection{Magnetic splitting of energy levels}\label{sec:magnetic_splitting}

To verify the reliability and accuracy of the computational methods employed in this work, we begin with the study of the influence of the nuclear magnetic field on the \textit{electronic} quasimolecular energy levels since such data are available in the literature~\cite{Rafelski_1976_Magnetic}. The Hamiltonian for the one-electron quasimolecule differs from the positronic Hamiltonian (\ref{eq:hamiltonian}) by the sign of the interaction terms. For direct comparison with the results of Ref.~\cite{Rafelski_1976_Magnetic}, we compute the matrix elements of the scaled term (\ref{eq310}) between the degenerate states with the angular momentum projections $J_z = \pm 1/2$ by setting $a\dot{\chi}=c$. We denote this scaled term as $\tilde{\mathsf{U}}_{\mathrm{nm}}$:
\begin{equation}
\begin{split}
\tilde{\mathsf{U}}_{\mathrm{nm}} &= \frac{1}{2}\left[U_{\mathrm{n}}(\sqrt{\rho^{2}+(z+a)^{2}})-U_{\mathrm{n}}(\sqrt{\rho^{2}+(z-a)^{2}})\right]\\
&\times\begin{Vmatrix}
0_{2}&-\sigma_{y}+i\sigma_{x}\\
\sigma_{y}-i\sigma_{x}&0_{2}
 \end{Vmatrix} .
\end{split}
\end{equation}
Calculations of matrix elements between low-lying bound electronic states are less demanding computationally than the calculations of unbound positron wave packet. Therefore we use fewer grid points, $N_\lambda = 160$ and $N_\eta = 16$, in the calculations of this subsection.

We note that the terms (\ref{eq295}) and (\ref{eq310}) are responsible for the energy splitting of degenerate states with $J_z = \pm 1/2$. The splitting only occurs if the internuclear axis rotates and $\dot{\chi}\ne 0$ (that is, for nonzero impact parameters). In head-on collisions, the magnetic interaction reduces to the term $\left[\mathsf{U}_{\mathrm{nm}}\right]_{mm}$ (\ref{eq:Unm_diagonal}), which is diagonal with respect to the angular momentum projection. In this case, the magnetic field of the moving nuclei does not split the states with $J_z = \pm 1/2$. Moreover, the energy levels of these states do not shift in the first order of the perturbation theory with respect to the magnetic interaction since all diagonal matrix elements of the operator (\ref{eq:Unm_diagonal}) vanish (the energy levels do shift in the second order of the perturbation theory but this shift is quite small). These implications follow from the symmetry properties of the Hamiltonians (\ref{eq195}) (see Appendix \ref{app}), which in turn are related to the symmetry of the original Dirac Hamiltonian (\ref{eq:hamiltonian}) with respect to the reflection in the plane containing the internuclear axis and time reversal.

\begin{figure}[]
    \centering
    \includegraphics[width=\linewidth]{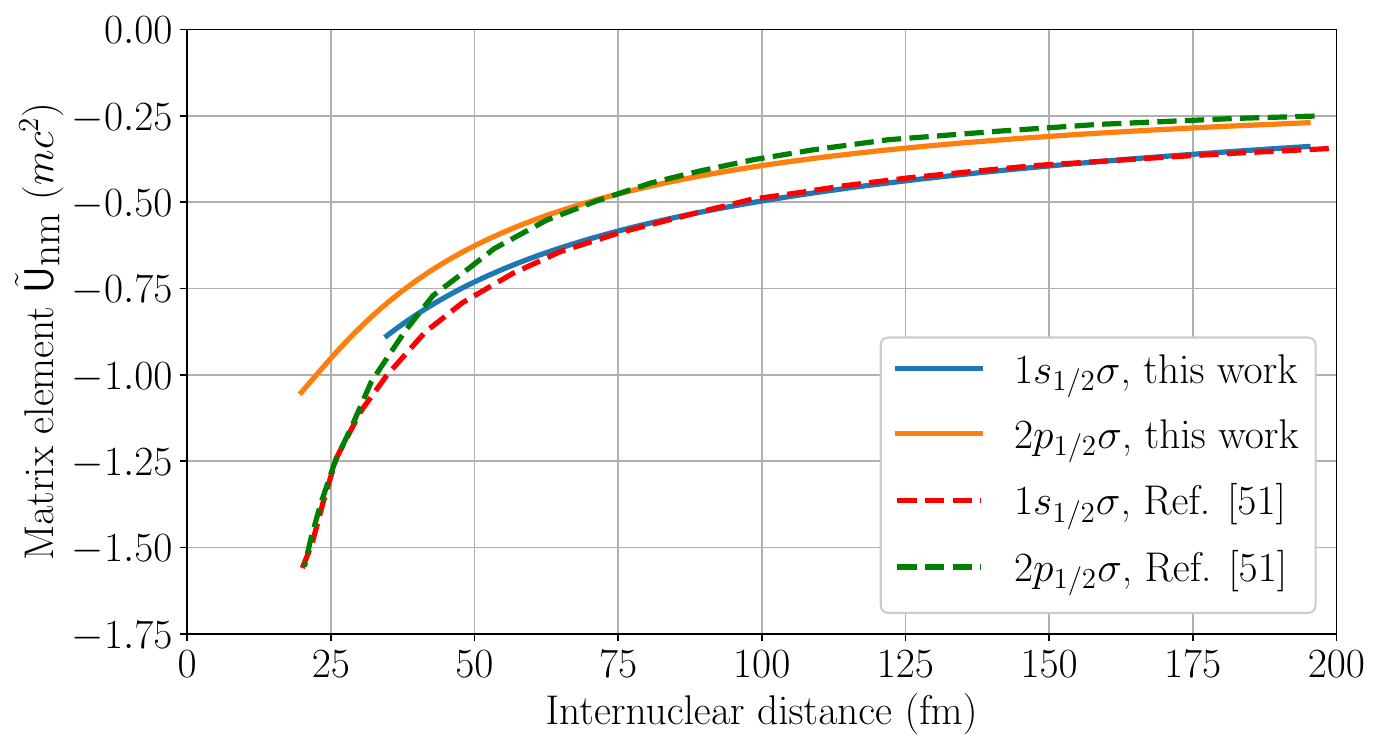}
    \caption{Dependence of matrix elements of the scaled magnetic interaction operator $\tilde{\mathsf{U}}_{\mathrm{nm}}$ on the internuclear distance for the $1s_{1/2}\sigma$ and $2p_{1/2}\sigma$ quasimolecular states in the U$^{92+}$--U$^{92+}$ system. For comparison, the corresponding results from Ref.~\cite{Rafelski_1976_Magnetic} are also shown.}
    \label{fig:mat_elements}
\end{figure}

Fig.~\ref{fig:mat_elements} shows the  matrix elements of the magnetic interaction operator $\tilde{\mathsf{U}}_{\mathrm{nm}}$ calculated between the degenerate states ($J_z = \pm 1/2$) of the $1s_{1/2}\sigma$ and $2p_{1/2}\sigma$ quasimolecular energy levels as functions of the internuclear distance $R$ for the U$^{92+}$-U$^{92+}$ system. The results from Ref.~\cite{Rafelski_1976_Magnetic} are also plotted for comparison. In that work, the calculations were carried out using a point-charge nuclear model, with Dirac wave functions expanded in spinor multipoles and truncated at sufficiently high angular momentum. 
As seen in the figure, our results for the $1s_{1/2}\sigma$ energy level agree well with those of Ref.~\cite{Rafelski_1976_Magnetic} for $R > 60$~fm. At smaller distances, the effects of finite nuclear size become increasingly important, causing our results to deviate from those obtained with the point-charge model, which grow more rapidly in absolute value. At $R \approx 35$~fm, where the $1s_{1/2}\sigma$ energy level dives into the continuum, our curve terminates. In contrast, Ref.~\cite{Rafelski_1976_Magnetic} treats the quasibound $1s_{1/2}\sigma$ state beyond the critical distance separately, allowing the curve to continue.
For the $2p_{1/2}\sigma$ energy level, our results are still in fair agreement with those of Ref.~\cite{Rafelski_1976_Magnetic} for $R > 60$~fm, although some visible differences appear. These discrepancies are probably due to the truncation of the spinor multipole expansion used in Ref.~\cite{Rafelski_1976_Magnetic}. 

The energy level splitting due to the nuclear magnetic field can be calculated from the matrix elements shown in Fig.~\ref{fig:mat_elements} by means of the first order degenerate perturbation theory. On the other hand, we can obtain the energy eigenvalues by direct diagonalization of the Dirac Hamiltonian and then calculate the energy splitting. Our calculations show that the results match with high accuracy. This confirms the reliability of the computational approach used in this work and indicates that the effect of the nuclear magnetic field on the low-lying electronic states is quite weak and well described by the first-order perturbation theory.

\begin{figure}[]
    \centering
    \includegraphics[width=\linewidth]{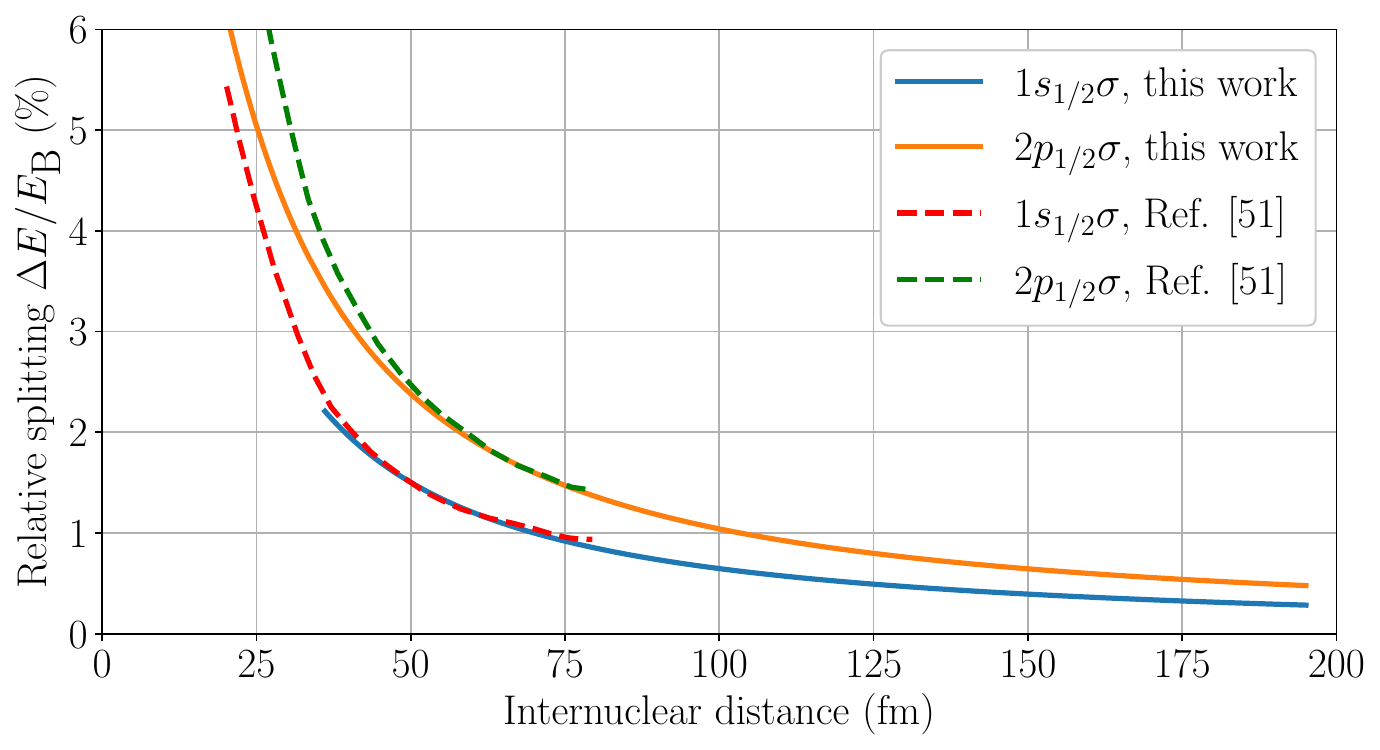}
    \caption{Relative splitting of quasimolecular electronic energy levels due to nuclear magnetic field as a function of internuclear distance for the $1s_{1/2}\sigma$ and $2p_{1/2}\sigma$ states in U$^{92+}$--U$^{92+}$ collisions at the collision energy of $9$~MeV/u and impact parameter of $13$~fm. For comparison, the corresponding results from Ref.~\cite{Rafelski_1976_Magnetic} are also shown.}
    \label{fig:splitting_rel}
\end{figure}
Next, we calculate the relative energy splitting of the $1s_{1/2}\sigma$ and $2p_{1/2}\sigma$ quasimolecular states induced by the nuclear magnetic field. Figure~\ref{fig:splitting_rel} shows the dependence of the relative splitting $\Delta E / E_{\mathrm{B}}$ on the internuclear distance $R$, where $E_{\mathrm{B}}$ is the binding energy of the corresponding state in the absence of the magnetic field. The results are presented for U$^{92+}$--U$^{92+}$ collisions at a collision energy of $9$~MeV/u and an impact parameter of $13$~fm. For comparison, the data from Ref.~\cite{Rafelski_1976_Magnetic} are also included.
As seen in the figure, our results for both $1s_{1/2}\sigma$ and $2p_{1/2}\sigma$ states show good quantitative agreement with those of Ref.~\cite{Rafelski_1976_Magnetic} for $R > 60$~fm. Notably, the deviations in the relative splitting values for the $2p_{1/2}\sigma$ state are smaller than those observed for the corresponding matrix elements.
At shorter internuclear distances, our results begin to deviate from those of Ref.~\cite{Rafelski_1976_Magnetic}. This behavior is attributed to the difference in the nuclear models employed in this work and in Ref.~\cite{Rafelski_1976_Magnetic}. We make use of the extended nucleus model with the Fermi charge distribution while the point-like nucleus is used in Ref.~\cite{Rafelski_1976_Magnetic}. At small internuclear separations, the difference in the binding energies due to different nuclear models is quite large. Overall, for the collisional system under consideration, the relative energy splitting induced by the nuclear magnetic field remains at the level of a few percent: up to about 2\% for the $1s_{1/2}\sigma$ state and about 6\% for the $2p_{1/2}\sigma$ state within the finite nuclear charge distribution model.

\begin{figure}[]
    \centering
    \includegraphics[width=\linewidth]{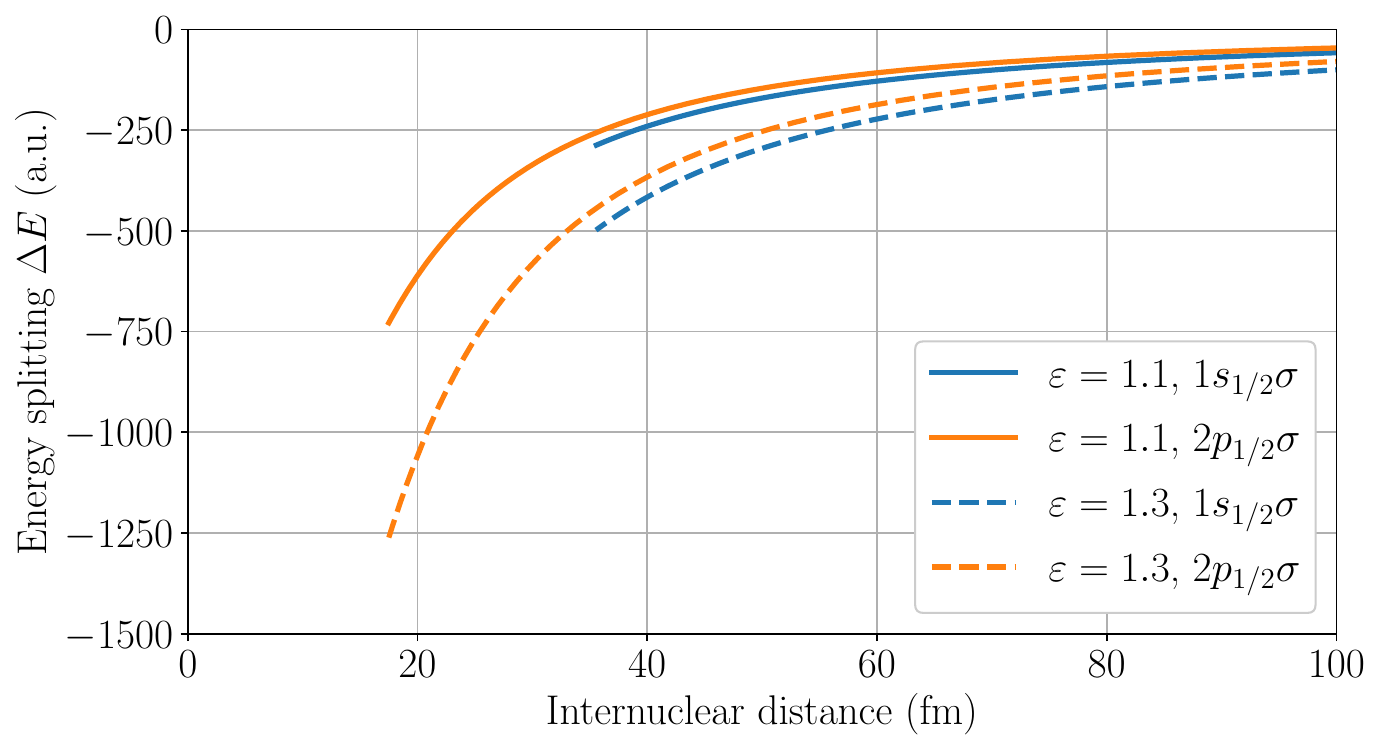}
    \caption{Splitting of vacuum positronic energy levels $1s_{1/2}\sigma$ and $2p_{1/2}\sigma$ due to nuclear magnetic field as a function of internuclear distance in U$^{92+}$--U$^{92+}$ collisions at scaled collision energies $\varepsilon=1.1$ and $1.3$ with $R_{\mathrm{min}}=17.5$~fm.}
    \label{fig:splitting}
\end{figure}
Further, we calculate the magnetic splitting of positronic energy levels for collisions with $R_{\mathrm{min}} = 17.5$~fm studied in this work. Figure~\ref{fig:splitting} presents the energy splitting of the quasimolecular $1s_{1/2}\sigma$ and $2p_{1/2}\sigma$ states for U$^{92+}$--U$^{92+}$ collisions at $\varepsilon = 1.1$ and $1.3$. The splitting is significantly larger in absolute value at $\varepsilon = 1.3$, owing to higher relative nuclear velocity. For the $1s_{1/2}\sigma$ state, the largest splitting before the level reaches the positive-energy continuum is approximately $290$~a.u. at $\varepsilon = 1.1$ and nearly $500$~a.u. at $\varepsilon = 1.3$. 
Since the $2p_{1/2}\sigma$ state does not enter the positive-energy continuum in U$^{92+}$--U$^{92+}$ collisions with $R_{\mathrm{min}} = 17.5$~fm, its splitting continues to increase up to about $730$~a.u. and $1260$~a.u. for $\varepsilon = 1.1$ and $1.3$, respectively, at $R_{\mathrm{min}}$.

\begin{figure}[]
    \centering
    \includegraphics[width=\linewidth]{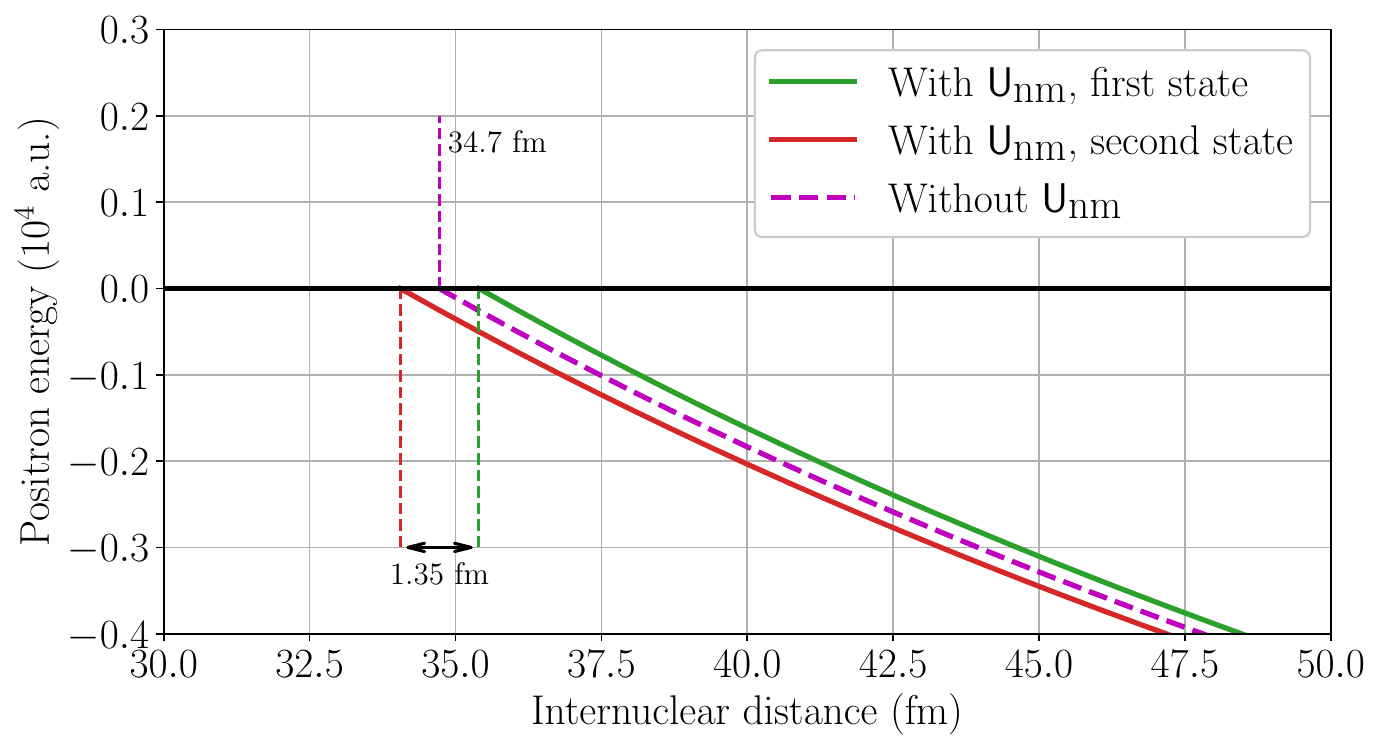}
    \caption{Energies of the first and second states induced by magnetic splitting of $1s_{1/2}\sigma$ state along with $1s_{1/2}\sigma$ energy without magnetic field, for U$^{92+}$--U$^{92+}$ collisions at scaled collision energy $\varepsilon = 1.3$ with $R_{\mathrm{min}}=17.5$~fm. The zero energy level is at the onset of the upper positron continuum.}
    \label{fig:en_levels}
\end{figure}

Figure~\ref{fig:en_levels} shows the energies of the first and second positron states, which originate from magnetic splitting of the $1s_{1/2}\sigma$ level, as functions of the internuclear distance in U$^{92+}$--U$^{92+}$ collisions at the scaled collision energy $\varepsilon = 1.3$. For comparison, the energy of the $1s_{1/2}\sigma$ state calculated without the magnetic interaction in the Hamiltonian is also shown. In the figure, the zero energy level corresponds to the onset of the positive-energy continuum. If the magnetic interaction with the nuclei is switched off, the $1s_{1/2}\sigma$ energy level enters the positive-energy continuum at the critical internuclear separation $R_{\mathrm{cr}} = 34.7$~fm. As one can see from Fig.~\ref{fig:en_levels}, with the magnetic interaction included, the critical internuclear distance for the first state is decreased by $0.68$~fm and for the second state increased by $0.68$~fm, that is $\sim2\%$ of $R_{\mathrm{cr}}$.

\subsection{Total positron creation probabilities}\label{sec:total_probabilities}
Table~\ref{table:total} presents the total positron creation probabilities for collisions of identical nuclei with the charges $Z = 83$, $87$, $92$, and $96$, at various scaled energy parameters $\varepsilon = 1.0$, $1.1$, $1.2$, and $1.3$. Results obtained with and without the magnetic interaction term $\mathsf{U}_{\mathrm{nm}}$ are shown, along with the absolute difference between them.

As shown, the contribution of the magnetic interaction increases with both nuclear charge and collision energy, ranging from $2.21 \times 10^{-6}$ for Bi$^{83+}$--Bi$^{83+}$ collisions at $\varepsilon = 1.0$ to $1.17 \times 10^{-3}$ for Cm$^{96+}$--Cm$^{96+}$ collisions at $\varepsilon = 1.3$. This trend is consistent with the fact that the magnetic field generated in the collision is proportional to the nuclear charge and velocity. For $Z = 83$ and $87$, the magnetic correction contributes less than $3$\% of the total probability, whereas for $Z = 92$ and $96$, the contribution reaches up to approximately $3.0-3.2$\%. Thus, the magnetic interaction has only a minor effect on the total positron creation probabilities at the collision energies $6-8$~MeV/u studied in the present paper.

In all cases, the inclusion of the magnetic interaction results in an increased total positron creation probability. A possible explanation for this behavior is that the energy splitting introduced by the magnetic field shifts the highest-lying discrete vacuum positron state closer to the positive-energy continuum during the collision. In subcritical collisions ($Z = 83$ and $87$), this reduces the energy gap between the vacuum state and the positive-energy continuum, enhancing dynamical pair creation. In supercritical collisions ($Z = 92$ and $96$), the highest-lying discrete vacuum positron state reaches the positive-energy continuum earlier and penetrates it deeper in the course of collision, thereby extending the time spent in the supercritical regime and increasing the positron creation rate. For this reason, spontaneous pair creation is enhanced, thus the total positron yield is increased.

Importantly, inclusion of the magnetic interaction does not qualitatively affect the dependence of the total pair-creation probability on the scaled collision energy. In the subcritical regime, the total probability increases with increasing $\varepsilon$, while in the supercritical regime it decreases. 
As the collision energy decreases, the contribution from dynamical pair creation diminishes, while the probability of spontaneous pair creation increases due to the nuclei spending more time at internuclear distances $R < R_{\mathrm{cr}}$. Consequently, a rising probability with decreasing $\varepsilon$ is a signature of the supercritical regime, in agreement with the previously found patterns \cite{Maltsev_2019_How, Popov_2020_How, Popov_2023_Spontaneous, Dulaev_2024_Angular, Dulaev_Three_2025}.
However, it should be noted that since the magnetic interaction enhances the total probability more strongly at higher collision energies, the relative contrast of the supercritical signatures is slightly reduced. For instance, in Cm$^{96+}$--Cm$^{96+}$ collisions the total probability decreases by approximately $12.6\%$ between $\varepsilon = 1.0$ and $1.3$ when the magnetic term is omitted, whereas the decrease is about $11.5\%$ when $\mathsf{U}_{\mathrm{nm}}$ is included. On the other hand, the enhancement of the probability in subcritical collisions becomes more pronounced: for Bi$^{83}$--Bi$^{83}$ collisions, the increase in total probability between $\varepsilon = 1.0$ and $1.3$ is about $37.4\%$ without the magnetic interaction and rises to approximately $39.4\%$ when this interaction is included.

It is instructive to analyze contributions of positrons with different angular momentum projections to the total positron population. In general, the unbound positron wave packet is a superposition of states with all possible angular momentum projections. However, for the collisions under consideration, populations of the positron continuum states with $J_z = \pm 1/2$ exceed by far those with other angular momentum projections~\cite{Dulaev_Three_2025}. Let us introduce the ratio $\gamma$ of the probabilities $P_{-}$ and $P_{+}$ to find the positron in the states with the angular momentum projections $J_z = -1/2$ and $J_z = 1/2$ on the internuclear axis $z$ after the collision:
\begin{equation}
 \gamma = \frac{P_{-}}{P_{+}}.
 \label{eq:ratio_gamma}
\end{equation}
For the initial vacuum positron states with $J_z = 1/2$, transitions to the states with $J_z = -1/2$ in the rotating frame of reference are caused by the rotational coupling and magnetic terms in Eq.~(\ref{eq:eq_rotated}). We note that the effect of the rotational coupling is not small even in the case of slow collisions; this effect is mostly kinematic and related to rotation of the coordinate frame. After the collision, the $z$ and $x$ axes are rotated by the angle $\chi_{\mathrm{max}}$ from their initial directions. In the case of Rutherford scattering, the angle $\chi_{\mathrm{max}}$ depends on the scaled collision energy only:
\begin{equation}
\chi_{\mathrm{max}} = 2\arctan\!\sqrt{4\varepsilon(\varepsilon - 1)}.
\label{eq:chimax}
\end{equation}
Suppose the system with the total angular momentum $1/2$ is prepared in the state with the angular momentum projection $1/2$ on the axis $z_0$ of the laboratory frame and there are no quantum transitions because of any interaction. Then only the kinematic effects determine the ratio (\ref{eq:ratio_gamma}); for the axis $z$ rotated by the angle $\chi_{\mathrm{max}}$ from the axis $z_0$ this ratio $\gamma_{\mathrm{kin}}$ is calculated as~\cite{Galitski_2013_Quantum}:
\begin{equation}
\gamma_{\mathrm{kin}} = \tan^2\!\left(\frac{\chi_{\mathrm{max}}}{2}\right) = 4\varepsilon(\varepsilon-1).
\label{eq:kin}
\end{equation}

In Table~\ref{table:projections} we list the ratio $\gamma$ calculated from the positron creation probabilities as well as the kinematic prediction $\gamma_{\mathrm{kin}}$ for the systems and collision parameters under consideration. For the listed scaled collision energies, the final rotation angles of the internuclear axis have following values: $\chi_{\mathrm{max}} = 0^\circ$ ($\varepsilon=1.0$), $\chi_{\mathrm{max}} = 67.1^\circ$ ($\varepsilon=1.1$), $\chi_{\mathrm{max}} = 88.8^\circ$ ($\varepsilon=1.2$), and $\chi_{\mathrm{max}} = 102.6^\circ$ ($\varepsilon=1.3$). 
Shown are both the results obtained with the Hamiltonian which contains the magnetic term $\mathsf{U}_{\mathrm{nm}}$ and those where the calculations neglect the magnetic interaction. 
As seen from Table~\ref{table:projections}, when the magnetic term $\mathsf{U}_{\mathrm{nm}}$ is neglected, the kinematic prediction $\gamma_{\mathrm{kin}}$ is quite accurate for the calculated value of $\gamma$ at each $\varepsilon$. As expected, the agreement between $\gamma$ and $\gamma_{\mathrm{kin}}$ is slightly worse at higher scaled collision energies where the dynamical rotational effects begin to show up. These results were reported in our previous work~\cite{Dulaev_Three_2025}. Now one can see that accounting for the magnetic interaction  breaks the kinematic prediction for all scaled collision energies except $\varepsilon=1.0$. Inclusion of the magnetic term in the Hamiltonian leads to systematically higher values of $\gamma$ compared with $\gamma_{\mathrm{kin}}$. 
In the case of the strongest magnetic interaction, namely collisions with $Z=96$ at $\varepsilon=1.3$, the magnetic field increases $\gamma$ by approximately $20\%$ relative to the case without $\mathsf{U}_{\mathrm{nm}}$. Consequently, for all collisions except the head-on case at $\varepsilon=1.0$, where the rotational coupling is absent, the contribution of the initially populated angular momentum projection $J_z = 1/2$ is reduced compared with the case without magnetic field, while the contributions of the $J_z = -1/2$ states are correspondingly increased.

\begin{table}[]
\caption{\label{table:total} Total positron creation probabilities in collisions of two identical nuclei with $R_{\mathrm{min}}=17.5$~fm. Results are shown both without (third column) and with (fourth column) inclusion of the magnetic interaction term $\mathsf{U}_{\mathrm{nm}}$. The last column shows the absolute difference between the probabilities with and without $\mathsf{U}_{\mathrm{nm}}$.}
\begin{ruledtabular}
\begin{tabular}{ccccc}
$Z$                   & $\varepsilon$ & Without $\mathsf{U}_{\mathrm{nm}}$  & With $\mathsf{U}_{\mathrm{nm}}$     & Difference        \\ \hline
                 83 & 1.0 & 3.94$\times 10^{-4}$ & 3.96$\times 10^{-4}$ & 2.21$\times 10^{-6}$                 \\
                    & 1.1 & 4.53$\times 10^{-4}$ & 4.58$\times 10^{-4}$ & 5.32$\times 10^{-6}$                 \\
                    & 1.2 & 4.98$\times 10^{-4}$ & 5.06$\times 10^{-4}$ & 8.06$\times 10^{-6}$                 \\
                    & 1.3 & 5.41$\times 10^{-4}$ & 5.52$\times 10^{-4}$ & 1.09$\times 10^{-5}$                 \\ \hline
                 87 & 1.0 & 1.91$\times 10^{-3}$ & 1.93$\times 10^{-3}$ & 2.36$\times 10^{-5}$                 \\
                    & 1.1 & 2.03$\times 10^{-3}$ & 2.06$\times 10^{-3}$ & 3.52$\times 10^{-5}$                 \\
                    & 1.2 & 2.12$\times 10^{-3}$ & 2.17$\times 10^{-3}$ & 4.64$\times 10^{-5}$                 \\
                    & 1.3 & 2.20$\times 10^{-3}$ & 2.26$\times 10^{-3}$ & 5.71$\times 10^{-5}$                 \\ \hline
                 92 & 1.0 & 1.14$\times 10^{-2}$ & 1.16$\times 10^{-2}$ & 2.01$\times 10^{-4}$                 \\
                    & 1.1 & 1.13$\times 10^{-2}$ & 1.16$\times 10^{-2}$ & 2.48$\times 10^{-4}$                 \\
                    & 1.2 & 1.12$\times 10^{-2}$ & 1.15$\times 10^{-2}$ & 2.93$\times 10^{-4}$                 \\
                    & 1.3 & 1.11$\times 10^{-2}$ & 1.15$\times 10^{-2}$ & 3.37$\times 10^{-4}$                 \\ \hline
                 96 & 1.0 & 4.15$\times 10^{-2}$ & 4.24$\times 10^{-2}$ & 8.21$\times 10^{-4}$                 \\
                    & 1.1 & 3.94$\times 10^{-2}$ & 4.03$\times 10^{-2}$ & 9.42$\times 10^{-4}$                 \\
                    & 1.2 & 3.77$\times 10^{-2}$ & 3.88$\times 10^{-2}$ & 1.05$\times 10^{-3}$                 \\
                    & 1.3 & 3.63$\times 10^{-2}$ & 3.75$\times 10^{-2}$ & 1.17$\times 10^{-3}$                
\end{tabular}
\end{ruledtabular}
\end{table}

\begin{table}[]
\begin{ruledtabular}
\caption{\label{table:projections} Parameter $\gamma$ defined by Eq.~(\ref{eq:ratio_gamma}) for collisions of two identical nuclei with $R_{\mathrm{min}}=17.5$~fm. and various scaled collision energies $\varepsilon$. Shown are the results obtained both without and with inclusion of the magnetic interaction term $\mathsf{U}_{\mathrm{nm}}$. The kinematically predicted parameter $\gamma_{\mathrm{kin}}$ is also listed.}
\begin{tabular}{ccccc}
\multicolumn{1}{c}{\multirow{2}{*}{$Z$}} & \multicolumn{1}{c}{\multirow{2}{*}{$\varepsilon$}}&\multicolumn{1}{c}{\multirow{2}{*}{$\gamma_{\mathrm{kin}}$}} & \multicolumn{2}{c}{$\gamma$}   \\ \cline{4-5}
&  & &Without $\mathsf{U}_{\mathrm{nm}}$ & With $\mathsf{U}_{\mathrm{nm}}$                    \\ \hline
83  & 1.0    & 0.00          & 0.00    & 0.00                  \\
    & 1.1    & 0.44          & 0.45    & 0.48                  \\
    & 1.2    & 0.96          & 0.95    & 1.02                  \\
    & 1.3    & 1.56          & 1.53    & 1.68                  \\ \hline
87  & 1.0    & 0.00          & 0.00    & 0.00                  \\
    & 1.1    & 0.44          & 0.44    & 0.48                  \\
    & 1.2    & 0.96          & 0.95    & 1.05                  \\
    & 1.3    & 1.56          & 1.54    & 1.73                  \\ \hline
92  & 1.0    & 0.00          & 0.00    & 0.00                  \\
    & 1.1    & 0.44          & 0.44    & 0.49                  \\
    & 1.2    & 0.96          & 0.95    & 1.08                  \\
    & 1.3    & 1.56          & 1.54    & 1.80                  \\ \hline
96  & 1.0    & 0.00          & 0.00    & 0.00                  \\
    & 1.1    & 0.44          & 0.44    & 0.50                  \\
    & 1.2    & 0.96          & 0.95    & 1.11                  \\
    & 1.3    & 1.56          & 1.54    & 1.86                                             
\end{tabular}
\end{ruledtabular}
\end{table}

\subsection{Energy distributions of outgoing positrons}\label{sec:energy_spectra}
\begin{figure}[]
    \centering
    \includegraphics[width=\linewidth]{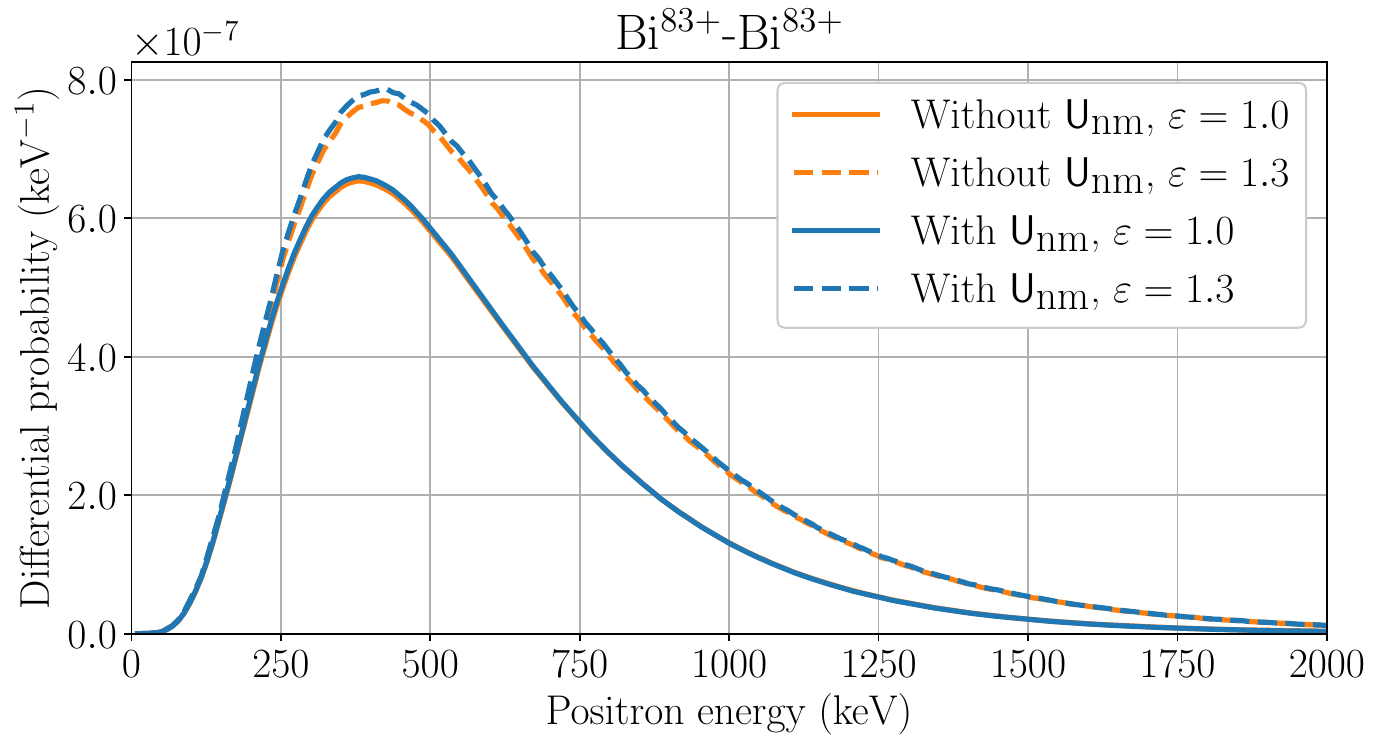}
    \caption{Energy spectra of positrons for symmetric collisions of nuclei with $Z=83$ at $R_{\mathrm{min}}=17.5$~fm and $\varepsilon=1.0$ and $1.3$. Results are shown both with and without inclusion of the magnetic interaction.}
    \label{fig:en_83}
\end{figure}

\begin{figure}[]
    \centering
    \includegraphics[width=\linewidth]{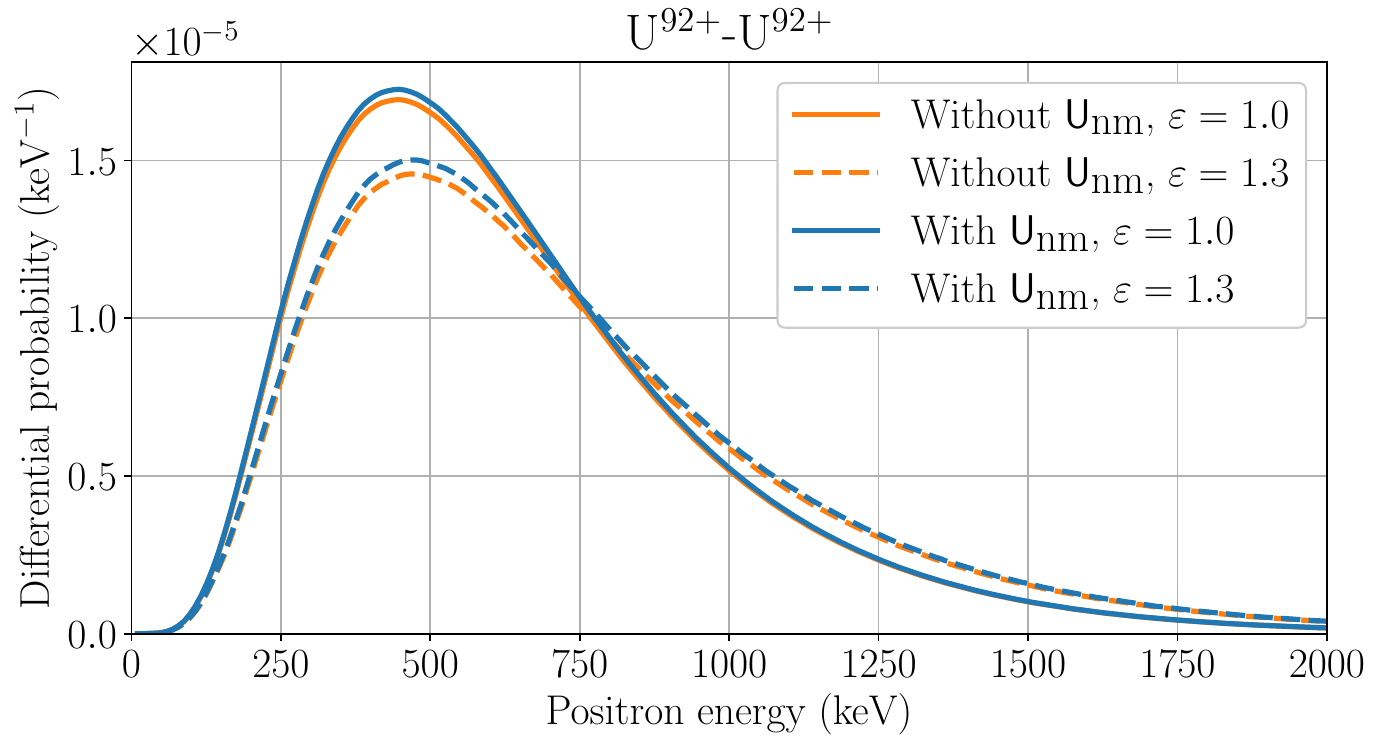}
    \caption{Energy spectra of positrons for symmetric collisions of nuclei with $Z=92$ at $R_{\mathrm{min}}=17.5$~fm and $\varepsilon=1.0$ and $1.3$. Results are shown both with and without inclusion of the magnetic interaction.}
    \label{fig:en_92}
\end{figure}

The energy spectra of outgoing positrons are obtained by integrating Eq.~(\ref{eq:distr}) over the emission angles. Figures~\ref{fig:en_83} and \ref{fig:en_92} show the resulting spectra for collisions of two identical bare nuclei with the charge numbers $Z = 83$ and $Z = 92$, respectively, at scaled collision energies $\varepsilon = 1.0$ and $1.3$. Results are presented for calculations both with and without the magnetic interaction term $\mathsf{U}_{\mathrm{nm}}$. The influence of the magnetic term on the energy distributions is consistent with its effect on the total positron creation probability: the differential probability is slightly increased across the spectrum, with the most noticeable difference occurring near the distribution maximum. As with the total probabilities, the effect becomes more pronounced for larger $Z$ and $\varepsilon$. For instance, in head-on Bi$^{83}$--Bi$^{83}$ collisions at $\varepsilon = 1.0$, the spectra with and without $\mathsf{U}_{\mathrm{nm}}$ are nearly indistinguishable, whereas at $\varepsilon = 1.3$ the peak value is approximately $2$\% higher when the magnetic interaction is included. In the case of U$^{92+}$--U$^{92+}$ collisions, the inclusion of $\mathsf{U}_{\mathrm{nm}}$ increases the peak value of the spectrum by about $2$\% for $\varepsilon = 1.0$ and $3$\% for $\varepsilon = 1.3$.

Previous studies~\cite{Popov_2020_How, Popov_2023_Spontaneous, Dulaev_2024_Angular} have shown, within both the monopole approximation and beyond, that a signature of the supercritical regime can be seen in the behavior of the angle-integrated positron energy distribution: in subcritical collisions, the peak of the distribution increases with increasing $\varepsilon$, while in the supercritical regime it decreases. This behavior reflects the trend observed in the total pair-creation probabilities. It was further demonstrated in Ref.~\cite{Dulaev_Three_2025} that inclusion of the rotational coupling does not alter this pattern. The spectra in Figs.~\ref{fig:en_83} and \ref{fig:en_92} confirm that the inclusion of the magnetic interaction term $\mathsf{U}_{\mathrm{nm}}$ does not qualitatively affect this characteristic behavior either. For subcritical collisions with $Z = 83$, the maximum of the distribution increases as $\varepsilon$ increases, with the highest peak observed at $\varepsilon = 1.3$, both with and without magnetic interaction.
For supercritical collisions with $Z = 92$, the energy spectrum peak is highest for head-on collisions at $\varepsilon = 1.0$, regardless of whether the magnetic interaction is included, consistent with previous findings. However, as with the total probabilities, the inclusion of the magnetic field slightly reduces the relative difference between the distribution maxima at $\varepsilon = 1.0$ and $1.3$ for supercritical collisions. Specifically, for $Z = 92$, the peak value decreases by approximately $14$\% without $\mathsf{U}_{\mathrm{nm}}$, and by about $13$\% when it is included.

\subsection{Angle-resolved energy distributions of outgoing positrons}\label{sec:angle_resolved}
\begin{figure}[]
    \centering
    \includegraphics[width=\linewidth]{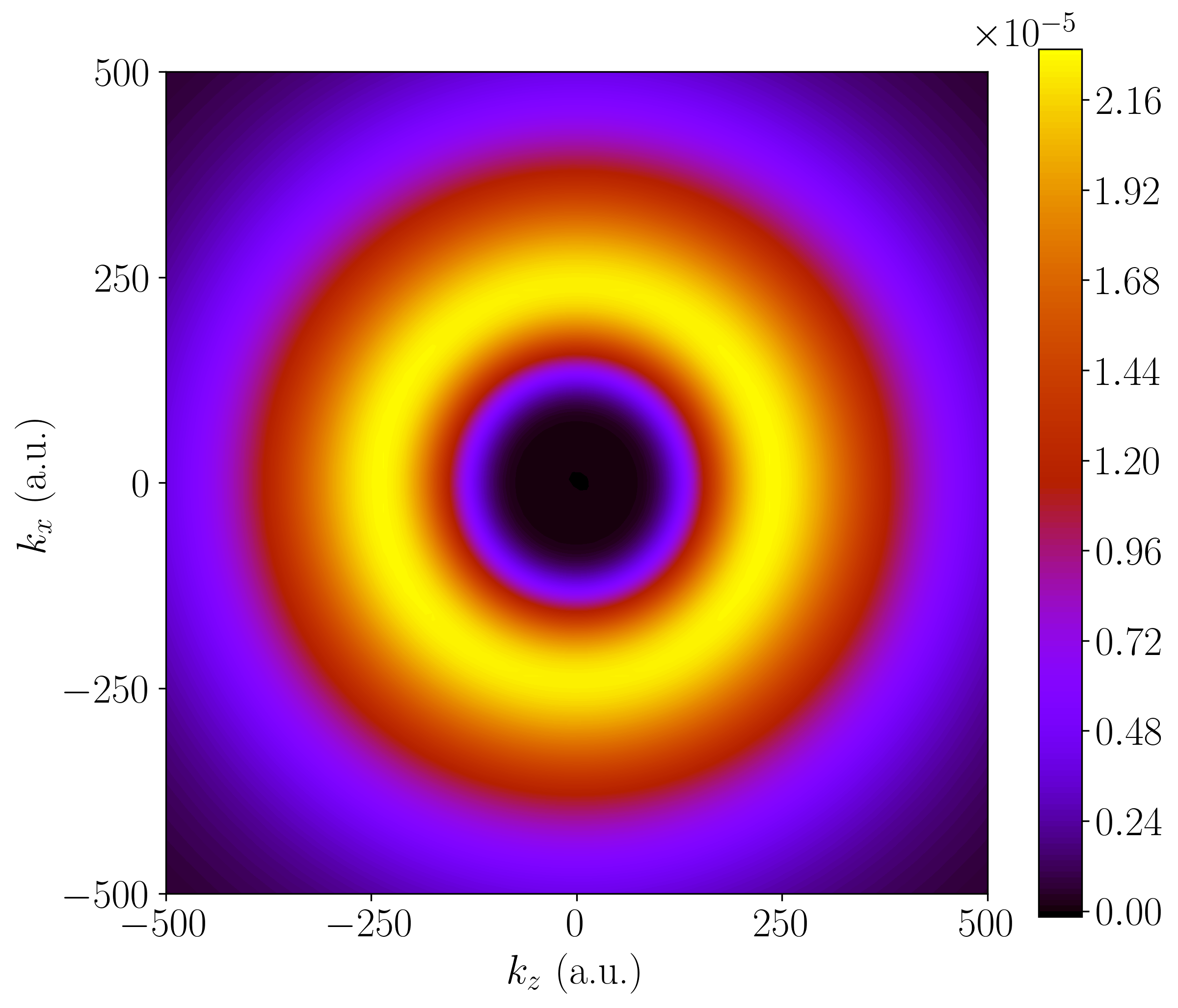}
    \caption{Pseudocolor plot of differential positron creation probability in the collision plane ($k_{z}$ and $k_{x}$ are the positron momentum projections onto the axes $z$ and $x$, respectively. The data shown are for collision with $Z = 96$, $R_{\mathrm{min}} = 17.5$~fm, and scaled collision energy $\varepsilon = 1.3$. The magnetic interaction term is included. The color scale is linear.}
    \label{fig:angular_distr}
\end{figure}

In our previous work~\cite{Dulaev_2024_Angular} in the framework of two-center approach we demonstrated that the angular distributions of outgoing positrons produced in the considered collisions are highly isotropic. In Ref.~\cite{Dulaev_Three_2025} it was further shown that the inclusion of rotational coupling does not significantly affect this low anisotropy. In the present study, to investigate the impact of the magnetic interaction on the angular distributions, we construct two-dimensional angle-resolved energy spectra, similar to those in Ref.~\cite{Dulaev_2024_Angular}. The resulting energy--angle distribution of outgoing positrons is shown in Fig.~\ref{fig:angular_distr} for a collision of nuclei with $Z=96$ at the scaled collision energy $\varepsilon = 1.3$, where the magnetic field has the strongest effect among the systems studied. The angle-resolved differential positron creation probabilities were calculated according to Eq.~(\ref{eq:distr}) at the end of the time evolution in the rotating frame and subsequently transformed into the initial center-of-mass frame. As a result, the momentum components in Fig.~\ref{fig:angular_distr} correspond to the coordinate system defined by the initial orientation of the axes.

As seen in Fig.~\ref{fig:angular_distr}, the distributions exhibit no significant anisotropy and remain nearly spherically symmetric. This behavior is expected. Electron–positron pair creation predominantly occurs at very short internuclear distances, where the electric field of the nuclei is the strongest. At such short internuclear distances, the system resembles a united atom with near-spherical symmetry, and the created positron wave packet in the positive-energy continuum is 
nearly spherically symmetric too. The rotational coupling term causes transitions between the states with different angular momentum projections within the same multiplet ($s_{1/2}$ or $p_{1/2}$) but it cannot alter the spatial symmetry because the $J_{y}$ operator preserves both the parity and total angular momentum. On the other hand, the magnetic interaction term can cause transitions between the states with different total angular momentum (for example, $s_{1/2}$ and $d_{3/2}$) thus changing the angular distributions of outgoing positrons, but the effect of such transitions in the positron continuum is small because continuum wave functions are suppressed in the vicinity of the nuclei where the magnetic field is strong.

\section{Conclusion}

In this work, we have analyzed the influence of the magnetic field due to the nuclear motion on electron–positron pair creation in low-energy collisions of bare heavy nuclei. Using a fully relativistic two-center approach, we include both the rotational coupling and the magnetic interaction terms in the time-dependent Dirac equation. The total pair-creation probabilities, along with the angle-integrated energy distributions of emitted positrons, are calculated. In addition, the angle-resolved energy distributions are obtained by projecting the outgoing positron wave packet onto relativistic plane waves. The influence of the magnetic field on the total positron creation probabilities, as well as on the angle-integrated and angle-resolved energy spectra, is investigated.

In all cases, accounting for the magnetic interaction results in an increased total positron creation probability. A possible explanation for this behavior is that the energy splitting introduced by the magnetic field shifts the highest-lying discrete vacuum positron state closer to the positive-energy continuum during the collision. In subcritical collisions ($Z = 83$ and $87$), this reduces the energy gap between the discrete state and the positive-energy continuum, enhancing dynamical pair creation. In supercritical collisions ($Z = 92$ and $96$), the highest-lying discrete vacuum positron state reaches the positive-energy continuum earlier in the course of collision, thereby extending the time spent in the supercritical regime. This prolongation enhances spontaneous pair creation and thus increases the total positron yield.

Our calculations show that the energy splitting of discrete vacuum positron  states, induced by the magnetic field, causes the highest-lying such state to shift closer to the positive-energy continuum. Consequently, in supercritical collisions, the critical internuclear distance $R_{\mathrm{cr}}$ at which the $1s_{1/2}\sigma$ state enters the positive-energy continuum is slightly increased. The positron yield may increase by several percent compared to the calculations that neglect the magnetic interaction for both subcritical and supercritical collisions. 

Accounting for the magnetic interaction does not qualitatively affect the shape of the energy and angular distributions of the emitted positrons. In particular, the characteristic features associated with the supercritical vacuum decay remain clearly visible. The angular distributions retain their nearly isotropic form. These findings demonstrate that the magnetic field effects do not obscure or diminish the signatures of the supercritical positron creation regime previously identified in the collisions of heavy bare nuclei.

\begin{acknowledgments}
This work was supported by the Russian Science Foundation (Grant No 22-62-00004, \cite{rscf}). Computations were held on the basis of the HybriLIT heterogeneous computing platform (LIT, JINR) \cite{HybriLIT}.
\end{acknowledgments}

\appendix
\section{Magnetic interaction in head-on collisions}\label{app}
In the case of head-on collisions Eq.~(\ref{eq:tdde_aft_trans}) becomes a set of uncoupled equations, thus the projection of the total angular momentum on the internuclear axis remains a good quantum number even if the magnetic interaction is included in the Hamiltonian. Consider the eigenvalue problem for the states with the angular momentum projection $J_{z}=m+1/2$:
\begin{equation}
  \mathsf{H}_{m}\psi_{m} = E_{m}\psi_{m},
  \label{app10}
\end{equation}
where the Hamiltonian $\mathsf{H}_{m}$ is defined in Eq.~(\ref{eq195}).
The unitary transformation with the operator
\begin{equation}
 P_{y} = \begin{Vmatrix}
 \sigma_{y} & 0_{2}\\
 0_{2} & -\sigma_{y}
 \end{Vmatrix}
 \label{app20}
\end{equation}
converts the Hamiltonian $\mathsf{H}_{m}$ into $\mathsf{H}_{-m-1}$:
\begin{equation}
 P_{y}\mathsf{H}_{m}P_{y}^{-1} = \mathsf{H}_{-m-1}.
 \label{app30}
\end{equation}
Applying this transformation to the equation (\ref{app10}), one arrives at the eigenvalue problem
\begin{equation}
  \mathsf{H}_{-m-1}[P_{y}\psi_{m}] = E_{m}[P_{y}\psi_{m}]
  \label{app40}
\end{equation}
for the states with the angular momentum projection $J_{z}=-m-1/2$ but with the same energy $E_{m}$ as for the states with $J_{z}=m+1/2$. That means the adiabatic states with  $J_{z}=\pm(m+1/2)$ remain degenerate in head-on collisions even if the magnetic interaction with the moving nuclei is included in the Hamiltonian.

Now let us introduce the ``unperturbed'' Hamiltonian $\mathsf{H}_{m}^{(0)}$, so the sum of $\mathsf{H}_{m}^{(0)}$ and the magnetic interaction $\left[\mathsf{U}_{\mathrm{nm}}\right]_{mm}$ is equal to the Hamiltonian $\mathsf{H}_{m}$ (\ref{eq195}):
\begin{equation}
 \mathsf{H}_{m} = \mathsf{H}_{m}^{(0)} +\left[\mathsf{U}_{\mathrm{nm}}\right]_{mm} .
 \label{app50}
\end{equation}
As one can see from Eqs.~(\ref{eq195}) and (\ref{eq:Unm_diagonal}), the operator $\mathsf{H}_{m}^{(0)}$ is real-valued while $\left[\mathsf{U}_{\mathrm{nm}}\right]_{mm}$ is pure imaginary. Therefore, the following commutation relations with the complex conjugation operator $C$ hold:
\begin{equation}
\begin{split}
 C\mathsf{H}_{m}^{(0)}C^{-1} &= \mathsf{H}_{m}^{(0)},\\ 
 C\left[\mathsf{U}_{\mathrm{nm}}\right]_{mm}C^{-1} &= -\left[\mathsf{U}_{\mathrm{nm}}\right]_{mm}.
 \end{split}
 \label{app60}
\end{equation}
Since the Hamiltonian $\mathsf{H}_{m}^{(0)}$ is real, the unperturbed eigenfunctions $\psi_{m}^{(0)}$,
\begin{equation}
 \mathsf{H}_{m}^{(0)}\psi_{m}^{(0)} = E_{m}^{(0)}\psi_{m}^{(0)},
 \label{app70}
\end{equation}
can be chosen real, too:
\begin{equation}
 C\psi_{m}^{(0)} = \psi_{m}^{(0)}.
 \label{app80}
\end{equation}
Then it is easy to see that all diagonal matrix elements of the magnetic interaction $\left[\mathsf{U}_{\mathrm{nm}}\right]_{mm}$ vanish:
\begin{equation}
\begin{split}
 \langle\psi_{m}^{(0)}|\left[\mathsf{U}_{\mathrm{nm}}\right]_{mm}|\psi_{m}^{(0)}\rangle = \langle\psi_{m}^{(0)}|\left[\mathsf{U}_{\mathrm{nm}}\right]_{mm}|\psi_{m}^{(0)}\rangle^{*}\\ 
 = -\langle\psi_{m}^{(0)}|\left[\mathsf{U}_{\mathrm{nm}}\right]_{mm}|\psi_{m}^{(0)}\rangle = 0.
 \end{split}
\end{equation}
It means there is no shift of the energy levels $E_{m}^{(0)}$ in the first order of the perturbation theory with respect to the magnetic interaction $\left[\mathsf{U}_{\mathrm{nm}}\right]_{mm}$.

\bibliography{main.bib}

\begin{thebibliography}{75}%
\makeatletter
\providecommand \@ifxundefined [1]{%
 \@ifx{#1\undefined}
}%
\providecommand \@ifnum [1]{%
 \ifnum #1\expandafter \@firstoftwo
 \else \expandafter \@secondoftwo
 \fi
}%
\providecommand \@ifx [1]{%
 \ifx #1\expandafter \@firstoftwo
 \else \expandafter \@secondoftwo
 \fi
}%
\providecommand \natexlab [1]{#1}%
\providecommand \enquote  [1]{``#1''}%
\providecommand \bibnamefont  [1]{#1}%
\providecommand \bibfnamefont [1]{#1}%
\providecommand \citenamefont [1]{#1}%
\providecommand \href@noop [0]{\@secondoftwo}%
\providecommand \href [0]{\begingroup \@sanitize@url \@href}%
\providecommand \@href[1]{\@@startlink{#1}\@@href}%
\providecommand \@@href[1]{\endgroup#1\@@endlink}%
\providecommand \@sanitize@url [0]{\catcode `\\12\catcode `\$12\catcode
  `\&12\catcode `\#12\catcode `\^12\catcode `\_12\catcode `\%12\relax}%
\providecommand \@@startlink[1]{}%
\providecommand \@@endlink[0]{}%
\providecommand \url  [0]{\begingroup\@sanitize@url \@url }%
\providecommand \@url [1]{\endgroup\@href {#1}{\urlprefix }}%
\providecommand \urlprefix  [0]{URL }%
\providecommand \Eprint [0]{\href }%
\providecommand \doibase [0]{https://doi.org/}%
\providecommand \selectlanguage [0]{\@gobble}%
\providecommand \bibinfo  [0]{\@secondoftwo}%
\providecommand \bibfield  [0]{\@secondoftwo}%
\providecommand \translation [1]{[#1]}%
\providecommand \BibitemOpen [0]{}%
\providecommand \bibitemStop [0]{}%
\providecommand \bibitemNoStop [0]{.\EOS\space}%
\providecommand \EOS [0]{\spacefactor3000\relax}%
\providecommand \BibitemShut  [1]{\csname bibitem#1\endcsname}%
\let\auto@bib@innerbib\@empty
\bibitem [{\citenamefont {Greiner}\ \emph {et~al.}(1985)\citenamefont
  {Greiner}, \citenamefont {Müller},\ and\ \citenamefont
  {Rafelski}}]{Greiner_1985_Quantum}%
  \BibitemOpen
  \bibfield  {author} {\bibinfo {author} {\bibfnamefont {W.}~\bibnamefont
  {Greiner}}, \bibinfo {author} {\bibfnamefont {B.}~\bibnamefont {Müller}},\
  and\ \bibinfo {author} {\bibfnamefont {J.}~\bibnamefont {Rafelski}},\
  }\href@noop {} {\emph {\bibinfo {title} {{Q}uantum {E}lectrodynamics of
  {S}trong {F}ields}}}\ (\bibinfo  {publisher} {Springer-Verlag},\ \bibinfo
  {address} {Berlin},\ \bibinfo {year} {1985})\BibitemShut {NoStop}%
\bibitem [{\citenamefont {Fedotov}\ \emph {et~al.}(2023)\citenamefont
  {Fedotov}, \citenamefont {Ilderton}, \citenamefont {Karbstein}, \citenamefont
  {King}, \citenamefont {Seipt}, \citenamefont {Taya},\ and\ \citenamefont
  {Torgrimsson}}]{Fedotov_2023_Advances}%
  \BibitemOpen
  \bibfield  {author} {\bibinfo {author} {\bibfnamefont {A.}~\bibnamefont
  {Fedotov}}, \bibinfo {author} {\bibfnamefont {A.}~\bibnamefont {Ilderton}},
  \bibinfo {author} {\bibfnamefont {F.}~\bibnamefont {Karbstein}}, \bibinfo
  {author} {\bibfnamefont {B.}~\bibnamefont {King}}, \bibinfo {author}
  {\bibfnamefont {D.}~\bibnamefont {Seipt}}, \bibinfo {author} {\bibfnamefont
  {H.}~\bibnamefont {Taya}},\ and\ \bibinfo {author} {\bibfnamefont
  {G.}~\bibnamefont {Torgrimsson}},\ }\bibfield  {title} {\bibinfo {title}
  {Advances in {QED} with intense background fields},\ }\href
  {https://doi.org/https://doi.org/10.1016/j.physrep.2023.01.003} {\bibfield
  {journal} {\bibinfo  {journal} {Phys. Rep.}\ }\textbf {\bibinfo {volume}
  {1010}},\ \bibinfo {pages} {1} (\bibinfo {year} {2023})}\BibitemShut
  {NoStop}%
\bibitem [{\citenamefont {Kostyukov}(2024)}]{Kostyukov_2024_Physics}%
  \BibitemOpen
  \bibfield  {author} {\bibinfo {author} {\bibfnamefont {I.~Y.}\ \bibnamefont
  {Kostyukov}},\ }\bibfield  {title} {\bibinfo {title} {Physics of extremely
  strong electromagnetic field: Status and prospects},\ }\href
  {https://doi.org/10.3103/S1068335624602206} {\bibfield  {journal} {\bibinfo
  {journal} {Bulletin of the Lebedev Physics Institute}\ }\textbf {\bibinfo
  {volume} {51}},\ \bibinfo {pages} {S653} (\bibinfo {year}
  {2024})}\BibitemShut {NoStop}%
\bibitem [{\citenamefont {Pomeranchuk}\ and\ \citenamefont
  {Smorodinsky}(1945)}]{Pomeranchuk_1945}%
  \BibitemOpen
  \bibfield  {author} {\bibinfo {author} {\bibfnamefont {I.}~\bibnamefont
  {Pomeranchuk}}\ and\ \bibinfo {author} {\bibfnamefont {J.}~\bibnamefont
  {Smorodinsky}},\ }\bibfield  {title} {\bibinfo {title} {On the energy levels
  of systems with ${Z}> 137$},\ }\href
  {https://ufn.ru/ufn13/ufn13_4/Russian/pomeranchuk.pdf} {\bibfield  {journal}
  {\bibinfo  {journal} {J. Phys. USSR}\ }\textbf {\bibinfo {volume} {9}},\
  \bibinfo {pages} {97} (\bibinfo {year} {1945})}\BibitemShut {NoStop}%
\bibitem [{\citenamefont {Gershtein}\ and\ \citenamefont
  {Zeldovich}(1969)}]{Gershtein_1969}%
  \BibitemOpen
  \bibfield  {author} {\bibinfo {author} {\bibfnamefont {S.~S.}\ \bibnamefont
  {Gershtein}}\ and\ \bibinfo {author} {\bibfnamefont {Y.~B.}\ \bibnamefont
  {Zeldovich}},\ }\bibfield  {title} {\bibinfo {title} {Positron production
  during the mutual approach of heavy nuclei and the polarization of the
  vacuum},\ }\href {http://jetp.ras.ru/cgi-bin/r/index/e/30/2/p358?a=list}
  {\bibfield  {journal} {\bibinfo  {journal} {Zh. Eksp. Teor. Fiz.}\ }\textbf
  {\bibinfo {volume} {57}},\ \bibinfo {pages} {654} (\bibinfo {year} {1969})},\
  \translation{Sov. Phys. JETP {\bf 30}, 358 (1970)}\BibitemShut {NoStop}%
\bibitem [{\citenamefont {Pieper}\ and\ \citenamefont
  {Greiner}(1969)}]{Pieper_1969_Interior}%
  \BibitemOpen
  \bibfield  {author} {\bibinfo {author} {\bibfnamefont {W.}~\bibnamefont
  {Pieper}}\ and\ \bibinfo {author} {\bibfnamefont {W.}~\bibnamefont
  {Greiner}},\ }\bibfield  {title} {\bibinfo {title} {Interior electron shells
  in superheavy nuclei},\ }\href {https://doi.org/10.1007/BF01670014}
  {\bibfield  {journal} {\bibinfo  {journal} {Z. Phys. A}\ }\textbf {\bibinfo
  {volume} {218}},\ \bibinfo {pages} {327} (\bibinfo {year}
  {1969})}\BibitemShut {NoStop}%
\bibitem [{\citenamefont {Popov}(1970{\natexlab{a}})}]{Popov_1970_1}%
  \BibitemOpen
  \bibfield  {author} {\bibinfo {author} {\bibfnamefont {V.~S.}\ \bibnamefont
  {Popov}},\ }\bibfield  {title} {\bibinfo {title} {Electron energy levels at
  ${Z} > 137$},\ }\href {http://jetpletters.ru/ps/1718/article_26105.shtml}
  {\bibfield  {journal} {\bibinfo  {journal} {Pis’ma Zh. Eksp. Teor. Fiz.}\
  }\textbf {\bibinfo {volume} {11}},\ \bibinfo {pages} {254} (\bibinfo {year}
  {1970}{\natexlab{a}})},\ \translation{JETP Lett. {\bf 11}, 162
  (1970)}\BibitemShut {NoStop}%
\bibitem [{\citenamefont {Popov}(1970{\natexlab{b}})}]{Popov_1970_2}%
  \BibitemOpen
  \bibfield  {author} {\bibinfo {author} {\bibfnamefont {V.~S.}\ \bibnamefont
  {Popov}},\ }\bibfield  {title} {\bibinfo {title} {Collapse to center at ${Z}$
  greater than 137 and critical nuclear charge},\ }\href@noop {} {\bibfield
  {journal} {\bibinfo  {journal} {Yad. Fiz.}\ }\textbf {\bibinfo {volume}
  {12}},\ \bibinfo {pages} {429} (\bibinfo {year} {1970}{\natexlab{b}})},\
  \translation{Sov. J. Nucl. Phys. {\bf 12}, 235 (1971)}\BibitemShut {NoStop}%
\bibitem [{\citenamefont {Popov}(1970{\natexlab{c}})}]{Popov_1970_3}%
  \BibitemOpen
  \bibfield  {author} {\bibinfo {author} {\bibfnamefont {V.~S.}\ \bibnamefont
  {Popov}},\ }\bibfield  {title} {\bibinfo {title} {Position production in a
  coulomb field with ${Z} > 137$},\ }\href
  {http://jetp.ras.ru/cgi-bin/r/index/e/32/3/p526?a=list} {\bibfield  {journal}
  {\bibinfo  {journal} {Zh. Eksp. Teor. Fiz.}\ }\textbf {\bibinfo {volume}
  {59}},\ \bibinfo {pages} {965} (\bibinfo {year} {1970}{\natexlab{c}})},\
  \translation{Sov. Phys. JETP {\bf 32}, 526 (1971)}\BibitemShut {NoStop}%
\bibitem [{\citenamefont {Popov}(1971)}]{Popov_1971_1}%
  \BibitemOpen
  \bibfield  {author} {\bibinfo {author} {\bibfnamefont {V.~S.}\ \bibnamefont
  {Popov}},\ }\bibfield  {title} {\bibinfo {title} {On the properties of the
  discrete spectrum for ${Z}$ close to 137},\ }\href
  {http://jetp.ras.ru/cgi-bin/r/index/e/33/4/p665?a=list} {\bibfield  {journal}
  {\bibinfo  {journal} {Zh. Eksp. Teor. Fiz.}\ }\textbf {\bibinfo {volume}
  {60}},\ \bibinfo {pages} {1228} (\bibinfo {year} {1971})},\ \translation{Sov.
  Phys. JETP {\bf 33}, 665 (1971)}\BibitemShut {NoStop}%
\bibitem [{\citenamefont {Zeldovich}\ and\ \citenamefont
  {Popov}(1971)}]{Zeldovich_1971_Electronic}%
  \BibitemOpen
  \bibfield  {author} {\bibinfo {author} {\bibfnamefont {Y.~B.}\ \bibnamefont
  {Zeldovich}}\ and\ \bibinfo {author} {\bibfnamefont {V.~S.}\ \bibnamefont
  {Popov}},\ }\href
  {https://iopscience.iop.org/article/10.1070/PU1972v014n06ABEH004735}
  {\bibfield  {journal} {\bibinfo  {journal} {Usp. Fiz. Nauk}\ }\textbf
  {\bibinfo {volume} {105}},\ \bibinfo {pages} {403} (\bibinfo {year}
  {1971})},\ \translation{Sov. Phys. Usp. {\bf 14}, 673 (1972)}\BibitemShut
  {NoStop}%
\bibitem [{\citenamefont {M{\"u}ller}\ \emph
  {et~al.}(1972{\natexlab{a}})\citenamefont {M{\"u}ller}, \citenamefont
  {Peitz}, \citenamefont {Rafelski},\ and\ \citenamefont
  {Greiner}}]{Muller_1972_Solution}%
  \BibitemOpen
  \bibfield  {author} {\bibinfo {author} {\bibfnamefont {B.}~\bibnamefont
  {M{\"u}ller}}, \bibinfo {author} {\bibfnamefont {H.}~\bibnamefont {Peitz}},
  \bibinfo {author} {\bibfnamefont {J.}~\bibnamefont {Rafelski}},\ and\
  \bibinfo {author} {\bibfnamefont {W.}~\bibnamefont {Greiner}},\ }\bibfield
  {title} {\bibinfo {title} {Solution of the {D}irac equation for strong
  external fields},\ }\href {https://doi.org/10.1103/PhysRevLett.28.1235}
  {\bibfield  {journal} {\bibinfo  {journal} {Phys. Rev. Lett.}\ }\textbf
  {\bibinfo {volume} {28}},\ \bibinfo {pages} {1235} (\bibinfo {year}
  {1972}{\natexlab{a}})}\BibitemShut {NoStop}%
\bibitem [{\citenamefont {M{\"u}ller}\ \emph
  {et~al.}(1972{\natexlab{b}})\citenamefont {M{\"u}ller}, \citenamefont
  {Rafelski},\ and\ \citenamefont {Greiner}}]{Muller_1972_Electron}%
  \BibitemOpen
  \bibfield  {author} {\bibinfo {author} {\bibfnamefont {B.}~\bibnamefont
  {M{\"u}ller}}, \bibinfo {author} {\bibfnamefont {J.}~\bibnamefont
  {Rafelski}},\ and\ \bibinfo {author} {\bibfnamefont {W.}~\bibnamefont
  {Greiner}},\ }\bibfield  {title} {\bibinfo {title} {Electron shells in
  over-critical external fields},\ }\href {https://doi.org/10.1007/BF01398198}
  {\bibfield  {journal} {\bibinfo  {journal} {Z. Phys. A}\ }\textbf {\bibinfo
  {volume} {257}},\ \bibinfo {pages} {62} (\bibinfo {year}
  {1972}{\natexlab{b}})}\BibitemShut {NoStop}%
\bibitem [{\citenamefont {Mur}\ and\ \citenamefont {Popov}(1976)}]{Mur_1976}%
  \BibitemOpen
  \bibfield  {author} {\bibinfo {author} {\bibfnamefont {V.~D.}\ \bibnamefont
  {Mur}}\ and\ \bibinfo {author} {\bibfnamefont {V.~S.}\ \bibnamefont
  {Popov}},\ }\bibfield  {title} {\bibinfo {title} {Bound states near the limit
  of the lower continuum},\ }\href
  {https://link.springer.com/article/10.1007/BF01051234} {\bibfield  {journal}
  {\bibinfo  {journal} {Teor. Mat. Fiz.}\ }\textbf {\bibinfo {volume} {27}},\
  \bibinfo {pages} {204} (\bibinfo {year} {1976})},\ \translation{Theor. Math.
  Phys. {\bf 27}, 429 (1976)}\BibitemShut {NoStop}%
\bibitem [{\citenamefont {Popov}\ \emph {et~al.}(1976)\citenamefont {Popov},
  \citenamefont {Eletsky},\ and\ \citenamefont {Mur}}]{Popov_1976}%
  \BibitemOpen
  \bibfield  {author} {\bibinfo {author} {\bibfnamefont {V.~S.}\ \bibnamefont
  {Popov}}, \bibinfo {author} {\bibfnamefont {V.~L.}\ \bibnamefont {Eletsky}},\
  and\ \bibinfo {author} {\bibfnamefont {V.~D.}\ \bibnamefont {Mur}},\
  }\bibfield  {title} {\bibinfo {title} {Properties of deep-lying levels in a
  strong electrostatic field},\ }\href
  {http://jetp.ras.ru/cgi-bin/r/index/e/44/3/p451?a=list} {\bibfield  {journal}
  {\bibinfo  {journal} {Zh. Eksp. Teor. Fiz.}\ }\textbf {\bibinfo {volume}
  {71}},\ \bibinfo {pages} {856} (\bibinfo {year} {1976})},\ \translation{Sov.
  Phys. JETP {\bf 44}, 451 (1976)}\BibitemShut {NoStop}%
\bibitem [{\citenamefont {M{\"u}ller}(1976)}]{Muller_1976_Positron}%
  \BibitemOpen
  \bibfield  {author} {\bibinfo {author} {\bibfnamefont {B.}~\bibnamefont
  {M{\"u}ller}},\ }\bibfield  {title} {\bibinfo {title} {Positron creation in
  superheavy quasi-molecules},\ }\href
  {https://doi.org/10.1146/annurev.ns.26.120176.002031} {\bibfield  {journal}
  {\bibinfo  {journal} {Annu. Rev. Nucl. Sci.}\ }\textbf {\bibinfo {volume}
  {26}},\ \bibinfo {pages} {351} (\bibinfo {year} {1976})}\BibitemShut
  {NoStop}%
\bibitem [{\citenamefont {Reinhardt}\ and\ \citenamefont
  {Greiner}(1977)}]{Reinhardt_1977_Quantum}%
  \BibitemOpen
  \bibfield  {author} {\bibinfo {author} {\bibfnamefont {J.}~\bibnamefont
  {Reinhardt}}\ and\ \bibinfo {author} {\bibfnamefont {W.}~\bibnamefont
  {Greiner}},\ }\bibfield  {title} {\bibinfo {title} {Quantum electrodynamics
  of strong fields},\ }\href {https://doi.org/10.1088/0034-4885/40/3/001}
  {\bibfield  {journal} {\bibinfo  {journal} {Rep. Prog. Phys.}\ }\textbf
  {\bibinfo {volume} {40}},\ \bibinfo {pages} {219} (\bibinfo {year}
  {1977})}\BibitemShut {NoStop}%
\bibitem [{\citenamefont {Soff}\ \emph {et~al.}(1977)\citenamefont {Soff},
  \citenamefont {Reinhardt}, \citenamefont {M\"uller},\ and\ \citenamefont
  {Greiner}}]{Soff_1977_Shakeoff}%
  \BibitemOpen
  \bibfield  {author} {\bibinfo {author} {\bibfnamefont {G.}~\bibnamefont
  {Soff}}, \bibinfo {author} {\bibfnamefont {J.}~\bibnamefont {Reinhardt}},
  \bibinfo {author} {\bibfnamefont {B.}~\bibnamefont {M\"uller}},\ and\
  \bibinfo {author} {\bibfnamefont {W.}~\bibnamefont {Greiner}},\ }\bibfield
  {title} {\bibinfo {title} {Shakeoff of the vacuum polarization in
  quasimolecular collisions of very heavy ions},\ }\href
  {https://doi.org/10.1103/PhysRevLett.38.592} {\bibfield  {journal} {\bibinfo
  {journal} {Phys. Rev. Lett.}\ }\textbf {\bibinfo {volume} {38}},\ \bibinfo
  {pages} {592} (\bibinfo {year} {1977})}\BibitemShut {NoStop}%
\bibitem [{\citenamefont {Rafelski}\ \emph {et~al.}(1978)\citenamefont
  {Rafelski}, \citenamefont {Fulcher},\ and\ \citenamefont
  {Klein}}]{Rafelski_1978_Fermions}%
  \BibitemOpen
  \bibfield  {author} {\bibinfo {author} {\bibfnamefont {J.}~\bibnamefont
  {Rafelski}}, \bibinfo {author} {\bibfnamefont {L.~P.}\ \bibnamefont
  {Fulcher}},\ and\ \bibinfo {author} {\bibfnamefont {A.}~\bibnamefont
  {Klein}},\ }\bibfield  {title} {\bibinfo {title} {Fermions and bosons
  interacting with arbitrarily strong external fields},\ }\href
  {https://doi.org/https://doi.org/10.1016/0370-1573(78)90116-3} {\bibfield
  {journal} {\bibinfo  {journal} {Phys. Rep.}\ }\textbf {\bibinfo {volume}
  {38}},\ \bibinfo {pages} {227} (\bibinfo {year} {1978})}\BibitemShut
  {NoStop}%
\bibitem [{\citenamefont {Popov}\ \emph {et~al.}(1979)\citenamefont {Popov},
  \citenamefont {Voskresensky}, \citenamefont {Eletskii},\ and\ \citenamefont
  {Mur}}]{Popov_1979}%
  \BibitemOpen
  \bibfield  {author} {\bibinfo {author} {\bibfnamefont {V.~S.}\ \bibnamefont
  {Popov}}, \bibinfo {author} {\bibfnamefont {D.~N.}\ \bibnamefont
  {Voskresensky}}, \bibinfo {author} {\bibfnamefont {V.~L.}\ \bibnamefont
  {Eletskii}},\ and\ \bibinfo {author} {\bibfnamefont {V.~D.}\ \bibnamefont
  {Mur}},\ }\bibfield  {title} {\bibinfo {title} {{WKB} method at ${Z}>137$ and
  its applications to the theory of supercritical atoms},\ }\href
  {http://jetp.ras.ru/cgi-bin/r/index/e/49/2/p218?a=list} {\bibfield  {journal}
  {\bibinfo  {journal} {Zh. Eksp. Teor. Fiz.}\ }\textbf {\bibinfo {volume}
  {76}},\ \bibinfo {pages} {431} (\bibinfo {year} {1979})},\ \translation{Sov.
  Phys. JETP {\bf 49}, 218 (1979)}\BibitemShut {NoStop}%
\bibitem [{\citenamefont {Smith}\ \emph {et~al.}(1974)\citenamefont {Smith},
  \citenamefont {Peitz}, \citenamefont {M\"uller},\ and\ \citenamefont
  {Greiner}}]{Smith_1974_Induced}%
  \BibitemOpen
  \bibfield  {author} {\bibinfo {author} {\bibfnamefont {K.}~\bibnamefont
  {Smith}}, \bibinfo {author} {\bibfnamefont {H.}~\bibnamefont {Peitz}},
  \bibinfo {author} {\bibfnamefont {B.}~\bibnamefont {M\"uller}},\ and\
  \bibinfo {author} {\bibfnamefont {W.}~\bibnamefont {Greiner}},\ }\bibfield
  {title} {\bibinfo {title} {Induced decay of the neutral vaccum in
  overcritical fields occurring in heavy-ion collisions},\ }\href
  {https://doi.org/10.1103/PhysRevLett.32.554} {\bibfield  {journal} {\bibinfo
  {journal} {Phys. Rev. Lett.}\ }\textbf {\bibinfo {volume} {32}},\ \bibinfo
  {pages} {554} (\bibinfo {year} {1974})}\BibitemShut {NoStop}%
\bibitem [{\citenamefont {Reinhardt}\ \emph {et~al.}(1981)\citenamefont
  {Reinhardt}, \citenamefont {M\"uller},\ and\ \citenamefont
  {Greiner}}]{Reinhardt_1981_Theory}%
  \BibitemOpen
  \bibfield  {author} {\bibinfo {author} {\bibfnamefont {J.}~\bibnamefont
  {Reinhardt}}, \bibinfo {author} {\bibfnamefont {B.}~\bibnamefont
  {M\"uller}},\ and\ \bibinfo {author} {\bibfnamefont {W.}~\bibnamefont
  {Greiner}},\ }\bibfield  {title} {\bibinfo {title} {Theory of positron
  production in heavy-ion collisions},\ }\href
  {https://doi.org/10.1103/PhysRevA.24.103} {\bibfield  {journal} {\bibinfo
  {journal} {Phys. Rev. A}\ }\textbf {\bibinfo {volume} {24}},\ \bibinfo
  {pages} {103} (\bibinfo {year} {1981})}\BibitemShut {NoStop}%
\bibitem [{\citenamefont {M{\"u}ller}\ \emph {et~al.}(1988)\citenamefont
  {M{\"u}ller}, \citenamefont {de~Reus}, \citenamefont {Reinhardt},
  \citenamefont {M\"uller}, \citenamefont {Greiner},\ and\ \citenamefont
  {Soff}}]{Muller_1988_Positron}%
  \BibitemOpen
  \bibfield  {author} {\bibinfo {author} {\bibfnamefont {U.}~\bibnamefont
  {M{\"u}ller}}, \bibinfo {author} {\bibfnamefont {T.}~\bibnamefont {de~Reus}},
  \bibinfo {author} {\bibfnamefont {J.}~\bibnamefont {Reinhardt}}, \bibinfo
  {author} {\bibfnamefont {B.}~\bibnamefont {M\"uller}}, \bibinfo {author}
  {\bibfnamefont {W.}~\bibnamefont {Greiner}},\ and\ \bibinfo {author}
  {\bibfnamefont {G.}~\bibnamefont {Soff}},\ }\bibfield  {title} {\bibinfo
  {title} {Positron production in crossed beams of bare uranium nuclei},\
  }\href {https://doi.org/10.1103/PhysRevA.37.1449} {\bibfield  {journal}
  {\bibinfo  {journal} {Phys. Rev. A}\ }\textbf {\bibinfo {volume} {37}},\
  \bibinfo {pages} {1449} (\bibinfo {year} {1988})}\BibitemShut {NoStop}%
\bibitem [{\citenamefont {Bosch}\ and\ \citenamefont
  {M{\"u}ller}(1986)}]{Bosch_1986_Positron}%
  \BibitemOpen
  \bibfield  {author} {\bibinfo {author} {\bibfnamefont {F.}~\bibnamefont
  {Bosch}}\ and\ \bibinfo {author} {\bibfnamefont {B.}~\bibnamefont
  {M{\"u}ller}},\ }\bibfield  {title} {\bibinfo {title} {Positron creation in
  heavy-ion collisions},\ }\href
  {https://doi.org/https://doi.org/10.1016/0146-6410(86)90005-0} {\bibfield
  {journal} {\bibinfo  {journal} {Prog. Part. Nucl. Phys.}\ }\textbf {\bibinfo
  {volume} {16}},\ \bibinfo {pages} {195} (\bibinfo {year} {1986})}\BibitemShut
  {NoStop}%
\bibitem [{\citenamefont {Müller-Nehler}\ and\ \citenamefont
  {Soff}(1994)}]{Muller_1994_Electron}%
  \BibitemOpen
  \bibfield  {author} {\bibinfo {author} {\bibfnamefont {U.}~\bibnamefont
  {Müller-Nehler}}\ and\ \bibinfo {author} {\bibfnamefont {G.}~\bibnamefont
  {Soff}},\ }\bibfield  {title} {\bibinfo {title} {Electron excitations in
  superheavy quasimolecules},\ }\href
  {https://doi.org/https://doi.org/10.1016/0370-1573(94)90068-X} {\bibfield
  {journal} {\bibinfo  {journal} {Phys. Rep.}\ }\textbf {\bibinfo {volume}
  {246}},\ \bibinfo {pages} {101} (\bibinfo {year} {1994})}\BibitemShut
  {NoStop}%
\bibitem [{\citenamefont {Reinhardt}\ and\ \citenamefont
  {Greiner}(2005)}]{Reinhardt_2005_Supercritical}%
  \BibitemOpen
  \bibfield  {author} {\bibinfo {author} {\bibfnamefont {J.}~\bibnamefont
  {Reinhardt}}\ and\ \bibinfo {author} {\bibfnamefont {W.}~\bibnamefont
  {Greiner}},\ }\bibinfo {title} {Supercritical fields and the decay of the
  vacuum},\ in\ \href@noop {} {\emph {\bibinfo {booktitle} {Proceeding of the
  Memorial Symposium for Gerhard Sof}}},\ \bibinfo {editor} {edited by\
  \bibinfo {editor} {\bibfnamefont {W.}~\bibnamefont {Greiner}}\ and\ \bibinfo
  {editor} {\bibfnamefont {J.}~\bibnamefont {Reinhardt}}}\ (\bibinfo
  {publisher} {EP Systema},\ \bibinfo {address} {Budapest},\ \bibinfo {year}
  {2005})\ pp.\ \bibinfo {pages} {181--192}\BibitemShut {NoStop}%
\bibitem [{\citenamefont {Rafelski}\ \emph {et~al.}(2017)\citenamefont
  {Rafelski}, \citenamefont {Kirsch}, \citenamefont {M{\"u}ller}, \citenamefont
  {Reinhardt},\ and\ \citenamefont {Greiner}}]{Rafelski_2017_Probing}%
  \BibitemOpen
  \bibfield  {author} {\bibinfo {author} {\bibfnamefont {J.}~\bibnamefont
  {Rafelski}}, \bibinfo {author} {\bibfnamefont {J.}~\bibnamefont {Kirsch}},
  \bibinfo {author} {\bibfnamefont {B.}~\bibnamefont {M{\"u}ller}}, \bibinfo
  {author} {\bibfnamefont {J.}~\bibnamefont {Reinhardt}},\ and\ \bibinfo
  {author} {\bibfnamefont {W.}~\bibnamefont {Greiner}},\ }\bibinfo {title}
  {Probing {QED} vacuum with heavy ions},\ in\ \href
  {https://doi.org/10.1007/978-3-319-44165-8_17} {\emph {\bibinfo {booktitle}
  {New Horizons in Fundamental Physics}}},\ \bibinfo {editor} {edited by\
  \bibinfo {editor} {\bibfnamefont {S.}~\bibnamefont {Schramm}}\ and\ \bibinfo
  {editor} {\bibfnamefont {M.}~\bibnamefont {Sch{\"a}fer}}}\ (\bibinfo
  {publisher} {Springer International Publishing},\ \bibinfo {address} {Cham},\
  \bibinfo {year} {2017})\ pp.\ \bibinfo {pages} {211--251}\BibitemShut
  {NoStop}%
\bibitem [{\citenamefont {Gumberidze}\ \emph {et~al.}(2009)\citenamefont
  {Gumberidze}, \citenamefont {Stöhlker}, \citenamefont {Beyer}, \citenamefont
  {Bosch}, \citenamefont {Bräuning-Demian}, \citenamefont {Hagmann},
  \citenamefont {Kozhuharov}, \citenamefont {Kühl}, \citenamefont {Mann},
  \citenamefont {Indelicato}, \citenamefont {Quint}, \citenamefont {Schuch},\
  and\ \citenamefont {Warczak}}]{Gumberidze_2009_X}%
  \BibitemOpen
  \bibfield  {author} {\bibinfo {author} {\bibfnamefont {A.}~\bibnamefont
  {Gumberidze}}, \bibinfo {author} {\bibfnamefont {T.}~\bibnamefont
  {Stöhlker}}, \bibinfo {author} {\bibfnamefont {H.}~\bibnamefont {Beyer}},
  \bibinfo {author} {\bibfnamefont {F.}~\bibnamefont {Bosch}}, \bibinfo
  {author} {\bibfnamefont {A.}~\bibnamefont {Bräuning-Demian}}, \bibinfo
  {author} {\bibfnamefont {S.}~\bibnamefont {Hagmann}}, \bibinfo {author}
  {\bibfnamefont {C.}~\bibnamefont {Kozhuharov}}, \bibinfo {author}
  {\bibfnamefont {T.}~\bibnamefont {Kühl}}, \bibinfo {author} {\bibfnamefont
  {R.}~\bibnamefont {Mann}}, \bibinfo {author} {\bibfnamefont {P.}~\bibnamefont
  {Indelicato}}, \bibinfo {author} {\bibfnamefont {W.}~\bibnamefont {Quint}},
  \bibinfo {author} {\bibfnamefont {R.}~\bibnamefont {Schuch}},\ and\ \bibinfo
  {author} {\bibfnamefont {A.}~\bibnamefont {Warczak}},\ }\bibfield  {title}
  {\bibinfo {title} {X-ray spectroscopy of highly-charged heavy ions at
  {FAIR}},\ }\href {https://doi.org/https://doi.org/10.1016/j.nimb.2008.10.079}
  {\bibfield  {journal} {\bibinfo  {journal} {Nucl. Instrum. Methods Phys.
  Res., Sect. B}\ }\textbf {\bibinfo {volume} {267}},\ \bibinfo {pages} {248}
  (\bibinfo {year} {2009})}\BibitemShut {NoStop}%
\bibitem [{\citenamefont {Lestinsky}\ \emph {et~al.}(2016)\citenamefont
  {Lestinsky}, \citenamefont {Andrianov}, \citenamefont {Aurand}, \citenamefont
  {Bagnoud}, \citenamefont {Bernhardt}, \citenamefont {Beyer}, \citenamefont
  {Bishop}, \citenamefont {Blaum}, \citenamefont {Bleile}, \citenamefont
  {Borovik}, \citenamefont {Bosch}, \citenamefont {Bostock}, \citenamefont
  {Brandau}, \citenamefont {Br{\"a}uning-Demian}, \citenamefont {Bray},
  \citenamefont {Davinson}, \citenamefont {Ebinger}, \citenamefont {Echler},
  \citenamefont {Egelhof}, \citenamefont {Ehresmann}, \citenamefont
  {Engstr{\"o}m}, \citenamefont {Enss}, \citenamefont {Ferreira}, \citenamefont
  {Fischer}, \citenamefont {Fleischmann}, \citenamefont {F{\"o}rster},
  \citenamefont {Fritzsche}, \citenamefont {Geithner}, \citenamefont {Geyer},
  \citenamefont {Glorius}, \citenamefont {G{\"o}bel}, \citenamefont {Gorda},
  \citenamefont {Goullon}, \citenamefont {Grabitz}, \citenamefont {Grisenti},
  \citenamefont {Gumberidze}, \citenamefont {Hagmann}, \citenamefont {Heil},
  \citenamefont {Heinz}, \citenamefont {Herfurth}, \citenamefont {He{\ss}},
  \citenamefont {Hillenbrand}, \citenamefont {Hubele}, \citenamefont
  {Indelicato}, \citenamefont {K{\"a}llberg}, \citenamefont {Kester},
  \citenamefont {Kiselev}, \citenamefont {Knie}, \citenamefont {Kozhuharov},
  \citenamefont {Kraft-Bermuth}, \citenamefont {K{\"u}hl}, \citenamefont
  {Lane}, \citenamefont {Litvinov}, \citenamefont {Liesen}, \citenamefont {Ma},
  \citenamefont {M{\"a}rtin}, \citenamefont {Moshammer}, \citenamefont
  {M{\"u}ller}, \citenamefont {Namba}, \citenamefont {Neumeyer}, \citenamefont
  {Nilsson}, \citenamefont {N{\"o}rtersh{\"a}user}, \citenamefont {Paulus},
  \citenamefont {Petridis}, \citenamefont {Reed}, \citenamefont {Reifarth},
  \citenamefont {Rei{\ss}}, \citenamefont {Rothhardt}, \citenamefont {Sanchez},
  \citenamefont {Sanjari}, \citenamefont {Schippers}, \citenamefont {Schmidt},
  \citenamefont {Schneider}, \citenamefont {Scholz}, \citenamefont {Schuch},
  \citenamefont {Schulz}, \citenamefont {Shabaev}, \citenamefont {Simonsson},
  \citenamefont {Sj{\"o}holm}, \citenamefont {Skeppstedt}, \citenamefont
  {Sonnabend}, \citenamefont {Spillmann}, \citenamefont {Stiebing},
  \citenamefont {Steck}, \citenamefont {St{\"o}hlker}, \citenamefont
  {Surzhykov}, \citenamefont {Torilov}, \citenamefont {Tr{\"a}bert},
  \citenamefont {Trassinelli}, \citenamefont {Trotsenko}, \citenamefont {Tu},
  \citenamefont {Uschmann}, \citenamefont {Walker}, \citenamefont {Weber},
  \citenamefont {Winters}, \citenamefont {Woods}, \citenamefont {Zhao},\ and\
  \citenamefont {Zhang}}]{Lestinsky_2016_Physics}%
  \BibitemOpen
  \bibfield  {author} {\bibinfo {author} {\bibfnamefont {M.}~\bibnamefont
  {Lestinsky}}, \bibinfo {author} {\bibfnamefont {V.}~\bibnamefont
  {Andrianov}}, \bibinfo {author} {\bibfnamefont {B.}~\bibnamefont {Aurand}},
  \bibinfo {author} {\bibfnamefont {V.}~\bibnamefont {Bagnoud}}, \bibinfo
  {author} {\bibfnamefont {D.}~\bibnamefont {Bernhardt}}, \bibinfo {author}
  {\bibfnamefont {H.}~\bibnamefont {Beyer}}, \bibinfo {author} {\bibfnamefont
  {S.}~\bibnamefont {Bishop}}, \bibinfo {author} {\bibfnamefont
  {K.}~\bibnamefont {Blaum}}, \bibinfo {author} {\bibfnamefont
  {A.}~\bibnamefont {Bleile}}, \bibinfo {author} {\bibfnamefont
  {A.}~\bibnamefont {Borovik}}, \bibinfo {author} {\bibfnamefont
  {F.}~\bibnamefont {Bosch}}, \bibinfo {author} {\bibfnamefont {C.~J.}\
  \bibnamefont {Bostock}}, \bibinfo {author} {\bibfnamefont {C.}~\bibnamefont
  {Brandau}}, \bibinfo {author} {\bibfnamefont {A.}~\bibnamefont
  {Br{\"a}uning-Demian}}, \bibinfo {author} {\bibfnamefont {I.}~\bibnamefont
  {Bray}}, \bibinfo {author} {\bibfnamefont {T.}~\bibnamefont {Davinson}},
  \bibinfo {author} {\bibfnamefont {B.}~\bibnamefont {Ebinger}}, \bibinfo
  {author} {\bibfnamefont {A.}~\bibnamefont {Echler}}, \bibinfo {author}
  {\bibfnamefont {P.}~\bibnamefont {Egelhof}}, \bibinfo {author} {\bibfnamefont
  {A.}~\bibnamefont {Ehresmann}}, \bibinfo {author} {\bibfnamefont
  {M.}~\bibnamefont {Engstr{\"o}m}}, \bibinfo {author} {\bibfnamefont
  {C.}~\bibnamefont {Enss}}, \bibinfo {author} {\bibfnamefont {N.}~\bibnamefont
  {Ferreira}}, \bibinfo {author} {\bibfnamefont {D.}~\bibnamefont {Fischer}},
  \bibinfo {author} {\bibfnamefont {A.}~\bibnamefont {Fleischmann}}, \bibinfo
  {author} {\bibfnamefont {E.}~\bibnamefont {F{\"o}rster}}, \bibinfo {author}
  {\bibfnamefont {S.}~\bibnamefont {Fritzsche}}, \bibinfo {author}
  {\bibfnamefont {R.}~\bibnamefont {Geithner}}, \bibinfo {author}
  {\bibfnamefont {S.}~\bibnamefont {Geyer}}, \bibinfo {author} {\bibfnamefont
  {J.}~\bibnamefont {Glorius}}, \bibinfo {author} {\bibfnamefont
  {K.}~\bibnamefont {G{\"o}bel}}, \bibinfo {author} {\bibfnamefont
  {O.}~\bibnamefont {Gorda}}, \bibinfo {author} {\bibfnamefont
  {J.}~\bibnamefont {Goullon}}, \bibinfo {author} {\bibfnamefont
  {P.}~\bibnamefont {Grabitz}}, \bibinfo {author} {\bibfnamefont
  {R.}~\bibnamefont {Grisenti}}, \bibinfo {author} {\bibfnamefont
  {A.}~\bibnamefont {Gumberidze}}, \bibinfo {author} {\bibfnamefont
  {S.}~\bibnamefont {Hagmann}}, \bibinfo {author} {\bibfnamefont
  {M.}~\bibnamefont {Heil}}, \bibinfo {author} {\bibfnamefont {A.}~\bibnamefont
  {Heinz}}, \bibinfo {author} {\bibfnamefont {F.}~\bibnamefont {Herfurth}},
  \bibinfo {author} {\bibfnamefont {R.}~\bibnamefont {He{\ss}}}, \bibinfo
  {author} {\bibfnamefont {P.-M.}\ \bibnamefont {Hillenbrand}}, \bibinfo
  {author} {\bibfnamefont {R.}~\bibnamefont {Hubele}}, \bibinfo {author}
  {\bibfnamefont {P.}~\bibnamefont {Indelicato}}, \bibinfo {author}
  {\bibfnamefont {A.}~\bibnamefont {K{\"a}llberg}}, \bibinfo {author}
  {\bibfnamefont {O.}~\bibnamefont {Kester}}, \bibinfo {author} {\bibfnamefont
  {O.}~\bibnamefont {Kiselev}}, \bibinfo {author} {\bibfnamefont
  {A.}~\bibnamefont {Knie}}, \bibinfo {author} {\bibfnamefont {C.}~\bibnamefont
  {Kozhuharov}}, \bibinfo {author} {\bibfnamefont {S.}~\bibnamefont
  {Kraft-Bermuth}}, \bibinfo {author} {\bibfnamefont {T.}~\bibnamefont
  {K{\"u}hl}}, \bibinfo {author} {\bibfnamefont {G.}~\bibnamefont {Lane}},
  \bibinfo {author} {\bibfnamefont {Y.}~\bibnamefont {Litvinov}}, \bibinfo
  {author} {\bibfnamefont {D.}~\bibnamefont {Liesen}}, \bibinfo {author}
  {\bibfnamefont {X.~W.}\ \bibnamefont {Ma}}, \bibinfo {author} {\bibfnamefont
  {R.}~\bibnamefont {M{\"a}rtin}}, \bibinfo {author} {\bibfnamefont
  {R.}~\bibnamefont {Moshammer}}, \bibinfo {author} {\bibfnamefont
  {A.}~\bibnamefont {M{\"u}ller}}, \bibinfo {author} {\bibfnamefont
  {S.}~\bibnamefont {Namba}}, \bibinfo {author} {\bibfnamefont
  {P.}~\bibnamefont {Neumeyer}}, \bibinfo {author} {\bibfnamefont
  {T.}~\bibnamefont {Nilsson}}, \bibinfo {author} {\bibfnamefont
  {W.}~\bibnamefont {N{\"o}rtersh{\"a}user}}, \bibinfo {author} {\bibfnamefont
  {G.}~\bibnamefont {Paulus}}, \bibinfo {author} {\bibfnamefont
  {N.}~\bibnamefont {Petridis}}, \bibinfo {author} {\bibfnamefont
  {M.}~\bibnamefont {Reed}}, \bibinfo {author} {\bibfnamefont {R.}~\bibnamefont
  {Reifarth}}, \bibinfo {author} {\bibfnamefont {P.}~\bibnamefont {Rei{\ss}}},
  \bibinfo {author} {\bibfnamefont {J.}~\bibnamefont {Rothhardt}}, \bibinfo
  {author} {\bibfnamefont {R.}~\bibnamefont {Sanchez}}, \bibinfo {author}
  {\bibfnamefont {M.~S.}\ \bibnamefont {Sanjari}}, \bibinfo {author}
  {\bibfnamefont {S.}~\bibnamefont {Schippers}}, \bibinfo {author}
  {\bibfnamefont {H.~T.}\ \bibnamefont {Schmidt}}, \bibinfo {author}
  {\bibfnamefont {D.}~\bibnamefont {Schneider}}, \bibinfo {author}
  {\bibfnamefont {P.}~\bibnamefont {Scholz}}, \bibinfo {author} {\bibfnamefont
  {R.}~\bibnamefont {Schuch}}, \bibinfo {author} {\bibfnamefont
  {M.}~\bibnamefont {Schulz}}, \bibinfo {author} {\bibfnamefont
  {V.}~\bibnamefont {Shabaev}}, \bibinfo {author} {\bibfnamefont
  {A.}~\bibnamefont {Simonsson}}, \bibinfo {author} {\bibfnamefont
  {J.}~\bibnamefont {Sj{\"o}holm}}, \bibinfo {author} {\bibfnamefont
  {{\"O}.}~\bibnamefont {Skeppstedt}}, \bibinfo {author} {\bibfnamefont
  {K.}~\bibnamefont {Sonnabend}}, \bibinfo {author} {\bibfnamefont
  {U.}~\bibnamefont {Spillmann}}, \bibinfo {author} {\bibfnamefont
  {K.}~\bibnamefont {Stiebing}}, \bibinfo {author} {\bibfnamefont
  {M.}~\bibnamefont {Steck}}, \bibinfo {author} {\bibfnamefont
  {T.}~\bibnamefont {St{\"o}hlker}}, \bibinfo {author} {\bibfnamefont
  {A.}~\bibnamefont {Surzhykov}}, \bibinfo {author} {\bibfnamefont
  {S.}~\bibnamefont {Torilov}}, \bibinfo {author} {\bibfnamefont
  {E.}~\bibnamefont {Tr{\"a}bert}}, \bibinfo {author} {\bibfnamefont
  {M.}~\bibnamefont {Trassinelli}}, \bibinfo {author} {\bibfnamefont
  {S.}~\bibnamefont {Trotsenko}}, \bibinfo {author} {\bibfnamefont {X.~L.}\
  \bibnamefont {Tu}}, \bibinfo {author} {\bibfnamefont {I.}~\bibnamefont
  {Uschmann}}, \bibinfo {author} {\bibfnamefont {P.~M.}\ \bibnamefont
  {Walker}}, \bibinfo {author} {\bibfnamefont {G.}~\bibnamefont {Weber}},
  \bibinfo {author} {\bibfnamefont {D.~F.~A.}\ \bibnamefont {Winters}},
  \bibinfo {author} {\bibfnamefont {P.~J.}\ \bibnamefont {Woods}}, \bibinfo
  {author} {\bibfnamefont {H.~Y.}\ \bibnamefont {Zhao}},\ and\ \bibinfo
  {author} {\bibfnamefont {Y.~H.}\ \bibnamefont {Zhang}},\ }\bibfield  {title}
  {\bibinfo {title} {Physics book: {CRYRING@ESR}},\ }\href
  {https://doi.org/10.1140/epjst/e2016-02643-6} {\bibfield  {journal} {\bibinfo
   {journal} {The European Physical Journal Special Topics}\ }\textbf {\bibinfo
  {volume} {225}},\ \bibinfo {pages} {797} (\bibinfo {year}
  {2016})}\BibitemShut {NoStop}%
\bibitem [{\citenamefont {Lestinsky}\ \emph {et~al.}(2022)\citenamefont
  {Lestinsky}, \citenamefont {Menz}, \citenamefont {Danared}, \citenamefont
  {Krantz}, \citenamefont {Lindroth}, \citenamefont {Andelkovic}, \citenamefont
  {Brandau}, \citenamefont {Bräuning-Demian}, \citenamefont {Fedotova},
  \citenamefont {Geithner}, \citenamefont {Herfurth}, \citenamefont {Kalinin},
  \citenamefont {Kraus}, \citenamefont {Spillmann}, \citenamefont {Vorobyev},\
  and\ \citenamefont {Stöhlker}}]{Lestinsky_2022_GSI}%
  \BibitemOpen
  \bibfield  {author} {\bibinfo {author} {\bibfnamefont {M.}~\bibnamefont
  {Lestinsky}}, \bibinfo {author} {\bibfnamefont {E.~B.}\ \bibnamefont {Menz}},
  \bibinfo {author} {\bibfnamefont {H.}~\bibnamefont {Danared}}, \bibinfo
  {author} {\bibfnamefont {C.}~\bibnamefont {Krantz}}, \bibinfo {author}
  {\bibfnamefont {E.}~\bibnamefont {Lindroth}}, \bibinfo {author}
  {\bibfnamefont {Z.}~\bibnamefont {Andelkovic}}, \bibinfo {author}
  {\bibfnamefont {C.}~\bibnamefont {Brandau}}, \bibinfo {author} {\bibfnamefont
  {A.}~\bibnamefont {Bräuning-Demian}}, \bibinfo {author} {\bibfnamefont
  {S.}~\bibnamefont {Fedotova}}, \bibinfo {author} {\bibfnamefont
  {W.}~\bibnamefont {Geithner}}, \bibinfo {author} {\bibfnamefont
  {F.}~\bibnamefont {Herfurth}}, \bibinfo {author} {\bibfnamefont
  {A.}~\bibnamefont {Kalinin}}, \bibinfo {author} {\bibfnamefont
  {I.}~\bibnamefont {Kraus}}, \bibinfo {author} {\bibfnamefont
  {U.}~\bibnamefont {Spillmann}}, \bibinfo {author} {\bibfnamefont
  {G.}~\bibnamefont {Vorobyev}},\ and\ \bibinfo {author} {\bibfnamefont
  {T.}~\bibnamefont {Stöhlker}},\ }\bibfield  {title} {\bibinfo {title} {First
  experiments with {CRYRING@ESR}},\ }\href
  {https://doi.org/10.3390/atoms10040141} {\bibfield  {journal} {\bibinfo
  {journal} {Atoms}\ }\textbf {\bibinfo {volume} {10}},\ \bibinfo {pages} {141}
  (\bibinfo {year} {2022})}\BibitemShut {NoStop}%
\bibitem [{\citenamefont {Ma}\ \emph {et~al.}(2017)\citenamefont {Ma},
  \citenamefont {Wen}, \citenamefont {Zhang}, \citenamefont {Yu}, \citenamefont
  {Cheng}, \citenamefont {Yang}, \citenamefont {Huang}, \citenamefont {Wang},
  \citenamefont {Zhu}, \citenamefont {Cai}, \citenamefont {Zhao}, \citenamefont
  {Mao}, \citenamefont {Yang}, \citenamefont {Zhou}, \citenamefont {Xu},
  \citenamefont {Yuan}, \citenamefont {Xia}, \citenamefont {Zhao},
  \citenamefont {Xiao},\ and\ \citenamefont {Zhan}}]{Ma_2017_HIAF}%
  \BibitemOpen
  \bibfield  {author} {\bibinfo {author} {\bibfnamefont {X.}~\bibnamefont
  {Ma}}, \bibinfo {author} {\bibfnamefont {W.}~\bibnamefont {Wen}}, \bibinfo
  {author} {\bibfnamefont {S.}~\bibnamefont {Zhang}}, \bibinfo {author}
  {\bibfnamefont {D.}~\bibnamefont {Yu}}, \bibinfo {author} {\bibfnamefont
  {R.}~\bibnamefont {Cheng}}, \bibinfo {author} {\bibfnamefont
  {J.}~\bibnamefont {Yang}}, \bibinfo {author} {\bibfnamefont {Z.}~\bibnamefont
  {Huang}}, \bibinfo {author} {\bibfnamefont {H.}~\bibnamefont {Wang}},
  \bibinfo {author} {\bibfnamefont {X.}~\bibnamefont {Zhu}}, \bibinfo {author}
  {\bibfnamefont {X.}~\bibnamefont {Cai}}, \bibinfo {author} {\bibfnamefont
  {Y.}~\bibnamefont {Zhao}}, \bibinfo {author} {\bibfnamefont {L.}~\bibnamefont
  {Mao}}, \bibinfo {author} {\bibfnamefont {J.}~\bibnamefont {Yang}}, \bibinfo
  {author} {\bibfnamefont {X.}~\bibnamefont {Zhou}}, \bibinfo {author}
  {\bibfnamefont {H.}~\bibnamefont {Xu}}, \bibinfo {author} {\bibfnamefont
  {Y.}~\bibnamefont {Yuan}}, \bibinfo {author} {\bibfnamefont {J.}~\bibnamefont
  {Xia}}, \bibinfo {author} {\bibfnamefont {H.}~\bibnamefont {Zhao}}, \bibinfo
  {author} {\bibfnamefont {G.}~\bibnamefont {Xiao}},\ and\ \bibinfo {author}
  {\bibfnamefont {W.}~\bibnamefont {Zhan}},\ }\bibfield  {title} {\bibinfo
  {title} {{HIAF}: New opportunities for atomic physics with highly charged
  heavy ions},\ }\href
  {https://doi.org/https://doi.org/10.1016/j.nimb.2017.03.129} {\bibfield
  {journal} {\bibinfo  {journal} {Nucl. Instrum. Methods Phys. Res., Sect. B}\
  }\textbf {\bibinfo {volume} {408}},\ \bibinfo {pages} {169} (\bibinfo {year}
  {2017})}\BibitemShut {NoStop}%
\bibitem [{\citenamefont {Zhou}\ \emph {et~al.}(2022)\citenamefont {Zhou},
  \citenamefont {Yang},\ and\ \citenamefont {{the HIAF project
  team}}}]{Zhou_2022_HIAF}%
  \BibitemOpen
  \bibfield  {author} {\bibinfo {author} {\bibfnamefont {X.}~\bibnamefont
  {Zhou}}, \bibinfo {author} {\bibfnamefont {J.}~\bibnamefont {Yang}},\ and\
  \bibinfo {author} {\bibnamefont {{the HIAF project team}}},\ }\bibfield
  {title} {\bibinfo {title} {Status of the high-intensity heavy-ion accelerator
  facility in {C}hina},\ }\href {https://doi.org/10.1007/s43673-022-00064-1}
  {\bibfield  {journal} {\bibinfo  {journal} {AAPPS Bulletin}\ }\textbf
  {\bibinfo {volume} {32}},\ \bibinfo {pages} {35} (\bibinfo {year}
  {2022})}\BibitemShut {NoStop}%
\bibitem [{\citenamefont {Ter-Akopian}\ \emph {et~al.}(2015)\citenamefont
  {Ter-Akopian}, \citenamefont {Greiner}, \citenamefont {Meshkov},
  \citenamefont {Oganessian}, \citenamefont {Reinhardt},\ and\ \citenamefont
  {Trubnikov}}]{Akopian_2015_Layout}%
  \BibitemOpen
  \bibfield  {author} {\bibinfo {author} {\bibfnamefont {G.~M.}\ \bibnamefont
  {Ter-Akopian}}, \bibinfo {author} {\bibfnamefont {W.}~\bibnamefont
  {Greiner}}, \bibinfo {author} {\bibfnamefont {I.~N.}\ \bibnamefont
  {Meshkov}}, \bibinfo {author} {\bibfnamefont {Y.~T.}\ \bibnamefont
  {Oganessian}}, \bibinfo {author} {\bibfnamefont {J.}~\bibnamefont
  {Reinhardt}},\ and\ \bibinfo {author} {\bibfnamefont {G.~V.}\ \bibnamefont
  {Trubnikov}},\ }\bibfield  {title} {\bibinfo {title} {Layout of new
  experiments on the observation of spontaneous electron–positron pair
  creation in supercritical {C}oulomb fields},\ }\href
  {https://doi.org/10.1142/S0218301315500160} {\bibfield  {journal} {\bibinfo
  {journal} {Int. J. Mod. Phys. E}\ }\textbf {\bibinfo {volume} {24}},\
  \bibinfo {pages} {1550016} (\bibinfo {year} {2015})}\BibitemShut {NoStop}%
\bibitem [{\citenamefont {Trubnikov}\ \emph {et~al.}(2023)\citenamefont
  {Trubnikov}, \citenamefont {Butenko}, \citenamefont {Golovatyuk},
  \citenamefont {Guskov}, \citenamefont {Kapishin}, \citenamefont {Kekelidze},
  \citenamefont {Lednicky}, \citenamefont {Meshkov},\ and\ \citenamefont
  {Sorin}}]{Trubnikov_2023_NICA}%
  \BibitemOpen
  \bibfield  {author} {\bibinfo {author} {\bibfnamefont {G.}~\bibnamefont
  {Trubnikov}}, \bibinfo {author} {\bibfnamefont {A.}~\bibnamefont {Butenko}},
  \bibinfo {author} {\bibfnamefont {V.}~\bibnamefont {Golovatyuk}}, \bibinfo
  {author} {\bibfnamefont {A.}~\bibnamefont {Guskov}}, \bibinfo {author}
  {\bibfnamefont {M.}~\bibnamefont {Kapishin}}, \bibinfo {author}
  {\bibfnamefont {V.}~\bibnamefont {Kekelidze}}, \bibinfo {author}
  {\bibfnamefont {R.}~\bibnamefont {Lednicky}}, \bibinfo {author}
  {\bibfnamefont {I.}~\bibnamefont {Meshkov}},\ and\ \bibinfo {author}
  {\bibfnamefont {A.}~\bibnamefont {Sorin}},\ }\bibfield  {title} {\bibinfo
  {title} {{NICA} heavy-ion collider at {JINR} ({D}ubna). {S}tatus of
  accelerator complex and first physics at {NICA}.},\ }\href
  {https://doi.org/10.1088/1742-6596/2586/1/012013} {\bibfield  {journal}
  {\bibinfo  {journal} {Journal of Physics: Conference Series}\ }\textbf
  {\bibinfo {volume} {2586}},\ \bibinfo {pages} {012013} (\bibinfo {year}
  {2023})}\BibitemShut {NoStop}%
\bibitem [{\citenamefont {Ackad}\ and\ \citenamefont
  {Horbatsch}(2007{\natexlab{a}})}]{Ackad_2007_Numerical}%
  \BibitemOpen
  \bibfield  {author} {\bibinfo {author} {\bibfnamefont {E.}~\bibnamefont
  {Ackad}}\ and\ \bibinfo {author} {\bibfnamefont {M.}~\bibnamefont
  {Horbatsch}},\ }\bibfield  {title} {\bibinfo {title} {Numerical calculation
  of supercritical {D}irac resonance parameters by analytic continuation
  methods},\ }\href {https://doi.org/10.1103/PhysRevA.75.022508} {\bibfield
  {journal} {\bibinfo  {journal} {Phys. Rev. A}\ }\textbf {\bibinfo {volume}
  {75}},\ \bibinfo {pages} {022508} (\bibinfo {year}
  {2007}{\natexlab{a}})}\BibitemShut {NoStop}%
\bibitem [{\citenamefont {Godunov}\ \emph {et~al.}(2017)\citenamefont
  {Godunov}, \citenamefont {Machet},\ and\ \citenamefont
  {Vysotsky}}]{Godunov_2017_Resonances}%
  \BibitemOpen
  \bibfield  {author} {\bibinfo {author} {\bibfnamefont {S.~I.}\ \bibnamefont
  {Godunov}}, \bibinfo {author} {\bibfnamefont {B.}~\bibnamefont {Machet}},\
  and\ \bibinfo {author} {\bibfnamefont {M.~I.}\ \bibnamefont {Vysotsky}},\
  }\bibfield  {title} {\bibinfo {title} {Resonances in positron scattering on a
  supercritical nucleus and spontaneous production of $e^+e^-$ pairs},\ }\href
  {https://doi.org/10.1140/epjc/s10052-017-5325-4} {\bibfield  {journal}
  {\bibinfo  {journal} {Eur. Phys. J. C}\ }\textbf {\bibinfo {volume} {77}},\
  \bibinfo {pages} {782} (\bibinfo {year} {2017})}\BibitemShut {NoStop}%
\bibitem [{\citenamefont {Ackad}\ and\ \citenamefont
  {Horbatsch}(2007{\natexlab{b}})}]{Ackad_2007_Supercritical}%
  \BibitemOpen
  \bibfield  {author} {\bibinfo {author} {\bibfnamefont {E.}~\bibnamefont
  {Ackad}}\ and\ \bibinfo {author} {\bibfnamefont {M.}~\bibnamefont
  {Horbatsch}},\ }\bibfield  {title} {\bibinfo {title} {Supercritical {D}irac
  resonance parameters from extrapolated analytic continuation methods},\
  }\href {https://doi.org/10.1103/PhysRevA.76.022503} {\bibfield  {journal}
  {\bibinfo  {journal} {Phys. Rev. A}\ }\textbf {\bibinfo {volume} {76}},\
  \bibinfo {pages} {022503} (\bibinfo {year} {2007}{\natexlab{b}})}\BibitemShut
  {NoStop}%
\bibitem [{\citenamefont {Marsman}\ and\ \citenamefont
  {Horbatsch}(2011)}]{Marsman_2011_Calculation}%
  \BibitemOpen
  \bibfield  {author} {\bibinfo {author} {\bibfnamefont {A.}~\bibnamefont
  {Marsman}}\ and\ \bibinfo {author} {\bibfnamefont {M.}~\bibnamefont
  {Horbatsch}},\ }\bibfield  {title} {\bibinfo {title} {Calculation of
  supercritical {D}irac resonance parameters for heavy-ion systems from a
  coupled-differential-equation approach},\ }\href
  {https://doi.org/10.1103/PhysRevA.84.032517} {\bibfield  {journal} {\bibinfo
  {journal} {Phys. Rev. A}\ }\textbf {\bibinfo {volume} {84}},\ \bibinfo
  {pages} {032517} (\bibinfo {year} {2011})}\BibitemShut {NoStop}%
\bibitem [{\citenamefont {Grashin}\ and\ \citenamefont
  {Sveshnikov}(2022)}]{Grashin_2022_Vacuum}%
  \BibitemOpen
  \bibfield  {author} {\bibinfo {author} {\bibfnamefont {P.}~\bibnamefont
  {Grashin}}\ and\ \bibinfo {author} {\bibfnamefont {K.}~\bibnamefont
  {Sveshnikov}},\ }\bibfield  {title} {\bibinfo {title} {Vacuum polarization
  energy decline and spontaneous positron emission in {QED} under coulomb
  supercriticality},\ }\href {https://doi.org/10.1103/PhysRevD.106.013003}
  {\bibfield  {journal} {\bibinfo  {journal} {Phys. Rev. D}\ }\textbf {\bibinfo
  {volume} {106}},\ \bibinfo {pages} {013003} (\bibinfo {year}
  {2022})}\BibitemShut {NoStop}%
\bibitem [{\citenamefont {Krasnov}\ and\ \citenamefont
  {Sveshnikov}(2022)}]{Krasnov_2022_Non}%
  \BibitemOpen
  \bibfield  {author} {\bibinfo {author} {\bibfnamefont {A.}~\bibnamefont
  {Krasnov}}\ and\ \bibinfo {author} {\bibfnamefont {K.}~\bibnamefont
  {Sveshnikov}},\ }\bibfield  {title} {\bibinfo {title} {Non-perturbative
  effects in the {QED}-vacuum energy exposed to the supercritical {C}oulomb
  field},\ }\href {https://doi.org/10.1142/S021773232250136X} {\bibfield
  {journal} {\bibinfo  {journal} {Mod. Phys. Lett. A}\ }\textbf {\bibinfo
  {volume} {37}},\ \bibinfo {pages} {2250136} (\bibinfo {year}
  {2022})}\BibitemShut {NoStop}%
\bibitem [{\citenamefont {Voskresensky}(2021)}]{Voskresensky_2021_Electron}%
  \BibitemOpen
  \bibfield  {author} {\bibinfo {author} {\bibfnamefont {D.~N.}\ \bibnamefont
  {Voskresensky}},\ }\bibfield  {title} {\bibinfo {title} {Electron-positron
  vacuum instability in strong electric fields. relativistic semiclassical
  approach},\ }\href {https://doi.org/10.3390/universe7040104} {\bibfield
  {journal} {\bibinfo  {journal} {Universe}\ }\textbf {\bibinfo {volume} {7}},\
  \bibinfo {pages} {104} (\bibinfo {year} {2021})}\BibitemShut {NoStop}%
\bibitem [{\citenamefont {Maltsev}\ \emph {et~al.}(2020)\citenamefont
  {Maltsev}, \citenamefont {Shabaev}, \citenamefont {Zaytsev}, \citenamefont
  {Popov}, \citenamefont {Kozhedub},\ and\ \citenamefont
  {Tumakov}}]{Maltsev_2020_Calculation}%
  \BibitemOpen
  \bibfield  {author} {\bibinfo {author} {\bibfnamefont {I.~A.}\ \bibnamefont
  {Maltsev}}, \bibinfo {author} {\bibfnamefont {V.~M.}\ \bibnamefont
  {Shabaev}}, \bibinfo {author} {\bibfnamefont {V.~A.}\ \bibnamefont
  {Zaytsev}}, \bibinfo {author} {\bibfnamefont {R.~V.}\ \bibnamefont {Popov}},
  \bibinfo {author} {\bibfnamefont {Y.~S.}\ \bibnamefont {Kozhedub}},\ and\
  \bibinfo {author} {\bibfnamefont {D.~A.}\ \bibnamefont {Tumakov}},\
  }\bibfield  {title} {\bibinfo {title} {Calculation of the energy and width of
  supercritical resonance in a uranium quasimolecule},\ }\href
  {https://doi.org/10.1134/S0030400X2008024X} {\bibfield  {journal} {\bibinfo
  {journal} {Opt. Spectrosc.}\ }\textbf {\bibinfo {volume} {128}},\ \bibinfo
  {pages} {1100} (\bibinfo {year} {2020})}\BibitemShut {NoStop}%
\bibitem [{\citenamefont {Telnov}\ \emph {et~al.}(2024)\citenamefont {Telnov},
  \citenamefont {Dulaev}, \citenamefont {Kozhedub}, \citenamefont {Maltsev},
  \citenamefont {Popov}, \citenamefont {Tupitsyn},\ and\ \citenamefont
  {Shabaev}}]{Telnov_positron_2024}%
  \BibitemOpen
  \bibfield  {author} {\bibinfo {author} {\bibfnamefont {D.~A.}\ \bibnamefont
  {Telnov}}, \bibinfo {author} {\bibfnamefont {N.~K.}\ \bibnamefont {Dulaev}},
  \bibinfo {author} {\bibfnamefont {Y.~S.}\ \bibnamefont {Kozhedub}}, \bibinfo
  {author} {\bibfnamefont {I.~A.}\ \bibnamefont {Maltsev}}, \bibinfo {author}
  {\bibfnamefont {R.~V.}\ \bibnamefont {Popov}}, \bibinfo {author}
  {\bibfnamefont {I.~I.}\ \bibnamefont {Tupitsyn}},\ and\ \bibinfo {author}
  {\bibfnamefont {V.~M.}\ \bibnamefont {Shabaev}},\ }\bibfield  {title}
  {\bibinfo {title} {Positron {Supercritical} {Resonances} and {Spontaneous}
  {Positron} {Creation} in {Slow} {Collisions} of {Heavy} {Nuclei}},\ }\href
  {https://doi.org/10.1134/S1063778824700261} {\bibfield  {journal} {\bibinfo
  {journal} {Phys. Atom. Nuclei}\ }\textbf {\bibinfo {volume} {87}},\ \bibinfo
  {pages} {319} (\bibinfo {year} {2024})}\BibitemShut {NoStop}%
\bibitem [{\citenamefont {Maltsev}\ \emph {et~al.}(2015)\citenamefont
  {Maltsev}, \citenamefont {Shabaev}, \citenamefont {Tupitsyn}, \citenamefont
  {Bondarev}, \citenamefont {Kozhedub}, \citenamefont {Plunien},\ and\
  \citenamefont {St\"ohlker}}]{Maltsev_2015_Electron}%
  \BibitemOpen
  \bibfield  {author} {\bibinfo {author} {\bibfnamefont {I.~A.}\ \bibnamefont
  {Maltsev}}, \bibinfo {author} {\bibfnamefont {V.~M.}\ \bibnamefont
  {Shabaev}}, \bibinfo {author} {\bibfnamefont {I.~I.}\ \bibnamefont
  {Tupitsyn}}, \bibinfo {author} {\bibfnamefont {A.~I.}\ \bibnamefont
  {Bondarev}}, \bibinfo {author} {\bibfnamefont {Y.~S.}\ \bibnamefont
  {Kozhedub}}, \bibinfo {author} {\bibfnamefont {G.}~\bibnamefont {Plunien}},\
  and\ \bibinfo {author} {\bibfnamefont {T.}~\bibnamefont {St\"ohlker}},\
  }\bibfield  {title} {\bibinfo {title} {Electron-positron pair creation in
  low-energy collisions of heavy bare nuclei},\ }\href
  {https://doi.org/10.1103/PhysRevA.91.032708} {\bibfield  {journal} {\bibinfo
  {journal} {Phys. Rev. A}\ }\textbf {\bibinfo {volume} {91}},\ \bibinfo
  {pages} {032708} (\bibinfo {year} {2015})}\BibitemShut {NoStop}%
\bibitem [{\citenamefont {Maltsev}\ \emph {et~al.}(2019)\citenamefont
  {Maltsev}, \citenamefont {Shabaev}, \citenamefont {Popov}, \citenamefont
  {Kozhedub}, \citenamefont {Plunien}, \citenamefont {Ma}, \citenamefont
  {St\"ohlker},\ and\ \citenamefont {Tumakov}}]{Maltsev_2019_How}%
  \BibitemOpen
  \bibfield  {author} {\bibinfo {author} {\bibfnamefont {I.~A.}\ \bibnamefont
  {Maltsev}}, \bibinfo {author} {\bibfnamefont {V.~M.}\ \bibnamefont
  {Shabaev}}, \bibinfo {author} {\bibfnamefont {R.~V.}\ \bibnamefont {Popov}},
  \bibinfo {author} {\bibfnamefont {Y.~S.}\ \bibnamefont {Kozhedub}}, \bibinfo
  {author} {\bibfnamefont {G.}~\bibnamefont {Plunien}}, \bibinfo {author}
  {\bibfnamefont {X.}~\bibnamefont {Ma}}, \bibinfo {author} {\bibfnamefont
  {T.}~\bibnamefont {St\"ohlker}},\ and\ \bibinfo {author} {\bibfnamefont
  {D.~A.}\ \bibnamefont {Tumakov}},\ }\bibfield  {title} {\bibinfo {title} {How
  to observe the vacuum decay in low-energy heavy-ion collisions},\ }\href
  {https://doi.org/10.1103/PhysRevLett.123.113401} {\bibfield  {journal}
  {\bibinfo  {journal} {Phys. Rev. Lett.}\ }\textbf {\bibinfo {volume} {123}},\
  \bibinfo {pages} {113401} (\bibinfo {year} {2019})}\BibitemShut {NoStop}%
\bibitem [{\citenamefont {Popov}\ \emph {et~al.}(2020)\citenamefont {Popov},
  \citenamefont {Shabaev}, \citenamefont {Telnov}, \citenamefont {Tupitsyn},
  \citenamefont {Maltsev}, \citenamefont {Kozhedub}, \citenamefont {Bondarev},
  \citenamefont {Kozin}, \citenamefont {Ma}, \citenamefont {Plunien},
  \citenamefont {St\"ohlker}, \citenamefont {Tumakov},\ and\ \citenamefont
  {Zaytsev}}]{Popov_2020_How}%
  \BibitemOpen
  \bibfield  {author} {\bibinfo {author} {\bibfnamefont {R.~V.}\ \bibnamefont
  {Popov}}, \bibinfo {author} {\bibfnamefont {V.~M.}\ \bibnamefont {Shabaev}},
  \bibinfo {author} {\bibfnamefont {D.~A.}\ \bibnamefont {Telnov}}, \bibinfo
  {author} {\bibfnamefont {I.~I.}\ \bibnamefont {Tupitsyn}}, \bibinfo {author}
  {\bibfnamefont {I.~A.}\ \bibnamefont {Maltsev}}, \bibinfo {author}
  {\bibfnamefont {Y.~S.}\ \bibnamefont {Kozhedub}}, \bibinfo {author}
  {\bibfnamefont {A.~I.}\ \bibnamefont {Bondarev}}, \bibinfo {author}
  {\bibfnamefont {N.~V.}\ \bibnamefont {Kozin}}, \bibinfo {author}
  {\bibfnamefont {X.}~\bibnamefont {Ma}}, \bibinfo {author} {\bibfnamefont
  {G.}~\bibnamefont {Plunien}}, \bibinfo {author} {\bibfnamefont
  {T.}~\bibnamefont {St\"ohlker}}, \bibinfo {author} {\bibfnamefont {D.~A.}\
  \bibnamefont {Tumakov}},\ and\ \bibinfo {author} {\bibfnamefont {V.~A.}\
  \bibnamefont {Zaytsev}},\ }\bibfield  {title} {\bibinfo {title} {How to
  access {QED} at a supercritical {C}oulomb field},\ }\href
  {https://doi.org/10.1103/PhysRevD.102.076005} {\bibfield  {journal} {\bibinfo
   {journal} {Phys. Rev. D}\ }\textbf {\bibinfo {volume} {102}},\ \bibinfo
  {pages} {076005} (\bibinfo {year} {2020})}\BibitemShut {NoStop}%
\bibitem [{\citenamefont {Maltsev}\ \emph {et~al.}(2017)\citenamefont
  {Maltsev}, \citenamefont {Shabaev}, \citenamefont {Tupitsyn}, \citenamefont
  {Kozhedub}, \citenamefont {Plunien},\ and\ \citenamefont
  {Stöhlker}}]{Maltsev_2017_Pair}%
  \BibitemOpen
  \bibfield  {author} {\bibinfo {author} {\bibfnamefont {I.}~\bibnamefont
  {Maltsev}}, \bibinfo {author} {\bibfnamefont {V.}~\bibnamefont {Shabaev}},
  \bibinfo {author} {\bibfnamefont {I.}~\bibnamefont {Tupitsyn}}, \bibinfo
  {author} {\bibfnamefont {Y.}~\bibnamefont {Kozhedub}}, \bibinfo {author}
  {\bibfnamefont {G.}~\bibnamefont {Plunien}},\ and\ \bibinfo {author}
  {\bibfnamefont {T.}~\bibnamefont {Stöhlker}},\ }\bibfield  {title} {\bibinfo
  {title} {Pair production in low-energy collisions of uranium nuclei beyond
  the monopole approximation},\ }\href
  {https://doi.org/https://doi.org/10.1016/j.nimb.2017.05.005} {\bibfield
  {journal} {\bibinfo  {journal} {Nucl. Instrum. Methods Phys. Res., Sect. B}\
  }\textbf {\bibinfo {volume} {408}},\ \bibinfo {pages} {97} (\bibinfo {year}
  {2017})}\BibitemShut {NoStop}%
\bibitem [{\citenamefont {Maltsev}\ \emph {et~al.}(2018)\citenamefont
  {Maltsev}, \citenamefont {Shabaev}, \citenamefont {Popov}, \citenamefont
  {Kozhedub}, \citenamefont {Plunien}, \citenamefont {Ma},\ and\ \citenamefont
  {St\"ohlker}}]{Maltsev_2018_Electron}%
  \BibitemOpen
  \bibfield  {author} {\bibinfo {author} {\bibfnamefont {I.~A.}\ \bibnamefont
  {Maltsev}}, \bibinfo {author} {\bibfnamefont {V.~M.}\ \bibnamefont
  {Shabaev}}, \bibinfo {author} {\bibfnamefont {R.~V.}\ \bibnamefont {Popov}},
  \bibinfo {author} {\bibfnamefont {Y.~S.}\ \bibnamefont {Kozhedub}}, \bibinfo
  {author} {\bibfnamefont {G.}~\bibnamefont {Plunien}}, \bibinfo {author}
  {\bibfnamefont {X.}~\bibnamefont {Ma}},\ and\ \bibinfo {author}
  {\bibfnamefont {T.}~\bibnamefont {St\"ohlker}},\ }\bibfield  {title}
  {\bibinfo {title} {Electron-positron pair production in slow collisions of
  heavy nuclei beyond the monopole approximation},\ }\href
  {https://doi.org/10.1103/PhysRevA.98.062709} {\bibfield  {journal} {\bibinfo
  {journal} {Phys. Rev. A}\ }\textbf {\bibinfo {volume} {98}},\ \bibinfo
  {pages} {062709} (\bibinfo {year} {2018})}\BibitemShut {NoStop}%
\bibitem [{\citenamefont {Popov}\ \emph {et~al.}(2018)\citenamefont {Popov},
  \citenamefont {Bondarev}, \citenamefont {Kozhedub}, \citenamefont {Maltsev},
  \citenamefont {Shabaev}, \citenamefont {Tupitsyn}, \citenamefont {Ma},
  \citenamefont {Plunien},\ and\ \citenamefont
  {St{\"o}hlker}}]{Popov_2018_One}%
  \BibitemOpen
  \bibfield  {author} {\bibinfo {author} {\bibfnamefont {R.~V.}\ \bibnamefont
  {Popov}}, \bibinfo {author} {\bibfnamefont {A.~I.}\ \bibnamefont {Bondarev}},
  \bibinfo {author} {\bibfnamefont {Y.~S.}\ \bibnamefont {Kozhedub}}, \bibinfo
  {author} {\bibfnamefont {I.~A.}\ \bibnamefont {Maltsev}}, \bibinfo {author}
  {\bibfnamefont {V.~M.}\ \bibnamefont {Shabaev}}, \bibinfo {author}
  {\bibfnamefont {I.~I.}\ \bibnamefont {Tupitsyn}}, \bibinfo {author}
  {\bibfnamefont {X.}~\bibnamefont {Ma}}, \bibinfo {author} {\bibfnamefont
  {G.}~\bibnamefont {Plunien}},\ and\ \bibinfo {author} {\bibfnamefont
  {T.}~\bibnamefont {St{\"o}hlker}},\ }\bibfield  {title} {\bibinfo {title}
  {One-center calculations of the electron-positron pair creation in low-energy
  collisions of heavy bare nuclei},\ }\href
  {https://doi.org/10.1140/epjd/e2018-90056-4} {\bibfield  {journal} {\bibinfo
  {journal} {Eur. Phys. J. D}\ }\textbf {\bibinfo {volume} {72}},\ \bibinfo
  {pages} {115} (\bibinfo {year} {2018})}\BibitemShut {NoStop}%
\bibitem [{\citenamefont {Popov}\ \emph {et~al.}(2023)\citenamefont {Popov},
  \citenamefont {Shabaev}, \citenamefont {Maltsev}, \citenamefont {Telnov},
  \citenamefont {Dulaev},\ and\ \citenamefont
  {Tumakov}}]{Popov_2023_Spontaneous}%
  \BibitemOpen
  \bibfield  {author} {\bibinfo {author} {\bibfnamefont {R.~V.}\ \bibnamefont
  {Popov}}, \bibinfo {author} {\bibfnamefont {V.~M.}\ \bibnamefont {Shabaev}},
  \bibinfo {author} {\bibfnamefont {I.~A.}\ \bibnamefont {Maltsev}}, \bibinfo
  {author} {\bibfnamefont {D.~A.}\ \bibnamefont {Telnov}}, \bibinfo {author}
  {\bibfnamefont {N.~K.}\ \bibnamefont {Dulaev}},\ and\ \bibinfo {author}
  {\bibfnamefont {D.~A.}\ \bibnamefont {Tumakov}},\ }\bibfield  {title}
  {\bibinfo {title} {Spontaneous vacuum decay in low-energy collisions of heavy
  nuclei beyond the monopole approximation},\ }\href
  {https://doi.org/10.1103/PhysRevD.107.116014} {\bibfield  {journal} {\bibinfo
   {journal} {Phys. Rev. D}\ }\textbf {\bibinfo {volume} {107}},\ \bibinfo
  {pages} {116014} (\bibinfo {year} {2023})}\BibitemShut {NoStop}%
\bibitem [{\citenamefont {Dulaev}\ \emph {et~al.}(2024)\citenamefont {Dulaev},
  \citenamefont {Telnov}, \citenamefont {Shabaev}, \citenamefont {Kozhedub},
  \citenamefont {Maltsev}, \citenamefont {Popov},\ and\ \citenamefont
  {Tupitsyn}}]{Dulaev_2024_Angular}%
  \BibitemOpen
  \bibfield  {author} {\bibinfo {author} {\bibfnamefont {N.~K.}\ \bibnamefont
  {Dulaev}}, \bibinfo {author} {\bibfnamefont {D.~A.}\ \bibnamefont {Telnov}},
  \bibinfo {author} {\bibfnamefont {V.~M.}\ \bibnamefont {Shabaev}}, \bibinfo
  {author} {\bibfnamefont {Y.~S.}\ \bibnamefont {Kozhedub}}, \bibinfo {author}
  {\bibfnamefont {I.~A.}\ \bibnamefont {Maltsev}}, \bibinfo {author}
  {\bibfnamefont {R.~V.}\ \bibnamefont {Popov}},\ and\ \bibinfo {author}
  {\bibfnamefont {I.~I.}\ \bibnamefont {Tupitsyn}},\ }\bibfield  {title}
  {\bibinfo {title} {Angular and energy distributions of positrons created in
  subcritical and supercritical slow collisions of heavy nuclei},\ }\href
  {https://doi.org/10.1103/PhysRevD.109.036008} {\bibfield  {journal} {\bibinfo
   {journal} {Phys. Rev. D}\ }\textbf {\bibinfo {volume} {109}},\ \bibinfo
  {pages} {036008} (\bibinfo {year} {2024})}\BibitemShut {NoStop}%
\bibitem [{\citenamefont {Dulaev}\ \emph
  {et~al.}(2025{\natexlab{a}})\citenamefont {Dulaev}, \citenamefont {Telnov},
  \citenamefont {Shabaev}, \citenamefont {Kozhedub}, \citenamefont {Ma},
  \citenamefont {Maltsev}, \citenamefont {Popov},\ and\ \citenamefont
  {Tupitsyn}}]{Dulaev_Three_2025}%
  \BibitemOpen
  \bibfield  {author} {\bibinfo {author} {\bibfnamefont {N.~K.}\ \bibnamefont
  {Dulaev}}, \bibinfo {author} {\bibfnamefont {D.~A.}\ \bibnamefont {Telnov}},
  \bibinfo {author} {\bibfnamefont {V.~M.}\ \bibnamefont {Shabaev}}, \bibinfo
  {author} {\bibfnamefont {Y.~S.}\ \bibnamefont {Kozhedub}}, \bibinfo {author}
  {\bibfnamefont {X.}~\bibnamefont {Ma}}, \bibinfo {author} {\bibfnamefont
  {I.~A.}\ \bibnamefont {Maltsev}}, \bibinfo {author} {\bibfnamefont {R.~V.}\
  \bibnamefont {Popov}},\ and\ \bibinfo {author} {\bibfnamefont {I.~I.}\
  \bibnamefont {Tupitsyn}},\ }\bibfield  {title} {\bibinfo {title}
  {Three-dimensional calculations of positron creation in supercritical
  collisions of heavy nuclei},\ }\href
  {https://doi.org/10.1103/PhysRevD.111.016018} {\bibfield  {journal} {\bibinfo
   {journal} {Phys. Rev. D}\ }\textbf {\bibinfo {volume} {111}},\ \bibinfo
  {pages} {016018} (\bibinfo {year} {2025}{\natexlab{a}})}\BibitemShut
  {NoStop}%
\bibitem [{\citenamefont {Popov}\ \emph {et~al.}(2025)\citenamefont {Popov},
  \citenamefont {Shabaev}, \citenamefont {Maltsev}, \citenamefont {Telnov},
  \citenamefont {Dulaev}, \citenamefont {Mironov},\ and\ \citenamefont
  {Ryzhkov}}]{Popov_2025_Influence}%
  \BibitemOpen
  \bibfield  {author} {\bibinfo {author} {\bibfnamefont {R.~V.}\ \bibnamefont
  {Popov}}, \bibinfo {author} {\bibfnamefont {V.~M.}\ \bibnamefont {Shabaev}},
  \bibinfo {author} {\bibfnamefont {I.~A.}\ \bibnamefont {Maltsev}}, \bibinfo
  {author} {\bibfnamefont {D.~A.}\ \bibnamefont {Telnov}}, \bibinfo {author}
  {\bibfnamefont {N.~K.}\ \bibnamefont {Dulaev}}, \bibinfo {author}
  {\bibfnamefont {A.~D.}\ \bibnamefont {Mironov}},\ and\ \bibinfo {author}
  {\bibfnamefont {A.~M.}\ \bibnamefont {Ryzhkov}},\ }\bibfield  {title}
  {\bibinfo {title} {Influence of the internuclear axis rotation on the
  probabilities of pair production in collisions of heavy nuclei},\ }\href
  {https://doi.org/10.1134/S1547477124702479} {\bibfield  {journal} {\bibinfo
  {journal} {Physics of Particles and Nuclei Letters}\ }\textbf {\bibinfo
  {volume} {22}},\ \bibinfo {pages} {374} (\bibinfo {year} {2025})}\BibitemShut
  {NoStop}%
\bibitem [{\citenamefont {Dulaev}\ \emph
  {et~al.}(2025{\natexlab{b}})\citenamefont {Dulaev}, \citenamefont {Telnov},
  \citenamefont {Popov}, \citenamefont {Shabaev}, \citenamefont {Kozhedub},
  \citenamefont {Ma}, \citenamefont {Maltsev}, \citenamefont {Mironov},\ and\
  \citenamefont {Tupitsyn}}]{Dulaev_2025_The}%
  \BibitemOpen
  \bibfield  {author} {\bibinfo {author} {\bibfnamefont {N.}~\bibnamefont
  {Dulaev}}, \bibinfo {author} {\bibfnamefont {D.}~\bibnamefont {Telnov}},
  \bibinfo {author} {\bibfnamefont {R.}~\bibnamefont {Popov}}, \bibinfo
  {author} {\bibfnamefont {V.}~\bibnamefont {Shabaev}}, \bibinfo {author}
  {\bibfnamefont {Y.}~\bibnamefont {Kozhedub}}, \bibinfo {author}
  {\bibfnamefont {X.-W.}\ \bibnamefont {Ma}}, \bibinfo {author} {\bibfnamefont
  {I.}~\bibnamefont {Maltsev}}, \bibinfo {author} {\bibfnamefont
  {A.}~\bibnamefont {Mironov}},\ and\ \bibinfo {author} {\bibfnamefont
  {I.}~\bibnamefont {Tupitsyn}},\ }\bibfield  {title} {\bibinfo {title}
  {{{{STORI'24}}}: The influence of electron–electron interaction on pair
  production in supercritical collisions of highly charged ions},\ }\href
  {https://doi.org/10.1088/1674-1137/ade12b} {\bibfield  {journal} {\bibinfo
  {journal} {Chinese Physics C}\ }\textbf {\bibinfo {volume} {49}},\ \bibinfo
  {pages} {094105} (\bibinfo {year} {2025}{\natexlab{b}})}\BibitemShut
  {NoStop}%
\bibitem [{\citenamefont {Rafelski}\ and\ \citenamefont
  {M\"uller}(1976)}]{Rafelski_1976_Magnetic}%
  \BibitemOpen
  \bibfield  {author} {\bibinfo {author} {\bibfnamefont {J.}~\bibnamefont
  {Rafelski}}\ and\ \bibinfo {author} {\bibfnamefont {B.}~\bibnamefont
  {M\"uller}},\ }\bibfield  {title} {\bibinfo {title} {Magnetic splitting of
  quasimolecular electronic states in strong fields},\ }\href
  {https://doi.org/10.1103/PhysRevLett.36.517} {\bibfield  {journal} {\bibinfo
  {journal} {Phys. Rev. Lett.}\ }\textbf {\bibinfo {volume} {36}},\ \bibinfo
  {pages} {517} (\bibinfo {year} {1976})}\BibitemShut {NoStop}%
\bibitem [{\citenamefont {Greiner}\ \emph {et~al.}(1978)\citenamefont
  {Greiner}, \citenamefont {Müller},\ and\ \citenamefont
  {Soff}}]{Greiner_1978_Delta}%
  \BibitemOpen
  \bibfield  {author} {\bibinfo {author} {\bibfnamefont {W.}~\bibnamefont
  {Greiner}}, \bibinfo {author} {\bibfnamefont {B.}~\bibnamefont {Müller}},\
  and\ \bibinfo {author} {\bibfnamefont {G.}~\bibnamefont {Soff}},\ }\bibfield
  {title} {\bibinfo {title} {Delta electrons as a probe of strong magnetic
  fields in heavy ion collisions},\ }\href
  {https://doi.org/https://doi.org/10.1016/0375-9601(78)90426-7} {\bibfield
  {journal} {\bibinfo  {journal} {Phys. Lett. A}\ }\textbf {\bibinfo {volume}
  {69}},\ \bibinfo {pages} {27} (\bibinfo {year} {1978})}\BibitemShut {NoStop}%
\bibitem [{\citenamefont {Soff}\ \emph
  {et~al.}(1981{\natexlab{a}})\citenamefont {Soff}, \citenamefont {Reinhardt},\
  and\ \citenamefont {Greiner}}]{Soff_1981_Spin}%
  \BibitemOpen
  \bibfield  {author} {\bibinfo {author} {\bibfnamefont {G.}~\bibnamefont
  {Soff}}, \bibinfo {author} {\bibfnamefont {J.}~\bibnamefont {Reinhardt}},\
  and\ \bibinfo {author} {\bibfnamefont {W.}~\bibnamefont {Greiner}},\
  }\bibfield  {title} {\bibinfo {title} {Spin polarization of electrons induced
  by strong collisional magnetic fields},\ }\href
  {https://doi.org/10.1103/PhysRevA.23.701} {\bibfield  {journal} {\bibinfo
  {journal} {Phys. Rev. A}\ }\textbf {\bibinfo {volume} {23}},\ \bibinfo
  {pages} {701} (\bibinfo {year} {1981}{\natexlab{a}})}\BibitemShut {NoStop}%
\bibitem [{\citenamefont {Soff}\ \emph
  {et~al.}(1981{\natexlab{b}})\citenamefont {Soff}, \citenamefont {Reinhardt},\
  and\ \citenamefont {Greiner}}]{Soff_1s_1981}%
  \BibitemOpen
  \bibfield  {author} {\bibinfo {author} {\bibfnamefont {G.}~\bibnamefont
  {Soff}}, \bibinfo {author} {\bibfnamefont {J.}~\bibnamefont {Reinhardt}},\
  and\ \bibinfo {author} {\bibfnamefont {W.}~\bibnamefont {Greiner}},\
  }\bibfield  {title} {\bibinfo {title} {$1s\sigma$ ionization in high-energy
  collisions of very heavy ions},\ }\href
  {https://doi.org/https://doi.org/10.1016/0375-9601(81)90874-4} {\bibfield
  {journal} {\bibinfo  {journal} {Physics Letters A}\ }\textbf {\bibinfo
  {volume} {83}},\ \bibinfo {pages} {158} (\bibinfo {year}
  {1981}{\natexlab{b}})}\BibitemShut {NoStop}%
\bibitem [{\citenamefont {Soff}\ and\ \citenamefont
  {Reinhardt}(1988)}]{Soff_1988_Positron}%
  \BibitemOpen
  \bibfield  {author} {\bibinfo {author} {\bibfnamefont {G.}~\bibnamefont
  {Soff}}\ and\ \bibinfo {author} {\bibfnamefont {J.}~\bibnamefont
  {Reinhardt}},\ }\bibfield  {title} {\bibinfo {title} {Positron creation in
  heavy ion collisions the influence of the magnetic field},\ }\href
  {https://doi.org/https://doi.org/10.1016/0370-2693(88)90829-5} {\bibfield
  {journal} {\bibinfo  {journal} {Phys. Lett. B}\ }\textbf {\bibinfo {volume}
  {211}},\ \bibinfo {pages} {179} (\bibinfo {year} {1988})}\BibitemShut
  {NoStop}%
\bibitem [{\citenamefont {Elton}(1961)}]{Elton_1961_Nuclear}%
  \BibitemOpen
  \bibfield  {author} {\bibinfo {author} {\bibfnamefont {L.~R.~B.}\
  \bibnamefont {Elton}},\ }\href@noop {} {\emph {\bibinfo {title} {{N}uclear
  {S}izes}}}\ (\bibinfo  {publisher} {Oxford University Press},\ \bibinfo
  {year} {1961})\BibitemShut {NoStop}%
\bibitem [{\citenamefont {Angeli}\ and\ \citenamefont
  {Marinova}(2013)}]{Angeli_2013_Table}%
  \BibitemOpen
  \bibfield  {author} {\bibinfo {author} {\bibfnamefont {I.}~\bibnamefont
  {Angeli}}\ and\ \bibinfo {author} {\bibfnamefont {K.}~\bibnamefont
  {Marinova}},\ }\bibfield  {title} {\bibinfo {title} {Table of experimental
  nuclear ground state charge radii: An update},\ }\href
  {https://doi.org/https://doi.org/10.1016/j.adt.2011.12.006} {\bibfield
  {journal} {\bibinfo  {journal} {At. Data Nucl. Data Tables}\ }\textbf
  {\bibinfo {volume} {99}},\ \bibinfo {pages} {69} (\bibinfo {year}
  {2013})}\BibitemShut {NoStop}%
\bibitem [{\citenamefont {M{\"u}ller}\ \emph
  {et~al.}(1972{\natexlab{c}})\citenamefont {M{\"u}ller}, \citenamefont
  {Rafelski},\ and\ \citenamefont {Greiner}}]{Muller_1972_Auto}%
  \BibitemOpen
  \bibfield  {author} {\bibinfo {author} {\bibfnamefont {B.}~\bibnamefont
  {M{\"u}ller}}, \bibinfo {author} {\bibfnamefont {J.}~\bibnamefont
  {Rafelski}},\ and\ \bibinfo {author} {\bibfnamefont {W.}~\bibnamefont
  {Greiner}},\ }\bibfield  {title} {\bibinfo {title} {Auto-ionization of
  positrons in heavy ion collisions},\ }\href
  {https://doi.org/10.1007/BF01401203} {\bibfield  {journal} {\bibinfo
  {journal} {Z. Phys.}\ }\textbf {\bibinfo {volume} {257}},\ \bibinfo {pages}
  {183} (\bibinfo {year} {1972}{\natexlab{c}})}\BibitemShut {NoStop}%
\bibitem [{\citenamefont {Heinz}\ \emph {et~al.}(1981)\citenamefont {Heinz},
  \citenamefont {Greiner},\ and\ \citenamefont
  {M\"uller}}]{Heinz_1981_Electron}%
  \BibitemOpen
  \bibfield  {author} {\bibinfo {author} {\bibfnamefont {U.}~\bibnamefont
  {Heinz}}, \bibinfo {author} {\bibfnamefont {W.}~\bibnamefont {Greiner}},\
  and\ \bibinfo {author} {\bibfnamefont {B.}~\bibnamefont {M\"uller}},\
  }\bibfield  {title} {\bibinfo {title} {Electron-translation effects in
  heavy-ion scattering},\ }\href {https://doi.org/10.1103/PhysRevA.23.562}
  {\bibfield  {journal} {\bibinfo  {journal} {Phys. Rev. A}\ }\textbf {\bibinfo
  {volume} {23}},\ \bibinfo {pages} {562} (\bibinfo {year} {1981})}\BibitemShut
  {NoStop}%
\bibitem [{\citenamefont {Jackson}(1999)}]{jackson1999}%
  \BibitemOpen
  \bibfield  {author} {\bibinfo {author} {\bibfnamefont {J.~D.}\ \bibnamefont
  {Jackson}},\ }\href@noop {} {\emph {\bibinfo {title} {Classical
  Electrodynamics}}},\ \bibinfo {edition} {3rd}\ ed.\ (\bibinfo  {publisher}
  {Wiley},\ \bibinfo {address} {New York},\ \bibinfo {year} {1999})\BibitemShut
  {NoStop}%
\bibitem [{\citenamefont {Landau}\ and\ \citenamefont
  {Lifshitz}(1975)}]{landau1980}%
  \BibitemOpen
  \bibfield  {author} {\bibinfo {author} {\bibfnamefont {L.~D.}\ \bibnamefont
  {Landau}}\ and\ \bibinfo {author} {\bibfnamefont {E.~M.}\ \bibnamefont
  {Lifshitz}},\ }\href@noop {} {\emph {\bibinfo {title} {Classical Theory of
  Fields}}},\ \bibinfo {edition} {4th}\ ed.\ (\bibinfo  {publisher}
  {Butterworth Heinemann},\ \bibinfo {address} {Amsterdam},\ \bibinfo {year}
  {1975})\BibitemShut {NoStop}%
\bibitem [{\citenamefont {M{\"u}ller}\ and\ \citenamefont
  {Greiner}(1976)}]{Muller_1976_The}%
  \BibitemOpen
  \bibfield  {author} {\bibinfo {author} {\bibfnamefont {B.}~\bibnamefont
  {M{\"u}ller}}\ and\ \bibinfo {author} {\bibfnamefont {W.}~\bibnamefont
  {Greiner}},\ }\bibfield  {title} {\bibinfo {title} {The two centre {D}irac
  equation},\ }\href {https://doi.org/doi:10.1515/zna-1976-0102} {\bibfield
  {journal} {\bibinfo  {journal} {Z. Naturforsch. A}\ }\textbf {\bibinfo
  {volume} {31}},\ \bibinfo {pages} {1} (\bibinfo {year} {1976})}\BibitemShut
  {NoStop}%
\bibitem [{\citenamefont {Telnov}\ and\ \citenamefont
  {Chu}(2007)}]{Telnov_2007_Ab}%
  \BibitemOpen
  \bibfield  {author} {\bibinfo {author} {\bibfnamefont {D.~A.}\ \bibnamefont
  {Telnov}}\ and\ \bibinfo {author} {\bibfnamefont {S.-I.}\ \bibnamefont
  {Chu}},\ }\bibfield  {title} {\bibinfo {title} {\textit{Ab initio} study of
  the orientation effects in multiphoton ionization and high-order harmonic
  generation from the ground and excited electronic states of
  $\mathrm{H}_{2}^{+}$},\ }\href {https://doi.org/10.1103/PhysRevA.76.043412}
  {\bibfield  {journal} {\bibinfo  {journal} {Phys. Rev. A}\ }\textbf {\bibinfo
  {volume} {76}},\ \bibinfo {pages} {043412} (\bibinfo {year}
  {2007})}\BibitemShut {NoStop}%
\bibitem [{\citenamefont {Telnov}\ and\ \citenamefont
  {Chu}(2009)}]{Telnov_2009_Effects}%
  \BibitemOpen
  \bibfield  {author} {\bibinfo {author} {\bibfnamefont {D.~A.}\ \bibnamefont
  {Telnov}}\ and\ \bibinfo {author} {\bibfnamefont {S.-I.}\ \bibnamefont
  {Chu}},\ }\bibfield  {title} {\bibinfo {title} {Effects of multiple
  electronic shells on strong-field multiphoton ionization and high-order
  harmonic generation of diatomic molecules with arbitrary orientation: An
  all-electron time-dependent density-functional approach},\ }\href
  {https://doi.org/10.1103/PhysRevA.80.043412} {\bibfield  {journal} {\bibinfo
  {journal} {Phys. Rev. A}\ }\textbf {\bibinfo {volume} {80}},\ \bibinfo
  {pages} {043412} (\bibinfo {year} {2009})}\BibitemShut {NoStop}%
\bibitem [{\citenamefont {Telnov}\ \emph {et~al.}(2018)\citenamefont {Telnov},
  \citenamefont {Krapivin}, \citenamefont {Heslar},\ and\ \citenamefont
  {Chu}}]{Telnov_2018_Multiphoton}%
  \BibitemOpen
  \bibfield  {author} {\bibinfo {author} {\bibfnamefont {D.~A.}\ \bibnamefont
  {Telnov}}, \bibinfo {author} {\bibfnamefont {D.~A.}\ \bibnamefont
  {Krapivin}}, \bibinfo {author} {\bibfnamefont {J.}~\bibnamefont {Heslar}},\
  and\ \bibinfo {author} {\bibfnamefont {S.-I.}\ \bibnamefont {Chu}},\
  }\bibfield  {title} {\bibinfo {title} {Multiphoton ionization of one-electron
  relativistic diatomic quasimolecules in strong laser fields},\ }\href
  {https://doi.org/10.1021/acs.jpca.8b07463} {\bibfield  {journal} {\bibinfo
  {journal} {J. Phys. Chem. A}\ }\textbf {\bibinfo {volume} {122}},\ \bibinfo
  {pages} {8026} (\bibinfo {year} {2018})}\BibitemShut {NoStop}%
\bibitem [{\citenamefont {Telnov}\ and\ \citenamefont
  {Chu}(2021)}]{Telnov_2021_Hyd}%
  \BibitemOpen
  \bibfield  {author} {\bibinfo {author} {\bibfnamefont {D.~A.}\ \bibnamefont
  {Telnov}}\ and\ \bibinfo {author} {\bibfnamefont {S.-I.}\ \bibnamefont
  {Chu}},\ }\bibfield  {title} {\bibinfo {title} {Relativistic ionization
  probabilities and photoelectron distributions of hydrogenlike ions in
  superstrong electromagnetic fields},\ }\href
  {https://doi.org/10.1103/PhysRevA.104.023111} {\bibfield  {journal} {\bibinfo
   {journal} {Phys. Rev. A}\ }\textbf {\bibinfo {volume} {104}},\ \bibinfo
  {pages} {023111} (\bibinfo {year} {2021})}\BibitemShut {NoStop}%
\bibitem [{\citenamefont {Crank}\ and\ \citenamefont
  {Nicolson}(1947)}]{Crank_1947_A}%
  \BibitemOpen
  \bibfield  {author} {\bibinfo {author} {\bibfnamefont {J.}~\bibnamefont
  {Crank}}\ and\ \bibinfo {author} {\bibfnamefont {P.}~\bibnamefont
  {Nicolson}},\ }\bibfield  {title} {\bibinfo {title} {A practical method for
  numerical evaluation of solutions of partial differential equations of the
  heat-conduction type},\ }\href {https://doi.org/10.1017/S0305004100023197}
  {\bibfield  {journal} {\bibinfo  {journal} {Math. Proc. Cambridge Phil.
  Soc.}\ }\textbf {\bibinfo {volume} {43}},\ \bibinfo {pages} {50–67}
  (\bibinfo {year} {1947})}\BibitemShut {NoStop}%
\bibitem [{\citenamefont {Berestetskii}\ \emph {et~al.}(1982)\citenamefont
  {Berestetskii}, \citenamefont {Lifshitz},\ and\ \citenamefont
  {Pitaevskii}}]{bere1982}%
  \BibitemOpen
  \bibfield  {author} {\bibinfo {author} {\bibfnamefont {V.~B.}\ \bibnamefont
  {Berestetskii}}, \bibinfo {author} {\bibfnamefont {E.~M.}\ \bibnamefont
  {Lifshitz}},\ and\ \bibinfo {author} {\bibfnamefont {L.~P.}\ \bibnamefont
  {Pitaevskii}},\ }\href {https://doi.org/10.1016/C2009-0-24486-2} {\emph
  {\bibinfo {title} {Quantum Electrodynamics}}},\ \bibinfo {edition} {2nd}\
  ed.\ (\bibinfo  {publisher} {Pergamon Press},\ \bibinfo {address} {Oxford},\
  \bibinfo {year} {1982})\BibitemShut {NoStop}%
\bibitem [{\citenamefont {Galitski}\ \emph {et~al.}(2013)\citenamefont
  {Galitski}, \citenamefont {Karnakov}, \citenamefont {Kogan},\ and\
  \citenamefont {Galitski~Jr}}]{Galitski_2013_Quantum}%
  \BibitemOpen
  \bibfield  {author} {\bibinfo {author} {\bibfnamefont {V.}~\bibnamefont
  {Galitski}}, \bibinfo {author} {\bibfnamefont {B.}~\bibnamefont {Karnakov}},
  \bibinfo {author} {\bibfnamefont {V.}~\bibnamefont {Kogan}},\ and\ \bibinfo
  {author} {\bibfnamefont {V.}~\bibnamefont {Galitski~Jr}},\ }\href
  {https://doi.org/10.1093/acprof:oso/9780199232710.001.0001} {\emph {\bibinfo
  {title} {Exploring Quantum Mechanics: A Collection of 700+ Solved Problems
  for Students, Lecturers, and Researchers}}}\ (\bibinfo  {publisher} {Oxford
  University Press},\ \bibinfo {address} {Oxford},\ \bibinfo {year}
  {2013})\BibitemShut {NoStop}%
\bibitem [{rsc()}]{rscf}%
  \BibitemOpen
  \href@noop {} {}\bibinfo {howpublished}
  {\url{https://rscf.ru/project/22-62-00004/}}\BibitemShut {NoStop}%
\bibitem [{\citenamefont {Anikina}\ \emph {et~al.}(2025)\citenamefont
  {Anikina}, \citenamefont {Belyakov}, \citenamefont {Bezhanyan}, \citenamefont
  {Kirakosyan}, \citenamefont {Kokorev}, \citenamefont {Lyubimova},
  \citenamefont {Matveev}, \citenamefont {Podgainy}, \citenamefont {Rahmonova},
  \citenamefont {Shadmehri}, \citenamefont {Streltsova}, \citenamefont
  {Torosyan}, \citenamefont {Vala},\ and\ \citenamefont {Zuev}}]{HybriLIT}%
  \BibitemOpen
  \bibfield  {author} {\bibinfo {author} {\bibfnamefont {A.}~\bibnamefont
  {Anikina}}, \bibinfo {author} {\bibfnamefont {D.}~\bibnamefont {Belyakov}},
  \bibinfo {author} {\bibfnamefont {T.}~\bibnamefont {Bezhanyan}}, \bibinfo
  {author} {\bibfnamefont {M.}~\bibnamefont {Kirakosyan}}, \bibinfo {author}
  {\bibfnamefont {A.}~\bibnamefont {Kokorev}}, \bibinfo {author} {\bibfnamefont
  {M.}~\bibnamefont {Lyubimova}}, \bibinfo {author} {\bibfnamefont
  {M.}~\bibnamefont {Matveev}}, \bibinfo {author} {\bibfnamefont
  {D.}~\bibnamefont {Podgainy}}, \bibinfo {author} {\bibfnamefont
  {A.}~\bibnamefont {Rahmonova}}, \bibinfo {author} {\bibfnamefont
  {S.}~\bibnamefont {Shadmehri}}, \bibinfo {author} {\bibfnamefont
  {O.}~\bibnamefont {Streltsova}}, \bibinfo {author} {\bibfnamefont
  {S.}~\bibnamefont {Torosyan}}, \bibinfo {author} {\bibfnamefont
  {M.}~\bibnamefont {Vala}},\ and\ \bibinfo {author} {\bibfnamefont
  {M.}~\bibnamefont {Zuev}},\ }\bibfield  {title} {\bibinfo {title} {Structure
  and features of the software and information environment of the {HybriLIT}
  heterogeneous platform},\ }in\ \href
  {https://doi.org/https://doi.org/10.1007/978-3-031-80853-1_33} {\emph
  {\bibinfo {booktitle} {Distributed Computer and Communication Networks}}},\
  \bibinfo {editor} {edited by\ \bibinfo {editor} {\bibfnamefont {V.~M.}\
  \bibnamefont {Vishnevsky}}, \bibinfo {editor} {\bibfnamefont {K.~E.}\
  \bibnamefont {Samouylov}},\ and\ \bibinfo {editor} {\bibfnamefont {D.~V.}\
  \bibnamefont {Kozyrev}}}\ (\bibinfo  {publisher} {Springer Nature
  Switzerland},\ \bibinfo {address} {Cham},\ \bibinfo {year} {2025})\ pp.\
  \bibinfo {pages} {444--457}\BibitemShut {NoStop}%
\end{thebibliography}%

\end{document}